\documentclass[onecolumn]{aastex631}
\pdfoutput=1 
\usepackage{amsmath,amstext}
\usepackage[T1]{fontenc}
\usepackage[figure,figure*]{hypcap}
\usepackage{CJKutf8}
\usepackage{longtable}
\usepackage{comment}
\usepackage{hyperref}
\usepackage{rotating}
\shorttitle{L\textit{i}DO: The Large \textit{i}nclination Distant Objects Survey}
\shortauthors{Alexandersen et al.}

\usepackage{xspace}
\newcommand{\lido}{L\textit{i}DO\xspace}

\graphicspath{{./}{images/}{images/Efficiency_curves}{images/Triplets}{images/Trombone_plots}}  

\begin{document}


\title{L\textit{i}DO: The Large \textit{i}nclination Distant Objects Survey}

\author[0000-0003-4143-8589]{Mike~Alexandersen} 
\affiliation{Center for Astrophysics | Harvard \& Smithsonian; 60 Garden Street, Cambridge, MA, 02138, USA}

\author[0000-0003-4797-5262]{Rosemary~E. Pike} 
\affiliation{Center for Astrophysics | Harvard \& Smithsonian; 60 Garden Street, Cambridge, MA, 02138, USA}

\author[0000-0001-7244-6069]{Ying-Tung Chen (\begin{CJK*}{UTF8}{bkai}陳英同\end{CJK*})} 
\affiliation{Institute of Astronomy and Astrophysics, Academia Sinica; 11F of AS/NTU Astronomy-Mathematics Building, No. 1 Roosevelt Rd., Sec. 4, Taipei 106319, Taiwan}

\author[0000-0001-5368-386X]{Samantha~M. Lawler} 
\affiliation{Campion College and the Department of Physics, University of Regina, Regina, SK S4S 0A2, Canada}

\author[0009-0007-9017-4010]{Mark~Comte} 
\affiliation{Campion College and the Department of Physics, University of Regina, Regina, SK S4S 0A2, Canada}

\author[0000-0002-9179-8323]{Lowell Peltier} 
\affiliation{Department of Physics and Astronomy, University of Victoria, Victoria, BC, Canada}

\author[0000-0001-7032-5255]{J.~J. Kavelaars} 
\affiliation{Herzberg Astronomy and Astrophysics Research Centre, National Research Council of Canada, 5071 West Saanich Rd, Victoria, BC V9E 2E7, Canada}
\affiliation{Department of Physics and Astronomy, University of Victoria, Victoria, BC, Canada}

\author[0000-0001-8736-236X]{Kathryn Volk} 
\affiliation{Planetary Science Institute, 1700 East Fort Lowell, Suite 106, Tucson, AZ 85719, USA}

\author[0009-0003-4483-4314]{Cameron Semenchuck} 
\affiliation{Department of Physics and Astronomy, University of Victoria, Victoria, BC, Canada}
\affiliation{Campion College and the Department of Physics, University of Regina, Regina, SK S4S 0A2, Canada}


\author[0009-0004-7149-5212]{Cameron Collyer} 
\affiliation{Florida Space Institute, University of Central Florida, 12354 Research Parkway, Orlando, FL 32826, USA}
\affiliation{Center for Astrophysics | Harvard \& Smithsonian; 60 Garden Street, Cambridge, MA, 02138, USA}

\author[0009-0004-3070-2772]{Jaidyn Draper} 
\affiliation{Campion College and the Department of Physics, University of Regina, Regina, SK S4S 0A2, Canada}

\author[0000-0001-8221-8406]{Stephen~Gwyn} 
\affiliation{Herzberg Astronomy and Astrophysics Research Centre, National Research Council of Canada, 5071 West Saanich Rd, Victoria, BC V9E 2E7, Canada}

\author[0000-0002-3376-7297]{Arcelia {Hermosillo Ruiz}} 
\affiliation{Department of Physics and Astronomy, University of Exeter, Stocker Road, Exeter EX4 4QL, UK}

\author[0000-0002-1139-4880]{Matthew~J.~Holman} 
\affiliation{Center for Astrophysics | Harvard \& Smithsonian; 60 Garden Street, Cambridge, MA, 02138, USA}

\author[0009-0007-2474-250X]{Katrina Kianpoor} 
\affiliation{Department of Astronomy and Astrophysics, University of California Santa Cruz, 95064, USA}

\author[0009-0002-4114-7861]{Cameron~J.~Morgan} 
\affiliation{Whiting School of Engineering, Johns Hopkins University, Baltimore, MD 21218, USA}
\affiliation{Center for Astrophysics | Harvard \& Smithsonian; 60 Garden Street, Cambridge, MA, 02138, USA}

\author[0000-0001-5061-0462]{Ruth Murray-Clay} 
\affiliation{Department of Astronomy and Astrophysics, University of California Santa Cruz, 95064, USA}

\author[0009-0004-4075-5691]{Jonathan Pond} 
\affiliation{Center for Astrophysics | Harvard \& Smithsonian; 60 Garden Street, Cambridge, MA, 02138, USA}

\author[0000-0002-1788-870X]{Benjamin Proudfoot} 
\affiliation{Florida Space Institute, University of Central Florida, 12354 Research Parkway, Orlando, FL 32826, USA}

\author[0009-0005-4944-9823]{AbdulRahman Rabaa} 
\affiliation{Campion College and the Department of Physics, University of Regina, Regina, SK S4S 0A2, Canada}

\author[0000-0003-1080-9770]{Darin Ragozzine} 
\affiliation{Brigham Young University Department of Physics \& Astronomy, N283 ESC, Brigham Young University, Provo, UT 84602, USA}

\author[0009-0008-2731-6618]{Emily Rieger} 
\affiliation{Campion College and the Department of Physics, University of Regina, Regina, SK S4S 0A2, Canada}

\author[0009-0007-3731-9073]{Mriana Yadkoo} 
\affiliation{Campion College and the Department of Physics, University of Regina, Regina, SK S4S 0A2, Canada}


\begin{abstract}

Previous trans-Neptunian objects (TNOs) discovery surveys targeted the ecliptic plane, where the overall highest density is seen. 
High-inclination objects spend little time near the plane, and are thus under-represented in previous TNO surveys. 
The Large \textit{i}nclination Distant Objects (\lido) survey targets these high-inclination populations in 34~sq.deg., located in two blocks 14-21$\degr$ north and south of the ecliptic plane, finding 148 objects, all with $i>14\degr$ due to survey design. 
A careful characterization of detection efficiency and other survey parameters results in a ``characterized sample'' of 125 objects detected within the survey's magnitude and rate limits, thus suitable for model comparison via survey simulations.
\lido detected 62 TNOs classified as resonant, including 21 Plutinos, of which 12 are securely Kozai resonators (always coming to pericenter off-ecliptic), as well as 9 objects in the 5:2 resonance, and several mixed mode inclination-eccentricity resonators. 
No objects beyond 63~au were detected, despite sensitivity to objects as far as 150~au. 
We found that the $i>14\degr$ orbital distribution of the previous model from OSSOS was rejectable with the inclusion of the 38 characterized hot classical TNOs from \lido.
We constructed a proof-of-concept orbital distribution model of hot classical TNOs based on long-term stability sculpting, which is statistically acceptable for the \lido and OSSOS samples, while producing population estimates (21,000$^{+3000}_{-3000}$ hot classicals with $H_r<8.3$) consistent with OSSOS estimates.
The discovery of 125 characterized mid-to-high inclination objects in a variety of orbital classifications provides useful constraints for population modeling and the migration history of the Solar System.

\end{abstract}

\section{Introduction}
\label{sec:intro}

The outer Solar System is populated by a large number of small, icy bodies.
The orbital distribution of these objects is a direct result of the formation and evolution of the Solar System \citep[e.g.,][]{Malhotra1993, Tsiganis2005, Dones2015, Nesvorny2016b, Gladman2021, Lawler2026}.
Many surveys have been carried out to discover and track thousands of TNOs (5,560 in the Minor Planet Center\footnote{https://www.minorplanetcenter.net/}, MPC, as of November 22, 2025). There are some significant biases in this sample, including a preference for observing near the ecliptic plane.
These surveys also likely detected additional objects they did not track and did not submit to the MPC, especially in the case of single night detections.
While all TNOs will pass through the ecliptic, the higher-inclination objects spend more time out of the ecliptic plane and are more efficiently discovered in off-ecliptic surveys.

Previous TNO discovery efforts have used a variety of observational techniques.
The Canada-France Ecliptic Plane survey \citep[CFEPS,][]{Petit2011}, the Alexandersen survey \citep{Alexandersen2016}, the CFEPS High Latitude extension \citep[HiLat,][]{Petit2017}, and the Outer Solar System Origins survey \citep[OSSOS,][]{Bannister2018} were all TNO discovery surveys by design and used three images taken in a single night for object discovery, followed by dedicated tracking observations.
Except for the HiLat survey, which had a depth of $m_R$ = 22.4–24.3 for the various pointings, and the smaller of the two Alexandersen survey blocks, these surveys were ecliptic focused.
The ecliptic focus of these surveys, hereafter referred to collectively as OSSOS++, resulted in a high number of total detections (1,142 characterized detections), but also biased those detections toward populations with lower average inclinations.
The Dark Energy Survey (DES) utilized a completely different survey cadence to link and discover 814 TNOs \citep{Bernardinelli2022} in a 5000 sq.deg. survey footprint that had the bulk of its on-sky area $\gtrsim$20 degrees off the ecliptic \citep{Bernardinelli2020}.
The DES catalog is described in \citet{Bernardinelli2022} as having a 50\% completeness limit of $m_r\simeq23.77$, and includes a large number of moderate to high inclination objects (504 with $i>14\degr$, 62\% of the total sample, according to their Table 3).
That work found that the CFEPS L7 model \citep{Petit2011} had to be modified in order to fit the inclination distribution of the DES detections.
A significant fraction ($\sim19\%$) of the DES detections are actually high inclination ($i>30\degr$), too hot for many of the excited TNO populations such as mean motion resonances near the classical belt.
The DECam Ecliptic Exploration Project \citep[DEEP][]{Trujillo2024, Bernardinelli2024, Smotherman2024} has 4 fields near the invariable and ecliptic planes.
The first DEEP multi-opposition discoveries include 110 objects with $m_{VR}\le26.0$, including 15 with $i>14\degr$ \citep{Smotherman2024}.
Other TNO discovery surveys have focused on particular populations, including distant objects \citep[e.g.][]{Sheppard2019} and Neptunian Trojans \citep[e.g.][]{Sheppard2010}.
There have also been other survey efforts, many of which predate the results discussed here, and have contributed a wide range of TNO discoveries to the MPC.
While previous surveys have found many mid-to-high inclination TNOs, this population is under-sampled relative to the low-$i$ population.
The OSSOS++ surveys found 280 objects with $i>14\degr$, with 112 from OSSOS.

The large-$i$ objects include several sub-populations, which are important for our understanding of the outer Solar System.
In particular, the distant resonators, large-pericenter ($q$) objects, objects in Kozai resonance \citep{Kozai1962,Lidov1962}, and Haumea family members are all large-$i$ objects.
Several of these subgroups have additional detection biases because of their orbital dynamics.
Considering these biases during survey design is critical for studying these populations.

\textbf{Distant Resonators:} The known distant $n$:1 and $n$:2 mean motion resonators, as well as the higher order resonances beyond the 5:2 at $\sim$55~au, are dynamically excited and include many large-$i$ and large-$e$ objects.
This is consistent with these objects being scattered from significantly sunward of their current location into the distant Solar System \citep{Nesvorny2016a, Nesvorny2016b,Kaib2016,Pike2017-DetailsResonantStructureNiceModelKuiperBelt-Predictions}.
These objects are also likely resonant sticking TNOs, so they are part of a scattering/transient population which extends their stability lifetimes by `sticking' temporarily (tens to hundreds of Myrs) in resonance \citep{Lykawka2007,Yu2018}.
In addition to having their inclinations dramatically increased during emplacement, the orbits of these objects can also evolve in resonance and increase their inclinations \citep{Gallardo2012, Lei2022}.
Many migration simulation end states show the distant scattered objects as nearly all large-$i$ and large-$e$ \citep{Kaib2016}.
These distant mean motion resonators will have \textit{some} longitudinal bias relative to Neptune, however, the temporary nature of the resonance indicates higher libration amplitudes are likely, and thus the importance of the libration centers is reduced compared to deeply captured resonators.

\textbf{Kozai Resonators:}  These objects have arguments of pericenter that librate within a confined range around $90\degr$ or $270\degr$, as well as evolving inclinations and eccentricities; however, their average inclinations are still typically quite high \citep{Gomes2003, Gallardo2012}.
This behavior can occur inside and outside of mean motion resonances \citep{Kozai1962,Lidov1962, Kozai1985,Gallardo2006, Gallardo2012, Lei2022}.
Objects of particular interest for this work are those displaying Kozai\footnote{
This particular secular resonance is sometimes called the ``von Zeipel-Lidov-Kozai resonance'' \citep{Ito2019}, or more properly a ``periodic orbit of the third kind'' \citep{Malhotra1997} or ``$g$ libration'' \citep{Malhotra2023}, but we will refer to it throughout this work as ``Kozai'' for brevity.} libration within the 3:2 and 5:2 mean motion resonances of Neptune.
These Kozai resonators within mean motion resonances such as the 3:2 are typically more long-term stable than non-Kozai resonators \citep{Balaji2023}.
Models of the inclination of the Kozai Plutinos are limited \citep[see][for an example]{Alexandersen2016}, and often assume many of the model parameters are the same as the non-Kozai Plutinos due to the small sample size.
Splitting both the 5:2 and 3:2 resonances into Kozai and non-Kozai components is likely also important for understanding the capture and evolutionary history of these populations.
The Plutinos are particularly sensitive to the migration history of Neptune \citep{Volk2016}.
Kozai TNOs within the 3:2 and 5:2 mean motion resonances (MMR) have smaller libration amplitudes than their non-Kozai counterparts \citep{Lawler2025}, and thus particular longitudes of pericenter with respect to Neptune due to their mean motion resonance.
Critically, these objects also have pericenter biases as a result of their Kozai libration, which ensures that they reach pericenter at latitudes numerically similar to their (usually high) inclinations, never close to the ecliptic.
This ensures that ecliptic surveys which target the non-Kozai 3:2 and 5:2 libration centers are \textit{significantly} less sensitive to Kozai objects in the same MMR, and measuring the relative intrinsic fraction of these populations using only ecliptic surveys would be challenging, \citep[e.g][]{Lawler2013}.

\textbf{Large-$q$ objects and Extreme TNOs:} Similar to the distant resonators, these objects had their $e$ and $i$ values pumped up during emplacement \citep[e.g.][]{Gomes2003, Brasser2006, Kaib2008, Brasser2013, Brasil2014, Lawler2019}.
In addition, these large-$q$ TNOs may have experienced Kozai oscillations during or after emplacement, which resulted in an even higher inclination and a slightly reduced eccentricity (and thus a higher $q$>40) \citep{Brasil2014, Kaib2016}.
The known extreme TNOs also have large-$i$  \citep[e.g.][]{Becker2018, Sheppard2019} and are thus more likely to be found off-ecliptic \citep[e.g.][]{Petit2017}.

\textbf{Haumea family members:} The Haumea family includes a group of objects with a narrow range of $a$-$e$-$i$ values, as well as additional members which have moved farther due to the effects of mean motion resonances.
The inclination range of 25-30$\degr$ includes all objects which have low $\delta$v values relative to the family center \citep{Ragozzine2007}, not corrected for any mean motion resonance interaction.

The \lido survey was designed to fill the gap in TNO discoveries in the large-inclination range, $15\degr<i<30\degr$, and search for the objects of particular interest.
We discuss the design and execution of the \lido survey in Section \ref{sec:design}.
The data analysis, TNO discovery process, orbital classifications, and survey characterization are described in Section \ref{sec:discoveries}.
We present our analysis of the orbital distribution of the hot classical TNOs in Section \ref{sec:science}.
We discuss the survey results in Section \ref{sec:Discussion} and summarize the \lido survey in Section \ref{sec:conclusions}.

\section{Survey Design and Execution}
\label{sec:design}

The \lido survey was designed to optimize for discovery of objects with large inclinations, as a complementary survey to the previous OSSOS++ survey efforts.
A similar survey design to OSSOS was used, with some minor changes to increase the discovery depth.
The Canada-France Hawaii Telescope (CFHT) MegaCam instrument \citep{Boulade1998, Boulade2003} was used as for all the OSSOS++ surveys, but with the wide $w$-band filter (also called $gri$), which reaches the same SNR level in about 60\% of the time of $r$-band; this allowed us to reach 0.28 magnitudes deeper in a fixed exposure time than conducting the survey in $r$-band as was done for most of OSSOS++.
This additional bandwidth is ideal, since exposure time is limited due to object motion. 
A slightly shorter discovery sequence was used than in the previous surveys, with 72-minute arcs for the discovery set; this was done by reducing the number of fields per repeat cycle, not by decreasing the exposure time.
This shorter sequence increases likelihood of observing conditions being ideal and stable over the entire sequence, while maintaining the needed sensitivity to the target populations, which are near the classical belt (3:2 and 5:2 resonators), and as far out as $\sim150$ au, well beyond the most distant likely detection.

\begin{figure}[tbp]\label{fig:blocks}
\centering
  \includegraphics[width=0.7\textwidth]{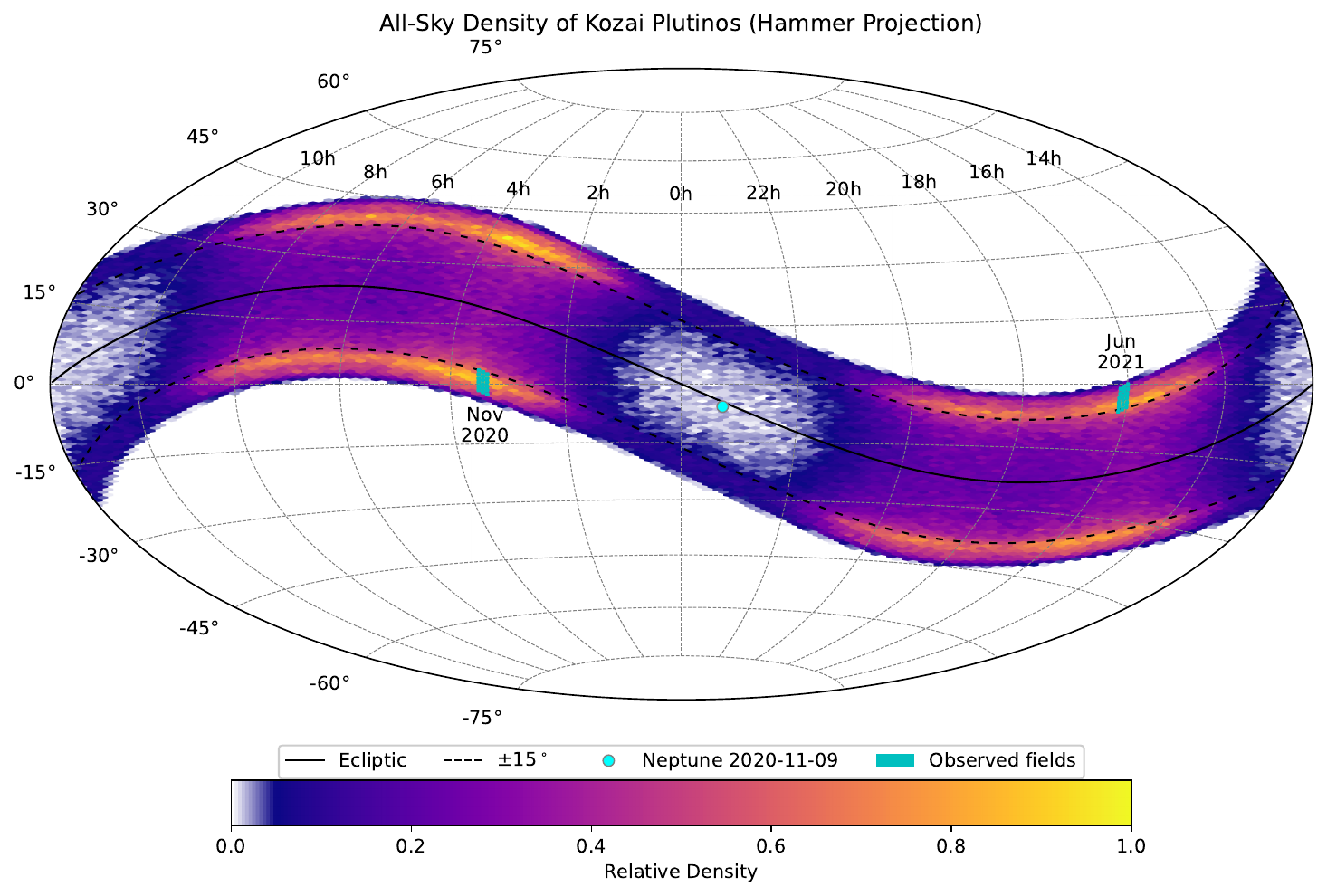}
  \caption{
  The \lido discovery blocks superimposed over a model of the detectable Kozai Plutinos. 
  The inclination of the equatorial plane combined with the height of the blocks (cyan polygons) resulted in both blocks coincidentally straddling Declination = 0\degr. 
  The ecliptic plane is marked in black, and dashed black lines show 15 degrees above and below the ecliptic plane. 
  The K90 librators all come to pericenter in the yellow areas near the $+15\degr$ line, while the K270 librators all come to pericenter in the yellow areas near the $-15\degr$ line.
  Non-Kozai Plutinos come to pericenter at the same RA ranges, but are clustered around the ecliptic.
  }
\end{figure}

The survey was designed with two blocks, one in each semester, in order to sample both Kozai Plutino islands.
Our survey depth goal was a limiting magnitude of $m_{gri}=25.5$ (or $m_r=25.35$), $>0.7$ magnitudes deeper than the \citet{Alexandersen2016} survey, with 3 times the area of their off-ecliptic block.
We required good conditions, including image quality $\leq$0.7$\arcsec$, airmass$<1.2$, 100\% transparency and dark sky.
The block placement was determined based on predictions of detectable Kozai Plutino populations, which as discussed in Section \ref{sec:intro} are more detectable near the two Plutino pericenter longitudes \textit{and} off-ecliptic.
We used a survey simulator \citep{Lawler2018} and a basic model of the Kozai Plutinos \citep{Alexandersen2016} with an assumed intrinsic Kozai fraction of 25\% in order to select our field placement.
For exact field placement, physical Palomar Sky Survey II overlays were consulted in the Wolbach Library, shifting fields by a few degrees to avoid bright starts.
We predicted similar numbers of Kozai and non-Kozai Plutino detections based on the inclination distributions of the two populations in the \citet{Alexandersen2016} models.
The survey depth and area was selected in order to discover enough Kozai Plutinos to provide significant constraints on the orbital distribution and intrinsic size of the population.
The on-sky position of the two \lido discovery blocks compared to the Kozai Plutino model is shown in \autoref{fig:blocks}.

The \lido survey observations spanned 2020-2023\footnote{CFHT programs 20AC02, 20BC019, 21AC013, 21BC003, 22AC021, 22BC018, 23AC016 (Canadian time PI: S.~Lawler) and 20AT02, 20BT005, 21AT004, 21BT003, 22AT011 (Taiwanese time PI: Y.~T.\ Chen)}. 
Our original timeline was for discovery in 2020A and 2020B, but COVID-19 shutdowns and weather loss resulted in insufficient images acquired in 2020A to allow discovery and tracking.
As a result, we acquired our spring-semester discovery observations in 2021A, with a field selected such that the majority of the objects would be trackable in the images we did obtain in 2020A.
Blind tracking was done in the discovery semesters and targeted tracking was done in the following semesters.
The 18 fields covered 19.6 sq.deg. (18.3 sq.deg. of actual pixel area when chip gaps are accounted for), beginning at 14$\degr$ off the ecliptic plane and extending an additional 6$\degr$ away from the plane. 
In 2021A weather was again a limiting factor, resulting in the two fields closest to the ecliptic never receiving useable discovery images. 
The total survey footprint is thus 34 fields (37.0~sq.deg., or 34.7~sq.deg. of pixel area) rather than the intended 36 fields (39.2~sq.deg.).

Discovery observations were done in three sets for each of the two semesters.
The discovery block is 3.3 degrees wide, for a total of 18 MegaCam fields.
These were sub-divided into three sets of 6 pointings, which we cycled through three times to acquire a triplet of three images of all 6 fields, with images of the same field spaced 36 minutes apart.
Our exposure times were 320 seconds because the motion of the objects ($\sim$3-5\arcsec/hour) results in trailing loss for longer exposures.
Including overheads, the 18-exposure discovery triplets required 108 minutes, and had a 72 minute arc lengths for each triplet image set.
We also acquired a single image 1-4 nights away from the discovery triplet, which provided valuable confirmation of the objects discovered in the triplet and significantly simplified the linking process to additional observations in other months.
The majority of our objects, 143, had at least 4 measurements over at least two nights in our discovery dark run (some had more since sub-par images were repeated).

Additional tracking was acquired in the discovery semester.
We acquired tracking data in 3 additional dark runs, with at least one being before discovery and at least one being after discovery.
This tracking within the discovery semester was done by shifting the block of fields in unison to account for the predicted average motion of TNOs, resulting in most objects being within the field in most images.
These objects were linked within the discovery semester, and the 3-5 month arcs were sufficient for pointed recoveries in the following semesters.
For the spring-semester discoveries, we were able to link 62 of the 82 2021A discoveries back to one or more 2020A images immediately, meaning that multi-year tracking of those objects was not substantially delayed by the delayed discovery. 
For tracking after the discovery semester (2021B-2023A), we used targeted pointings, with locations optimized to recover the most objects possible in the fewest exposures.
The targets had also sheared considerably due to their varying distances, so slightly more observing time was required per recovery than in the discovery year.
In the first tracking semester (2021B and 2022A) we acquired at least two tracking observations targeting each object; bright objects in fields that had to be repeated in order to recover nearby faint objects received as many as 22 tracking observations in the first tracking semester.
In the second tracking semesters (2022B and 2023A) the objects had sheared farther; we prioritized tracking the near-resonant objects, large-$a$ objects, and objects with the most uncertain orbits (large $\sigma_a/a$).
For the spring-semester targets, 2023A tracking also prioritized objects which were not pre-covered in 2020A.
We tracked all our discoveries to a $\sigma_a/a<0.0025$, except for seven objects that were outside the characterized part of the survey (fainter than the magnitude limit or fast Centaurs beyond our rate limit); for all of our resonant TNOs we reached $\sigma_a/a\le0.001$.
These orbital uncertainties were sufficient to ensure both easy long-term followup observations and conclusive orbital classification.

The images acquired from CFHT were processed and searched for moving objects.
The pipeline starts with the raw data.
The de-trending (bias-subtraction, flat-fielding) was done with the Pitcairn software package. 
The flat fields were built for each run using the science images acquired during that run. This approach has been shown to give a 0.05 magnitude depth advantage over using twilight flats.
The astrometric/photometric calibration was done using the MegaPipe pipeline \citep{Gwyn2008}.
The astrometric calibration was done on an exposure-by-exposure basis using GAIA DR2 \citep{GaiaCollaboration2018} as a reference.
The GAIA positions were corrected for proper motions to the epoch of the exposure.
Typical astrometric residuals are on the order of 10-20 mas, varying depending on the observing conditions.
The wide filter used by \lido leads to a small amount of differential chromatic refraction; this effect is included in the residuals above.
The photometric calibration used Pan-STARRS 3pi \citep{Magnier2020} as a reference.
The Pan-STARRS $griz$ photometry was converted into the MegaCam system using a third order polynomial in $g-i$ over a restricted color range.
The residuals around this transformation are 0.02 magnitudes.
The transformed photometric catalog is used as in-field standards.
Each exposure was calibrated separately, with typically a few thousand standard stars per exposure. 
The photometric accuracy is limited by the accuracy of the transformation, ie. 0.02 magnitudes. 
This accuracy is confirmed by cross-matching source catalogs from the independently calibrated images.
These images were then ready to be used as input for the discovery search using the OSSOS moving object pipeline and the tracking analysis.

\section{TNO Discovery} \label{sec:discoveries}

We searched the discovery triplet images for moving objects within our rate sensitivity limits.
To identify moving objects, we utilized the OSSOS Moving Object Pipeline \citep[MOP,][]{Bannister2016}, which is based on the previous MOP \cite{Petit2004} used by the Canada-France Ecliptic Plane Survey \citep[CFEPS,][]{Jones2006, Petit2011} and the \citet{Alexandersen2016} survey.
The MOP identifies sources in each image of a discovery triplet, rejects sources that appear stationary in the three images, and then identifies combinations of three detections that feature roughly linear motion at a roughly constant rate.
See the cited sources for a more in depth explanation of the MOP.
The MOP is available on GitHub: \url{https://github.com/OSSOS/MOP}.
In order to limit the false candidate rate, the MOP was used to search for objects with motion between 0.5$\arcsec$~hr$^{-1}$ ($\sim300$ au) and 15$\arcsec$~hr$^{-1}$ ($\sim10$ au), moving at angles plausibly consistent with real objects (since the apparent motion of distant objects is dominated by the reflex motion caused by the Earth's actual motion, their direction of apparent motion is confined). 
\lido is not designed to be able to track objects faster than about 10$\arcsec$~hr$^{-1}$ ($\sim15$ au), but we searched with the MOP to identify objects up to 15$\arcsec$~hr$^{-1}$ nonetheless; any centaurs/comets discovered in this rate range are simply not part of the characterized (modelable) survey sample. 

When the MOP is run, it produces both true positives (detections of real objects) and false positives.
We run the MOP on both the real data and copies with implanted synthetic sources, and produce a detection list for human vetting.
Synthetic sources are implanted in copies of the real data where the header timestamps of the three images have been transposed such that real Solar System bodies will not exhibit linear motion. 
In this way only synthetic sources and false positives appear as moving sources in the planted images.
The planted objects are used to characterize the detection efficiency and the non-planted detections from the time scrambled images that pass vetting steps are used to characterize the false positive rate, as described in \autoref{sec:characterization}.
All detections in the forward time series and a small fraction in the time scrambled series are visually inspected.
Each possible detection was vetted visually by two team members.
Any object identified by either of the two vetters was further inspected by a group of at least four team members to determine whether it should be accepted.
As discussed in more detail in \ref{sec:characterization}, at this stage the vetters do not know whether they are checking a source in the forward data or the time scrambled data.
This process produces a list of astrometry for real objects detected in the discovery triplet, artificial sources detected in the scrambled data, and false positives.

\begin{figure}[tbp]\label{fig:da_a}
\centering
  \includegraphics[width=0.7\textwidth]{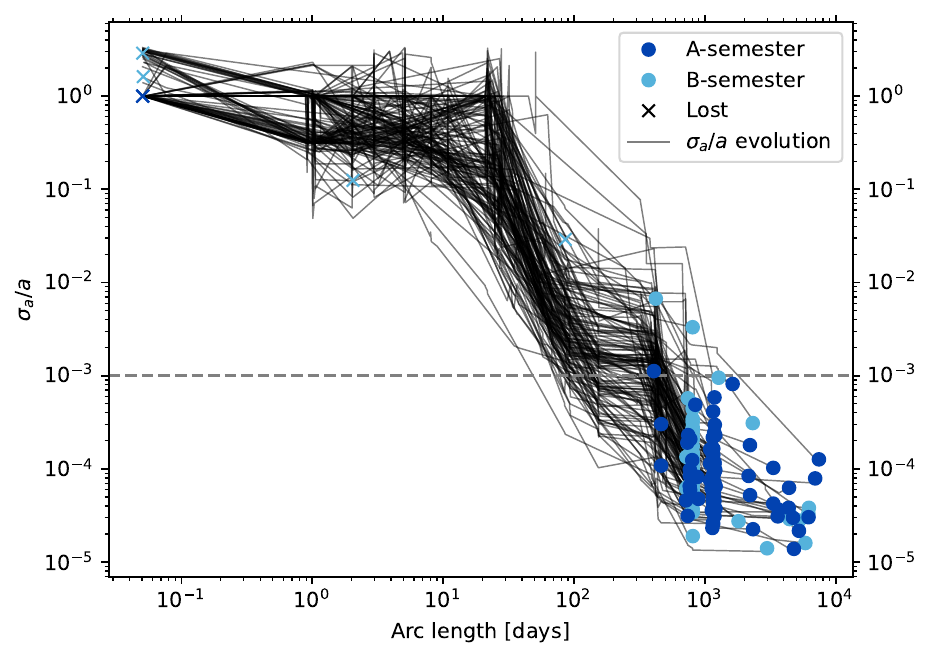}
  \caption{Time evolution of the fractional semi-major axis uncertainty, $\sigma_a/a$.
  As time goes on and additional observations get added, lengthening the arc, the orbit becomes more and more constrained, represented here by generally decreasing $\sigma_a/a$.
  It should be noted that the arc length is the largest factor that influences how well an orbit is constrained, but not the only factor: the quality of the measurements (both accuracy and precision) as well as how evenly the measurements sample the entire arc are also important factors.
  ``A-semester'' shows the objects discovered in the spring 2021 semester, and ``B-semester'' shows the objects discovered in the autumn 2020 semester.
  (For the four previously-known objects, the MPC astrometry was not included in the construction of this plot).
  }
\end{figure}

Once a discovery triplet was accepted as a real object, we tracked it across all available images.
This was easiest for sources where the object was found within the error ellipse in data taken within the same dark run, within a few days to a week of discovery.
Some objects were not recovered in these images (they could have been in chip gaps, obscured by stars, fainter due to light curve variation, etc.).
Some of the objects not recovered in the discovery dark run were successfully linked to data 1-2 months before or after discovery.
We were unable to track 7 objects.
These include one Centaur which sheared off the block soon after discovery, and was not in the blind tracking images.
This is not unexpected, as the survey tracking was optimized for Plutinos, and we do not include Centaurs moving faster than 10\arcsec/hr (roughly $<15$ au) in the characterized sample.
There were also six objects which we were unable to track due to their large $m_w$.
Two of these were \textit{also} Centaurs moving at >10\arcsec/hr.
These objects were all determined to be beyond the characterized limit of the respective block that they were discovered in.
The remaining 141 objects were tracked within the blind recovery/precovery images in the discovery year, and then tracked with targeted recoveries in the following 1-2 years as planned.
We monitored the $\sigma_a/a$ of the discoveries (see \autoref{fig:da_a}) and prioritized additional tracking of objects with $\sigma_a/a>0.001$ and objects near mean motion resonances (MMR).
The end product was high-quality astrometric arcs of 141 TNOs, with 100\% of the characterized sample tracked.

\subsection{Discovery triplets}
\label{sec:triplets}

In order for the survey to go as deep as possible, we aimed for image quality (IQ) < 0.7\arcsec for all discovery images. 
In June 2021, most images of the spring fields had IQ < 0.7\arcsec, although all triplets for the T-block fields had one or two images with IQ slightly above 0.7\arcsec, and all triplets for U-block had one image with IQ = 0.7-0.8\arcsec, while for V-block all but two images had IQ < 0.6\arcsec, which will result in that block being significantly deeper than the other blocks.
In November of 2020, variable weather forced us to reobserved the fall triplets on multiple nights, in hopes to get better image quality. 
However, none of the obtained triplets were close to our IQ < 0.7\arcsec goal.
For this reason, we opted to search (and characterize) two sets of triplets for eight of the fields. 
Block A and B both cover the same fields, while D block is a partial repeat of C block (some fields did not get three images on the second night due to a communication error).
Most images for the fall-block triplets have IQ<0.8, although several (particularly every triplet in C-block) have an image with IQ as high as 1.0\arcsec.

The logic behind searching and characterizing two sets of heavily-overlapping fields was to recoup losses in total number of detections due to the shallower detection limit. 
Since the fall blocks would be a couple of tenths of a magnitude shallower than hoped, we would detect 10-20\% fewer objects. 
This loss could have been recouped by adding additional fields to the survey footprint, but that would also have increased the time budget needed for tracking, as the objects would be more spread out from the start. 
Increasing the sample size from a fixed area has a much smaller effect on the tracking budget, as the objects are often trackable multiple objects per field.
On any given night, $\sim7-15\%$ of the brightest objects (more for fainter objects) that are on the CCD won't be detectable due to being on top of stars, galaxies, chip defects, etc. 
Additionally, the field of view is only about 93\% filled with pixels due to gaps between CCD chips.
Thus a total of $14-22\%$ of objects within a field of view won't be detected in a given night, and the objects move around relative to the background sources and chip gaps/defects.
So the $14-22\%$ of objects not detectable on one night will to a large extent be different objects than the night before/after.
By searching the triplets from multiple nights, we would mostly recoup the loss from the missed depth.

\subsection{Extending Observational Arcs}\label{sec:arc-extend}

The original \lido program tracked objects using CFHT in 2020-2023, but we also extended the observed arcs of some objects with additional observations from Magellan and Gemini North, archival Subaru and CFHT data, and data from the Minor Planet Center (MPC) for objects that were previously known.
The survey length was selected in order to be sufficient for secure classification of the Plutinos, so for some larger-$a$ objects the arc length was insufficient for secure classification as resonant or non-resonant.
In addition to these larger-$a$ objects that required longer arcs for classification, a fast moving object 4A12E12 (MPC designation 2020 VJ46) was observed to ensure it was not lost.
A few other objects were also observed for color measurements, and astrometry was measured on these images \citep[for details, see][]{Collyer2025}.
Additional astrometry on some objects was acquired from Magellan, depending on sky conditions in our classically scheduled time.
For the four previously-known objects we incorporated astrometry from the MPC and checked the residuals of those data compared to our orbital fits.
This astrometry frequently has a larger uncertainty than the \lido data, but still provides useful constraints on the orbital fits.
The origin of any astrometry beyond the base \lido survey for each object is detailed in the comments of Table \ref{tab:objects}.

The images acquired from Magellan (PI Pike) were processed with our team's pipeline and measured.
The Magellan images were acquired with Inamori-Magellan Areal Camera and Spectrograph \citep[IMACS,][]{Dressler2011} using f/4 and f/2 modes using the $WB4800-7800$ filter and the Sloan $r$ filter.
The exposure times were tailored to the expected TNO brightness.
Images were flat and bias corrected using our processing software \citep[see][for additional details on processing Magellan data]{Collyer2025}.
The TNO positions were measured using SourceExtractor for objects where only astrometry was measured or TRIPPy \citep{Fraser2016} in the cases where photometry was also measured.
This astrometry has precision comparable to the main survey astrometry measured in the CFHT images.

We acquired additional astrometry of one of our targets, 0W14B00 (MPC designation 2020 VN40), using Gemini \citep[GMOS][]{Hook2004}, GN-2023B-DD-110 (PI Pike).
This is a large semi-major axis object near a resonance boundary, so additional astrometry was needed.
Data was acquired using GMOS in $r$-band and pre-processed using the Gemini IRAF package.
The WCS was refined using astrometry.net \citep{Lang2010} and the Pan-STARRS PS1PV3 catalog \citep{Schlafly2012, Tonry2012, Magnier2013}, as with Magellan, and the TNO position was measured using SourceExtractor \citep{Bertin1996}.

We used the Solar System Object Image Search\footnote{https://www.cadc-ccda.hia-iha.nrc-cnrc.gc.ca/en/ssois/} \citep[SSOIS,][]{Gwyn2012} to check for archival data of our survey objects.
We checked the CFHT images with longer exposure times, but did not find any convincing recoveries of our objects in data from 2013 onward.
We were able to identify 0W14B00 in five archival Subaru images from September 2017.
We processed these images using HSCPipe 8.3 \citep{Bosch2018} using Pan-STARRS PS1PV3 and measured the positions of the object with SourceExtractor.
This astrometry is included in our survey data, see Table \ref{tab:objects}.

\subsection{Discovery List}

The \lido survey detected 148 moving objects.
This included some Centaurs moving faster than the survey was designed to track, as well as objects which were fainter than our characterization limit.
Many of the Centaurs were not tracked in our blind tracking, and as a result we do not include Centaurs in our characterized sample.
Our final characterized sample is thus  125 objects.
All of our characterized targets have an arc length of at least 407 days (median 1129 days), and $\sigma_a/a\leq0.23$\%.
The objects have $12.96\leq a\leq139.9$~au and $14.3\degr\leq i\leq46.0\degr$.
There are no low-$i$ objects because all of our fields were off-ecliptic and it is thus physically impossible to detect low-$i$ objects.
Table \ref{tab:objects} lists our survey objects.

\begin{longrotatetable}
\setlength{\tabcolsep}{2pt}
\tablewidth{10in}
\tabletypesize{\footnotesize}
\begin{longtable}{ll|ccccccccccc | l}
\caption{All \lido Survey Objects} \label{tab:objects}\\
\hline\hline\\
Survey & MPC & Classification & S/I & $m_w$ & $H_w$ & $a$ & $e$ & $i$ & $q$ & dist & arc length & nobs &Notes\\
ID & designation &  &  & mag & mag & au &  & $\degr$ & au & au & years &  & \\\hline\hline
\endhead
0109A01 & 2024 XA48 & Centaur & Ss & 22.47 & 11.33 & 12.9691 & 0.143344 & 22.741 & 11.1101 & 12.879 & 8.1841 & 30 & CFHT (Archival)  \\ 
1Q02U11 & 2021 LU58 & Centaur & Ss & 24.63 & 13.17 & 13.8358 & 0.20326 & 16.887 & 11.0235 & 13.848 & 2.0237 & 20 & Magellan  \\ 
2K34D10 & 2020 VJ45 & Centaur & Ss & 25.02 & 11.68 & 28.333 & 0.24930 & 19.417 & 21.270 & 21.457 & 2.2123 & 29 & \\ 
5M00B02 & 2013 VX30 & 1:1 & Ss & 22.78 & 8.35 & 30.0853 & 0.08369 & 31.258 & 27.5675 & 27.573 & 16.0553 & 144 & CFHT (Archival), MPC  \\ 
2K35C08 & 2020 VK45 & 1:1 & Ss & 24.91 & 10.16 & 30.175 & 0.0545 & 24.09 & 28.530 & 29.94 & 2.2091 & 22 & \\ 
3B00E15 & 2020 VN45 & 1:1 & Ss & 24.29 & 9.49 & 30.1952 & 0.0447 & 20.511 & 28.847 & 30.284 & 2.2150 & 44 & \\ 
0J16C11 & 2020 VP44 & 4:3 & Ss & 24.93 & 9.46 & 36.427 & 0.0580 & 20.046 & 34.314 & 34.996 & 2.2121 & 28 & \\ 
5A05B02 & 2020 VS45 & 4:3 & Ss & 24.90 & 9.64 & 36.487 & 0.08327 & 21.918 & 33.449 & 33.449 & 2.2120 & 31 & \\ 
0932B05 & 2020 VN44 & 4:3 & Ss & 24.50 & 9.59 & 36.527 & 0.19775 & 21.549 & 29.303 & 30.407 & 10.0400 & 47 & CFHT (Archival)  \\ 
5G22B02 & 2020 VV45 & Scattering & Ss & 23.64 & 9.09 & 37.513 & 0.25230 & 25.29 & 28.048 & 28.139 & 2.2120 & 30 & \\ 
1124C07 & 2020 VU44 & 10:7:1:2 & Is & 24.16 & 8.56 & 38.157 & 0.0678 & 32.929 & 35.570 & 35.712 & 2.2092 & 26 & \\ 
0309T04 & 2021 LY56 & Inner Hot Classical & Ss & 24.82 & 9.10 & 38.506 & 0.04888 & 31.566 & 36.624 & 36.912 & 3.0940 & 20 & \\ 
3R36D11 & 2020 VP45 & 16:11:3:2 & Ss & 25.40 & 9.79 & 38.611 & 0.0855 & 24.598 & 35.309 & 36.071 & 2.2092 & 26 & \\ 
0838V16 & 2021 LG57 & Inner Hot Classical & Ss & 25.33 & 9.45 & 38.831 & 0.0936 & 22.805 & 35.198 & 39.327 & 3.0830 & 33 & \\ 
z136E15 & 2020 VN46 & 3:2 & Ss & 24.67 & 8.29 & 39.229 & 0.19609 & 29.458 & 31.537 & 43.432 & 16.9234 & 29 & Magellan, CFHT (Archival)  \\ 
4V07E14 & 2020 VL46 & 3:2 & Ss & 24.57 & 8.94 & 39.285 & 0.2230 & 16.155 & 30.53 & 36.538 & 2.2093 & 25 & Magellan  \\ 
0A07U07 & (556130) 2014 KV101 & 3:2 & Ss & 22.33 & 7.31 & 39.3207 & 0.258773 & 16.816 & 29.1456 & 31.419 & 13.1283 & 115 & Magellan, MPC  \\ 
1H17U11 & 2021 LT58 & 3:2 & Ss & 25.09 & 10.14 & 39.327 & 0.21615 & 17.515 & 30.827 & 30.938 & 9.1251 & 39 & CFHT (Archival)  \\ 
0538U07 & 2021 LE57 & 3:2 & Si & 24.17 & 9.21 & 39.335 & 0.27306 & 16.833 & 28.594 & 30.923 & 12.9236 & 49 & CFHT (Archival)  \\ 
2O12C08 & 2020 VG46 & 3:2 & Ss & 24.42 & 10.14 & 39.340 & 0.2841 & 18.442 & 28.163 & 28.601 & 2.2064 & 24 & Magellan  \\ 
2A14V15 & 2021 LZ58 & 3:2 & Ss & 24.32 & 9.69 & 39.350 & 0.27566 & 27.533 & 28.502 & 28.54 & 11.9621 & 31 & CFHT (Archival)  \\ 
4J29E13 & 2020 VK46 & 3:2 & Ss & 25.03 & 9.97 & 39.393 & 0.2267 & 22.332 & 30.461 & 31.577 & 2.2119 & 27 & Magellan  \\ 
1Y25C06 & 2020 VD45 & 3:2 & Ss & 24.48 & 9.38 & 39.394 & 0.2281 & 18.297 & 30.408 & 31.931 & 2.2092 & 25 & \\ 
0133T03 & 2021 LQ56 & 3:2 & Ss & 24.34 & 8.46 & 39.394 & 0.2614 & 27.497 & 29.097 & 38.101 & 3.3099 & 26 & \\ 
2022V14 & 2021 LW58 & 3:2 & Ss & 25.35 & 9.64 & 39.440 & 0.1125 & 18.963 & 35.001 & 36.758 & 3.1702 & 27 & \\ 
0A10T05 & 2021 LK54 & 3:2 & Ss & 23.92 & 8.98 & 39.4432 & 0.21931 & 23.402 & 30.7930 & 30.848 & 6.3896 & 41 & Magellan, CFHT (Archival)  \\ 
3I37C09 & 2020 VT40 & 3:2 & Ss & 23.65 & 8.69 & 39.455 & 0.3445 & 29.897 & 25.864 & 30.934 & 2.2119 & 30 & \\ 
2112C06 & 2020 VR40 & 3:2 & Ss & 23.91 & 8.69 & 39.460 & 0.2343 & 30.315 & 30.213 & 33.385 & 2.2092 & 21 & \\ 
1W10B04 & 2020 VE46 & 3:2 & Ii & 25.31 & 9.13 & 39.484 & 0.03696 & 26.324 & 38.025 & 40.943 & 2.0259 & 28 & Magellan  \\ 
0403U07 & 2021 LB57 & 3:2 & Ss & 25.22 & 10.13 & 39.503 & 0.31402 & 23.235 & 27.099 & 31.723 & 3.1701 & 32 & \\ 
t806E15 & 2020 VQ40 & 3:2 & Ss & 23.46 & 7.96 & 39.519 & 0.20709 & 20.55 & 31.335 & 35.069 & 4.9507 & 41 & Magellan, CFHT (Archival)  \\ 
0G01C11 & 2020 VA46 & 3:2 & Ss & 24.41 & 10.05 & 39.520 & 0.32330 & 23.976 & 26.744 & 26.979 & 2.2014 & 29 & Magellan  \\ 
3C05C09 & 2020 VS40 & 3:2 & Ss & 23.88 & 9.48 & 39.550 & 0.32565 & 19.4 & 26.671 & 27.431 & 2.2149 & 37 & \\ 
1K03U11 & 2021 LL54 & Main Hot Classical & Ss & 23.84 & 7.91 & 39.577 & 0.02247 & 18.729 & 38.688 & 38.692 & 3.1702 & 33 & Magellan  \\ 
0U16V17 & 2021 LH58 & 3:2 & Ss & 25.14 & 9.29 & 39.578 & 0.2767 & 21.955 & 28.625 & 38.047 & 3.0038 & 29 & \\ 
3H00V16 & 2021 LJ59 & 3:2 & Ss & 24.79 & 8.55 & 39.587 & 0.2749 & 23.224 & 28.704 & 42.136 & 3.1675 & 28 & \\ 
0332U07 & 2021 LZ56 & Main Hot Classical & Is & 24.97 & 8.65 & 40.219 & 0.1126 & 16.263 & 35.691 & 42.647 & 3.2439 & 40 & Some clones show transient 17:11:4:2 resonance  \\ 
0832B05 & 2020 VM44 & Main Hot Classical & Ss & 24.49 & 8.63 & 40.497 & 0.12548 & 20.977 & 35.416 & 38.228 & 12.1343 & 69 & CFHT (Archival)  \\ 
2536B05 & 2020 VE45 & Main Hot Classical & Ss & 24.97 & 9.04 & 40.559 & 0.0577 & 20.744 & 38.22 & 38.769 & 2.2120 & 27 & \\ 
2X08C08 & 2020 VM45 & Main Hot Classical & Ss & 23.73 & 7.72 & 40.784 & 0.1047 & 18.228 & 36.52 & 39.23 & 1.9801 & 17 & \\ 
z429E15 & 2020 VY45 & Main Hot Classical & Ss & 24.59 & 8.57 & 41.331 & 0.0319 & 15.449 & 40.01 & 40.126 & 2.2015 & 29 & \\ 
0O06T04 & 2021 LC58 & Main Hot Classical & Ss & 24.70 & 8.24 & 41.494 & 0.1221 & 20.218 & 36.428 & 43.988 & 3.2958 & 34 & \\ 
0I21U06 & 2021 LV57 & Main Hot Classical & Ss & 23.06 & 6.57 & 41.648 & 0.09791 & 22.679 & 37.570 & 44.468 & 16.9698 & 42 & CFHT (Archival)  \\ 
0L13B05 & 2020 VQ44 & Main Hot Classical & Ss & 24.78 & 8.95 & 41.703 & 0.1035 & 28.994 & 37.39 & 38.098 & 2.2092 & 26 & \\ 
0K33T00 & 2021 LY57 & Main Hot Classical & Ss & 24.55 & 8.71 & 41.734 & 0.1223 & 29.031 & 36.629 & 37.991 & 1.9657 & 30 & Magellan  \\ 
0H15V12 & 2021 LT57 & Main Hot Classical & Is & 25.35 & 8.82 & 41.769 & 0.1193 & 23.36 & 36.79 & 44.45 & 2.1027 & 26 & Some sticking in 18:11:5:2  \\ 
0137T02 & 2021 LR56 & Main Hot Classical & Ss & 25.07 & 9.03 & 41.802 & 0.1115 & 14.324 & 37.141 & 40.268 & 3.1647 & 32 & \\ 
3L31D11 & 2020 VO45 & Main Hot Classical & Ss & 24.06 & 7.55 & 41.805 & 0.0856 & 20.594 & 38.22 & 44.396 & 2.2092 & 32 & \\ 
0T22U10 & 2021 LG58 & Main Hot Classical & Ss & 24.28 & 8.33 & 41.824 & 0.07381 & 19.674 & 38.737 & 38.747 & 3.1267 & 37 & \\ 
0F29T05 & 2021 LP57 & Main Hot Classical & Ss & 24.70 & 8.35 & 41.904 & 0.0733 & 25.98 & 38.83 & 42.597 & 2.0972 & 20 & Magellan  \\ 
0K10U10 & 2021 LX57 & Scattering & Ss & 23.82 & 9.55 & 42.084 & 0.41309 & 26.944 & 24.699 & 26.245 & 2.3157 & 19 & \\ 
2420C10 & 2020 VF46 & Main Hot Classical & Ss & 24.53 & 8.63 & 42.516 & 0.1048 & 17.545 & 38.060 & 38.896 & 2.2093 & 30 & Magellan  \\ 
0911T02 & 2021 LH57 & Main Hot Classical & Ss & 25.21 & 9.37 & 42.533 & 0.1231 & 23.209 & 37.297 & 38.108 & 2.1135 & 29 & \\ 
0Y33C07 & 2014 RM86 & Main Hot Classical & Ss & 23.41 & 7.85 & 43.008 & 0.17918 & 19.135 & 35.302 & 35.558 & 14.3868 & 55 & MPC, CFHT (Archival)  \\ 
0N14U10 & 2021 LB58 & Main Hot Classical & Ss & 24.42 & 8.69 & 43.010 & 0.13970 & 19.934 & 37.002 & 37.084 & 3.1485 & 35 & \\ 
2Y18V16 & 2021 LG59 & Main Hot Classical & Ss & 24.78 & 8.72 & 43.101 & 0.07510 & 20.681 & 39.864 & 39.903 & 3.2439 & 37 & \\ 
0A01T01 & 2021 LJ57 & Main Hot Classical & Ss & 25.09 & 9.41 & 43.152 & 0.14499 & 14.997 & 36.896 & 36.902 & 3.1701 & 26 & \\ 
0O27U06 & 2021 LE58 & Main Hot Classical & Ss & 24.70 & 8.98 & 43.251 & 0.15271 & 21.81 & 36.646 & 36.856 & 3.2165 & 32 & \\ 
1S01A03 & 2020 VB45 & Main Hot Classical & Ss & 24.69 & 7.99 & 43.483 & 0.0876 & 20.048 & 39.68 & 47.128 & 2.2091 & 32 & \\ 
0Y36A00 & 2020 VT44 & 7:4 & Ss & 25.05 & 10.07 & 43.639 & 0.3001 & 19.642 & 30.544 & 30.979 & 2.2095 & 24 & \\ 
0418A01 & 2020 VK44 & Main Hot Classical & Ss & 24.33 & 7.27 & 43.983 & 0.1984 & 21.974 & 35.26 & 50.327 & 2.2120 & 34 & \\ 
0F36T02 & 2021 LQ57 & Main Hot Classical & Ss & 24.83 & 8.63 & 44.289 & 0.1406 & 27.906 & 38.06 & 41.214 & 2.1926 & 24 & \\ 
4P22E14 & 2020 VQ45 & Main Hot Classical & Ss & 24.93 & 9.00 & 44.308 & 0.1996 & 25.295 & 35.46 & 38.906 & 2.2093 & 27 & Occasional sticking in 16:9:5:2   \\ 
0A21T02 & 2021 LK57 & 11:6 & Ss & 24.57 & 8.60 & 45.100 & 0.1811 & 14.555 & 36.93 & 39.034 & 2.4003 & 31 & \\ 
1R03A05 & 2020 VA45 & Main Hot Classical & Ss & 23.95 & 8.10 & 45.301 & 0.16350 & 28.034 & 37.894 & 37.895 & 11.9812 & 30 & CFHT (Archival)  \\ 
1207U08 & 2021 LM58 & 13:7:4:2 & Ss & 24.86 & 8.38 & 45.438 & 0.1429 & 21.747 & 38.94 & 44.035 & 3.1949 & 37 & \\ 
0S33B00 & 2020 VR44 & Main Hot Classical & Ss & 24.86 & 7.97 & 45.459 & 0.0679 & 22.662 & 42.375 & 48.481 & 2.2120 & 26 & \\ 
0108V17 & 2021 LM56 & Main Hot Classical & Ss & 23.87 & 7.05 & 45.513 & 0.1158 & 20.017 & 40.241 & 47.679 & 3.2439 & 29 & \\ 
5D11B02 & 2020 VU45 & Main Hot Classical & Ss & 25.15 & 8.28 & 45.565 & 0.1175 & 21.156 & 40.21 & 48.369 & 2.2090 & 29 & \\ 
2O15V15 & 2021 LD59 & Main Hot Classical & Ss & 25.33 & 9.22 & 45.605 & 0.1471 & 31.625 & 38.89 & 40.454 & 2.1026 & 28 & \\ 
2U05C08 & 2020 VL45 & Main Hot Classical & Ss & 24.54 & 8.10 & 45.654 & 0.1590 & 21.576 & 38.39 & 43.602 & 2.2092 & 26 & \\ 
1730U11 & 2021 LP58 & 15:8:5:2 & Is & 24.92 & 8.90 & 45.73 & 0.1737 & 28.937 & 37.79 & 40.251 & 1.2700 & 33 & \\ 
0A24A01 & 2020 VO44 & Main Hot Classical & Ss & 24.60 & 8.04 & 45.778 & 0.02121 & 23.854 & 44.807 & 44.831 & 17.0705 & 35 & CFHT (Archival)  \\ 
2D38V13 & 2021 LA59 & Main Hot Classical & Ss & 23.87 & 7.49 & 45.989 & 0.0740 & 18.885 & 42.586 & 43.109 & 3.1702 & 30 & \\ 
1H25D10 & 2020 VY44 & 19:10:7:2 & Is & 24.90 & 7.62 & 46.11 & 0.1946 & 33.377 & 37.14 & 53.029 & 2.2121 & 29 & \\ 
2812D10 & 2020 VF45 & 21:11 & Is & 24.76 & 8.75 & 46.283 & 0.2237 & 17.875 & 35.93 & 39.628 & 2.2093 & 35 & Choppy, Intermittent resonant behavior  \\ 
1V10A03 & 2020 VC45 & Main Hot Classical & Ss & 25.02 & 8.05 & 46.74 & 0.2157 & 31.329 & 36.66 & 49.765 & 2.2094 & 25 & \\ 
z715E15 & 2020 VZ45 & Main Hot Classical & Ss & 24.64 & 7.39 & 46.99 & 0.2038 & 15.803 & 37.41 & 52.197 & 2.2121 & 27 & \\ 
2435V13 & 2021 LY58 & Main Hot Classical & Ss & 24.62 & 8.12 & 47.141 & 0.2241 & 18.445 & 36.57 & 44.097 & 3.0066 & 19 & \\ 
0U37T04 & 2021 LJ58 & Main Hot Classical & Ss & 24.39 & 8.17 & 47.249 & 0.14962 & 22.255 & 40.180 & 41.429 & 3.2905 & 36 & \\ 
0O17V17 & 2021 LD58 & Main Hot Classical & Ss & 25.52 & 9.18 & 47.320 & 0.1664 & 26.199 & 39.45 & 42.832 & 3.0092 & 25 & \\ 
0130T05 & 2021 LP56 & 2:1 & Ss & 24.91 & 8.33 & 47.531 & 0.1618 & 16.455 & 39.84 & 44.847 & 2.1188 & 23 & \\ 
3X12D11 & 2020 VH46 & 2:1 & Ss & 24.50 & 9.33 & 47.628 & 0.3053 & 18.19 & 33.090 & 33.383 & 2.2093 & 34 & Magellan  \\ 
2G39V13 & 2021 LC59 & 2:1 & Ss & 24.70 & 8.98 & 47.652 & 0.27063 & 18.161 & 34.756 & 36.711 & 10.0315 & 34 & CFHT (Archival)  \\ 
1A18D10 & 2020 VC46 & 2:1 & Ss & 24.89 & 8.65 & 47.688 & 0.2108 & 17.264 & 37.64 & 41.833 & 2.2093 & 39 & Magellan  \\ 
1B06A00 & 2020 VX44 & 2:1 & Ss & 24.72 & 8.28 & 47.79 & 0.1948 & 19.361 & 38.48 & 43.766 & 2.2120 & 27 & \\ 
2L32V15 & 2015 JH13 & Scattering & Ss & 23.79 & 10.23 & 47.836 & 0.56026 & 22.429 & 21.036 & 22.265 & 6.0644 & 34 & Magellan, CFHT (Archival)  \\ 
0B09T03 & 2021 LL57 & Detached & Ss & 24.97 & 9.26 & 48.774 & 0.2793 & 23.853 & 35.150 & 36.647 & 3.2905 & 22 & \\ 
0421V17 & 2021 LC57 & Detached & Ss & 24.05 & 7.83 & 49.181 & 0.2838 & 33.393 & 35.22 & 41.406 & 2.0154 & 25 & Magellan  Some short-term sticking 23:11, \\  &  & & &  & & &  & & & &  &  &   possible mixed mode  \\ 
0H08U10 & 2021 LS57 & Detached & Ss & 25.23 & 9.81 & 50.051 & 0.27251 & 18.365 & 36.412 & 36.564 & 3.1702 & 31 & \\ 
0I33U09 & 2021 LW57 & Detached & Is & 24.69 & 7.35 & 50.51 & 0.3089 & 26.167 & 34.90 & 53.677 & 3.1485 & 30 & Many clones temporarily sticking in 24:11  \\ 
0X04V17 & 2021 LK58 & Detached & Ss & 24.90 & 8.68 & 51.329 & 0.2814 & 39.508 & 36.88 & 41.532 & 3.1648 & 20 & \\ 
0I19T02 & 2021 LU57 & Detached & Is & 23.48 & 6.92 & 51.92 & 0.2975 & 23.076 & 36.47 & 44.611 & 2.0427 & 32 & Many clones temporarily sticking in 25:11:12:2  \\ 
3B23V16 & 2021 LH59 & Detached & Ss & 24.25 & 8.34 & 52.007 & 0.27098 & 19.184 & 37.914 & 38.466 & 9.8570 & 34 & CFHT (Archival)  \\ 
0433B03 & 2020 VL44 & Detached & Ss & 25.25 & 9.41 & 52.596 & 0.27778 & 22.792 & 37.986 & 38.125 & 2.2092 & 24 & \\ 
1E29U11 & 2021 LR58 & Detached & Ss & 24.34 & 8.51 & 53.043 & 0.31000 & 18.584 & 36.599 & 38.119 & 3.1486 & 30 & Some sticking in 7:3  \\ 
0237V12 & 2021 LX56 & Detached & Is & 25.44 & 8.07 & 53.44 & 0.338 & 23.021 & 35.39 & 54.003 & 4.4641 & 19 & Magellan  \\ 
1706V12 & 2021 LO58 & Detached & Is & 24.38 & 8.05 & 54.596 & 0.3450 & 23.008 & 35.759 & 42.124 & 3.1675 & 35 & Some sticking in 22:9  \\ 
0P18V12 & 2021 LF58 & Extended Hot Classical & Ss & 24.84 & 6.82 & 54.921 & 0.1634 & 29.585 & 45.95 & 62.372 & 3.1675 & 33 & \\ 
0233T03 & 2021 LW56 & 5:2 & Ss & 23.98 & 8.53 & 55.209 & 0.43092 & 16.475 & 31.418 & 34.722 & 3.2360 & 29 & \\ 
0123T02 & 2021 LO56 & 5:2 & Ss & 24.93 & 9.94 & 55.320 & 0.44103 & 14.38 & 30.922 & 31.212 & 3.2439 & 31 & \\ 
1X38V14 & 2021 LV58 & 5:2 & Ss & 25.14 & 9.87 & 55.330 & 0.41861 & 20.616 & 32.168 & 33.012 & 3.2441 & 24 & \\ 
2S31V13 & 2021 LE59 & 5:2 & Ss & 25.33 & 9.74 & 55.354 & 0.4282 & 18.624 & 31.652 & 35.681 & 3.1702 & 35 & \\ 
2B38B05 & 2020 VH45 & 5:2 & Ss & 25.23 & 8.09 & 55.40 & 0.1865 & 46.024 & 45.07 & 51.686 & 2.2091 & 28 & \\ 
1P13A03 & 2020 VZ44 & 5:2 & Si & 25.15 & 8.93 & 55.42 & 0.379 & 27.592 & 34.40 & 41.622 & 2.2091 & 24 & \\ 
2G20V15 & 2021 LB59 & 5:2 & Ss & 24.99 & 8.99 & 55.442 & 0.3425 & 41.774 & 36.454 & 39.207 & 3.2440 & 27 & \\ 
5P33B02 & 2014 WJ517 & 5:2 & Ss & 24.41 & 9.00 & 55.564 & 0.3970 & 20.07 & 33.508 & 34.523 & 2.2120 & 30 & \\ 
0113U07 & 2021 LN56 & 17:6 & Is & 24.46 & 8.52 & 60.161 & 0.38922 & 30.143 & 36.745 & 38.821 & 3.2220 & 32 & MPC  \\ 
0821V12 & (674118) 2015 KH162 & Detached & Ss & 22.06 & 4.22 & 61.795 & 0.33199 & 28.86 & 41.280 & 60.505 & 14.3323 & 225 & CFHT (Archival), MPC  \\ 
0224T01 & 2021 LU56 & 28:9 & Is & 24.52 & 7.88 & 64.07 & 0.4150 & 17.542 & 37.49 & 46.032 & 3.2304 & 42 & \\ 
2U01V15 & 2021 LF59 & Detached & Is & 24.87 & 8.88 & 66.73 & 0.420 & 33.615 & 38.69 & 39.012 & 1.1168 & 16 & Some sticking in 33:10:21:2 and 43:13  \\ 
0F17U06 & 2021 LO57 & 7:2 & Ss & 25.27 & 9.88 & 69.129 & 0.51303 & 21.659 & 33.664 & 34.647 & 5.9142 & 28 & CFHT (Archival)  \\ 
0229T03 & 2021 LV56 & 7:2 & Ss & 24.77 & 9.39 & 69.152 & 0.50930 & 17.775 & 33.933 & 33.968 & 3.2144 & 25 & \\ 
2814C06 & 2016 SV56 & Scattering & Ss & 24.07 & 10.53 & 69.5 & 0.7663 & 18.955 & 16.23 & 22.522 & 1.1553 & 20 & \\ 
0L07V12 & 2021 LZ57 & Scattering & Ss & 24.02 & 8.54 & 73.300 & 0.56444 & 20.933 & 31.926 & 34.506 & 12.0471 & 35 & \\ 
1B33U11 & 2021 LQ58 & Scattering & Ss & 25.43 & 10.18 & 77.630 & 0.62157 & 17.911 & 29.378 & 33.116 & 9.1250 & 29 & CFHT (Archival)  \\ 
1F32V14 & 2021 LS58 & 21:5 & Ss & 23.80 & 8.17 & 78.253 & 0.54494 & 27.879 & 35.609 & 36.005 & 19.0725 & 40 & Magellan  \\ 
1529B01 & 2020 VW44 & Detached & Ss & 24.65 & 7.70 & 79.58 & 0.4518 & 37.399 & 43.63 & 49.609 & 2.0340 & 29 & \\ 
1107U08 & 2021 LL58 & Scattering & Ss & 24.37 & 9.02 & 84.57 & 0.60991 & 16.652 & 32.991 & 34.075 & 3.1483 & 28 & \\ 
0737U09 & 2021 LF57 & Detached & Is & 25.33 & 9.74 & 90.57 & 0.60564 & 23.576 & 35.719 & 35.719 & 6.0565 & 30 & CFHT (Archival), MPC  \\ 
1438U11 & 2021 LN58 & 13:2 & Ss & 24.49 & 7.26 & 104.75 & 0.67256 & 26.823 & 34.301 & 52.454 & 20.3771 & 24 & Magellan, CFHT (Archival)  \\ 
0E03U10 & 2021 LN57 & Scattering & Ss & 25.22 & 10.52 & 110.62 & 0.74293 & 18.819 & 28.436 & 29.415 & 3.2903 & 21 & \\ 
4Y00E14 & 2020 VM46 & Scattering & Ss & 24.81 & 9.21 & 124.46 & 0.71319 & 24.44 & 35.70 & 35.932 & 2.2121 & 30 & Magellan  \\ 
0H08T00 & 2021 LR57 & Scattering & Ss & 25.42 & 10.10 & 137.87 & 0.75002 & 15.123 & 34.47 & 34.595 & 3.2627 & 30 & Magellan  \\ 
0W14B00 & 2020 VN40 & 10:1 & Ss & 24.69 & 8.23 & 139.94 & 0.72660 & 33.407 & 38.26 & 43.858 & 6.3598 & 33 & Subaru (Archival), Gemini\footnote{GN-2023B-DD-110-4-001}  \\ 
\hline
u4A12E12 & 2020 VJ46 & Centaur & Ss & 23.86 & 15.42 & 29.898 & 0.77027 & 20.695 & 6.8683 & 7.044 & 1.9749 & 28 & \\ 
u0202U07 & 2021 LS56 & Inner Hot Classical & Ss & 25.52 & 9.46 & 38.492 & 0.0323 & 24.208 & 37.247 & 39.695 & 2.4442 & 32 & \\ 
u1234A04 & 2020 VV44 & Main Hot Classical & Ss & 25.39 & 9.37 & 41.49 & 0.1073 & 20.648 & 37.03 & 39.669 & 2.2094 & 23 & \\ 
u0C06U06 & 2021 LM57 & 5:3 & Si & 25.50 & 8.57 & 42.166 & 0.2188 & 17.843 & 32.94 & 48.718 & 3.1486 & 26 & \\ 
u2A04A02 & 2020 VG45 & Main Hot Classical & Ss & 25.34 & 9.08 & 42.533 & 0.1421 & 26.939 & 36.49 & 41.786 & 2.0310 & 18 & \\ 
u0210V16 & 2021 LT56 & Main Hot Classical & Ss & 25.69 & 9.60 & 43.531 & 0.0849 & 20.0 & 39.84 & 40.053 & 1.2698 & 24 & \\ 
u0107U09 & 2021 LL56 & Main Hot Classical & Ss & 25.71 & 9.77 & 45.274 & 0.1645 & 25.617 & 37.825 & 39.14 & 3.2139 & 26 & \\ 
u5429B02 & 2020 VR45 & 13:7 & Is & 25.50 & 9.26 & 45.47 & 0.2676 & 21.946 & 33.31 & 43.577 & 2.2090 & 25 & \\ 
u0S31C07 & 2020 VB46 & Main Hot Classical & Ss & 25.62 & 9.13 & 46.197 & 0.0883 & 19.659 & 42.12 & 42.894 & 2.2091 & 28 & \\ 
uf703E12 & 2020 VW45 & Main Hot Classical & Ss & 25.27 & 8.74 & 47.17 & 0.1618 & 20.683 & 39.54 & 45.661 & 2.2093 & 24 & \\ 
u0V28C07 & 2020 VS44 & 17:8:7:2 & Ss & 25.29 & 9.37 & 49.71 & 0.2437 & 34.346 & 37.60 & 39.351 & 2.2120 & 25 & \\ 
u2036U11 & 2021 LX58 & 5:2 & Ss & 25.50 & 9.93 & 55.549 & 0.36223 & 17.928 & 35.427 & 35.605 & 3.2441 & 19 & \\ 
u0522V12 & 2021 LD57 & Detached & Ss & 25.78 & 8.83 & 59.78 & 0.3755 & 19.392 & 37.33 & 49.363 & 3.1676 & 31 & \\ 
u0402T05 & 2021 LA57 & Scattering & Ss & 25.44 & 10.14 & 60.555 & 0.44052 & 19.817 & 33.879 & 33.932 & 3.2904 & 24 & \\ 
u0L15U09 & 2021 LA58 & 5:1 & Si & 25.50 & 7.48 & 87.81 & 0.3548 & 38.084 & 56.65 & 63.191 & 3.2440 & 28 & \\ 
u1E26A04 & 2020 VD46 & 7:1 & Ss & 25.19 & 8.73 & 109.9 & 0.6162 & 32.183 & 42.18 & 44.569 & 3.4894 & 31 & \\ 
\hline
u1H04A00 &  & Untracked & x & 25.53 & 15.87 & 10 & 0.3 & 30 & 10 & 9 & 0.0001 & 3 & \\ 
u0D18E15 &  & Untracked & x & 25.46 & 9.82 & 30 & 0.7 & 20 & 10 & 40 & 0.0001 & 3 & \\ 
u0W18U10 &  & Untracked & x & 24.83 & 15.52 & 30 & 0.8 & 20 & 5 & 9 & 0.0001 & 3 & \\ 
u2R04C08 & 2020 VX45 & Untracked & x & 25.24 & 16.12 & 5.7 & 0.4 & 16.3 & 3 & 8.2 & 0.0056 & 6 & \\ 
u2Q10D10 & 2020 VT45 & Untracked & x & 25.71 & 9.56 & 52 & 0.38 & 18.141 & 32 & 40.811 & 0.2349 & 14 & \\ 
u0L22U06 &  & Untracked & x & 25.39 & 12.50 & 70 & 0.8 & 40 & 10 & 20 & 0.0001 & 3 & \\ 
u1N35U11 &  & Untracked & x & 25.48 & 9.28 & 90 & 0.9 & 20 & 10 & 40 & 0.0001 & 3 & \\ 
\hline\hline
\end{longtable}
\footnotesize{
Orbital parameters and other information about every object detected in \lido.
The uncertainty for orbital elements is on the final digit reported.
`Survey ID' is the internal designation that the \lido team gave the object.
`MPC designation' is the provisional designation assigned by the Minor Planet Center, as well as the permanent IAU designation in the two cases where that is available.
`Classification` is the objects class, as determined in \autoref{sec:classify}.
`S/I' indicates whether an object is Securely or Insecurely classified; the first (capital) letter refers to the primary classification, and the second (lower case) letter refers to whether the classification of Kozai or non-Kozai is secure or insecure, while an X indicates a lost object that could not be classified.
`$m_w$ mag' is the average of the three apparent magnitude measurements from the discovery triplets.
`$H_w$ mag' is the absolute magnitude of the object on the discovery night.
`$a$', `$e$', `$i$', `$q$', `dist' are the barycentric semi-major axis, eccentricity, inclination, pericenter distance and distance at discovery. 
`arc lenth' is the observed arc length and `nobs' is the number of measurements.
`Notes' contain any additional notes, such as when objects were observed with telescopes other than CFHT.
}
\end{longrotatetable}

\begin{longrotatetable}
\setlength{\tabcolsep}{2pt}
\tablewidth{10in}
\tabletypesize{\footnotesize}
\begin{longtable}{ll|ccccccccc | l}
\caption{Resonant \lido Survey Objects} \label{tab:resobjects}\\
\hline\hline\\
Survey & MPC & Resonance & S/I & Res Frac& RC & RA &Island Frac& KC & KA & KFrac  &Notes \\
ID & designation &  &  & \% & $\degr$ & $\degr$ &\%& $\degr$ & $\degr$ & \% &  \\\hline \hline
\endhead
5M00B02 & 2013 VX30 & 1:1 & Ss & 100 & 58.8$_{-0.1}^{+0.1}$ & 12.1$_{-0.7}^{+0.5}$ & 100 & -- & -- & -- & \\ 
2K35C08 & 2020 VK45 & 1:1 & Ss & 100 & 63.1$_{-0.7}^{+0.7}$ & 40$_{-4}^{+3}$ & 100 & -- & -- & -- & \\ 
3B00E15 & 2020 VN45 & 1:1 & Ss & 100 & 63.4$_{-1.9}^{+0.7}$ & 40$_{-10}^{+3}$ & 100 & -- & -- & -- & \\ 
0J16C11 & 2020 VP44 & 4:3 & Ss & 100 & 180.0$_{-0.5}^{+0.5}$ & 70$_{-10}^{+10}$ & 100 & -- & -- & -- & \\ 
5A05B02 & 2020 VS45 & 4:3 & Ss & 100 & 180.0$_{-0.3}^{+0.6}$ & 120$_{-5}^{+6}$ & 100 & -- & -- & -- & \\ 
0932B05 & 2020 VN44 & 4:3 & Ss & 100 & 179.9$_{-0.4}^{+0.5}$ & 138$_{-16}^{+8}$ & 100 & 269.94$_{-0.01}^{+0.02}$ & 29.1$_{-1.4}^{+0.9}$ & 100 & Omega oscillation  \\ 
1124C07 & 2020 VU44 & 10:7:1:2 & Is & 65 & 180$_{-9}^{+7}$ & 320$_{-70}^{+30}$ & 100 & -- & -- & -- & \\ 
3R36D11 & 2020 VP45 & 16:11:3:2 & Ss & 67 & 179$_{-7}^{+6}$ & 210$_{-110}^{+70}$ & 100 & -- & -- & -- & \\ 
z136E15 & 2020 VN46 & 3:2 & Ss & 100 & 180$_{-1}^{+1}$ & 230$_{-30}^{+30}$ & 100 & -- & -- & -- & \\ 
4V07E14 & 2020 VL46 & 3:2 & Ss & 100 & 180$_{-1}^{+1}$ & 169$_{-4}^{+5}$ & 100 & 93$_{-5}^{+3}$ & 159$_{-8}^{+33}$ & 40 & \\ 
0A07U07 & (556130) 2014 KV101 & 3:2 & Ss & 100 & 180.0$_{-0.4}^{+0.5}$ & 112$_{-5}^{+6}$ & 100 & 89.68$_{-0.05}^{+0.06}$ & 70$_{-1}^{+1}$ & 100 & \\ 
1H17U11 & 2021 LT58 & 3:2 & Ss & 100 & 180.0$_{-0.7}^{+0.6}$ & 122$_{-7}^{+6}$ & 100 & 90.01$_{-0.02}^{+0.02}$ & 81.9$_{-0.5}^{+0.7}$ & 100 & \\ 
0538U07 & 2021 LE57 & 3:2 & Si & 100 & 180$_{-2}^{+2}$ & 131$_{-9}^{+11}$ & 100 & 94.9$_{-3.1}^{+0.4}$ & 152$_{-9}^{+10}$ & 76 & \\ 
2O12C08 & 2020 VG46 & 3:2 & Ss & 100 & 180$_{-1}^{+1}$ & 163$_{-3}^{+3}$ & 100 & 270.0$_{-0.1}^{+0.2}$ & 94.2$_{-0.4}^{+0.8}$ & 100 & \\ 
2A14V15 & 2021 LZ58 & 3:2 & Ss & 100 & 180$_{-2}^{+2}$ & 113$_{-4}^{+5}$ & 100 & 90.44$_{-0.10}^{+0.17}$ & 88.8$_{-0.3}^{+0.3}$ & 100 & \\ 
4J29E13 & 2020 VK46 & 3:2 & Ss & 100 & 180.0$_{-0.6}^{+0.6}$ & 113$_{-4}^{+5}$ & 100 & 269.808$_{-0.005}^{+0.006}$ & 65.8$_{-0.2}^{+0.1}$ & 100 & \\ 
1Y25C06 & 2020 VD45 & 3:2 & Ss & 100 & 179.9$_{-0.9}^{+1.0}$ & 114$_{-3}^{+2}$ & 100 & -- & -- & -- & \\ 
0133T03 & 2021 LQ56 & 3:2 & Ss & 100 & 180.0$_{-0.7}^{+1.1}$ & 59$_{-2}^{+3}$ & 100 & 270.054$_{-0.002}^{+0.003}$ & 52.93$_{-0.02}^{+0.02}$ & 100 & \\ 
2022V14 & 2021 LW58 & 3:2 & Ss & 100 & 180.0$_{-0.6}^{+0.3}$ & 87$_{-2}^{+2}$ & 100 & -- & -- & -- & \\ 
0A10T05 & 2021 LK54 & 3:2 & Ss & 100 & 180$_{-2}^{+1}$ & 92$_{-4}^{+4}$ & 100 & 88.50$_{-0.08}^{+0.10}$ & 114$_{-1}^{+2}$ & 100 & \\ 
3I37C09 & 2020 VT40 & 3:2 & Ss & 100 & 180$_{-2}^{+2}$ & 92$_{-5}^{+6}$ & 100 & 271.60$_{-0.08}^{+0.09}$ & 93.8$_{-0.5}^{+0.5}$ & 100 & \\ 
2112C06 & 2020 VR40 & 3:2 & Ss & 100 & 180.0$_{-1.0}^{+0.9}$ & 122$_{-5}^{+7}$ & 100 & 269.59$_{-0.04}^{+0.06}$ & 38.1$_{-0.4}^{+0.5}$ & 100 & \\ 
1W10B04 & 2020 VE46 & 3:2 & Ii & 70 & 180$_{-3}^{+3}$ & 270$_{-50}^{+40}$ & 100 & -- & -- & -- & Repeated slipping, but resonant behavior dominates  \\ 
0403U07 & 2021 LB57 & 3:2 & Ss & 100 & 180$_{-2}^{+2}$ & 123$_{-6}^{+9}$ & 100 & 91.1$_{-0.4}^{+0.5}$ & 107$_{-2}^{+1}$ & 100 & \\ 
t806E15 & 2020 VQ40 & 3:2 & Ss & 100 & 180$_{-1}^{+1}$ & 172$_{-3}^{+3}$ & 100 & 269.5$_{-0.2}^{+0.2}$ & 101.7$_{-0.9}^{+0.4}$ & 100 & \\ 
0G01C11 & 2020 VA46 & 3:2 & Ss & 100 & 179.9$_{-0.6}^{+0.7}$ & 180$_{-20}^{+10}$ & 100 & -- & -- & -- & \\ 
3C05C09 & 2020 VS40 & 3:2 & Ss & 100 & 180$_{-2}^{+2}$ & 160$_{-20}^{+20}$ & 100 & -- & -- & -- & \\ 
0U16V17 & 2021 LH58 & 3:2 & Ss & 100 & 180$_{-2}^{+2}$ & 206$_{-13}^{+7}$ & 100 & -- & -- & -- & \\ 
3H00V16 & 2021 LJ59 & 3:2 & Ss & 99 & 180.0$_{-0.9}^{+0.9}$ & 230$_{-10}^{+40}$ & 100 & -- & -- & -- & \\ 
0Y36A00 & 2020 VT44 & 7:4 & Ss & 100 & 180$_{-1}^{+1}$ & 128$_{-8}^{+6}$ & 100 & -- & -- & -- & \\ 
0A21T02 & 2021 LK57 & 11:6 & Ss & 63 & 180$_{-4}^{+4}$ & 270$_{-20}^{+20}$ & 100 & -- & -- & -- & \\ 
1207U08 & 2021 LM58 & 13:7:4:2 & Ss & 100 & 180$_{-7}^{+7}$ & 80$_{-20}^{+20}$ & 100 & -- & -- & -- & \\ 
1730U11 & 2021 LP58 & 15:8:5:2 & Is & -- & -- & -- & -- & -- & -- & -- & Majority of clones sticking in resonance  \\ 
1H25D10 & 2020 VY44 & 19:10:7:2 & Is & 42 & 180$_{-9}^{+10}$ & 250$_{-100}^{+50}$ & 100 & -- & -- & -- & \\ 
2812D10 & 2020 VF45 & 21:11 & Is & 5 & 180$_{-8}^{+7}$ & 220$_{-50}^{+60}$ & 100 & -- & -- & -- & Choppy, Intermittent resonant behavior  \\ 
0130T05 & 2021 LP56 & 2:1 & Ss & 80 & 180$_{-4}^{+4}$ & 330$_{-30}^{+20}$ & 100 & -- & -- & -- & \\ 
3X12D11 & 2020 VH46 & 2:1 & Ss & 100 & 78$_{-3}^{+4}$ & 54$_{-5}^{+8}$ & 100 & -- & -- & -- & \\ 
2G39V13 & 2021 LC59 & 2:1 & Ss & 100 & 280.6$_{-1.0}^{+1.0}$ & 67$_{-2}^{+2}$ & 100 & -- & -- & -- & \\ 
1A18D10 & 2020 VC46 & 2:1 & Ss & 100 & 180$_{-3}^{+3}$ & 292$_{-8}^{+6}$ & 100 & -- & -- & -- & \\ 
1B06A00 & 2020 VX44 & 2:1 & Ss & 100 & 82$_{-1}^{+2}$ & 52$_{-4}^{+5}$ & 100 & -- & -- & -- & \\ 
0233T03 & 2021 LW56 & 5:2 & Ss & 100 & 180.0$_{-0.8}^{+0.9}$ & 190$_{-10}^{+10}$ & 100 & -- & -- & -- & \\ 
0123T02 & 2021 LO56 & 5:2 & Ss & 100 & 179.9$_{-0.3}^{+0.5}$ & 110$_{-20}^{+40}$ & 100 & -- & -- & -- & \\ 
1X38V14 & 2021 LV58 & 5:2 & Ss & 100 & 179.8$_{-0.9}^{+0.9}$ & 100$_{-40}^{+40}$ & 100 & 220$_{-16}^{+7}$ & 290$_{-60}^{+50}$ & 100 & \\ 
2S31V13 & 2021 LE59 & 5:2 & Ss & 100 & 180.0$_{-0.9}^{+1.0}$ & 110$_{-30}^{+30}$ & 100 & 183$_{-5}^{+5}$ & 333$_{-3}^{+2}$ & 76 & \\ 
2B38B05 & 2020 VH45 & 5:2 & Ss & 45 & 180$_{-40}^{+40}$ & 290$_{-60}^{+50}$ & 100 & 270.2$_{-1.9}^{+0.4}$ & 43$_{-6}^{+7}$ & 74 & \\ 
1P13A03 & 2020 VZ44 & 5:2 & Si & 99 & 180$_{-10}^{+10}$ & 200$_{-120}^{+80}$ & 100 & 190$_{-20}^{+60}$ & 200$_{-100}^{+200}$ & 57 & \\ 
2G20V15 & 2021 LB59 & 5:2 & Ss & 100 & 180$_{-3}^{+2}$ & 90$_{-20}^{+20}$ & 100 & -- & -- & -- & \\ 
5P33B02 & 2014 WJ517 & 5:2 & Ss & 100 & 180$_{-4}^{+2}$ & 200$_{-30}^{+20}$ & 100 & 154$_{-10}^{+23}$ & 335$_{-4}^{+3}$ & 34 & \\ 
0113U07 & 2021 LN56 & 17:6 & Is & 28 & 180$_{-20}^{+20}$ & 230$_{-60}^{+70}$ & 100 & -- & -- & -- & \\ 
0224T01 & 2021 LU56 & 28:9 & Is & 34 & 180$_{-4}^{+5}$ & 160$_{-80}^{+110}$ & 100 & -- & -- & -- & \\ 
0F17U06 & 2021 LO57 & 7:2 & Ss & 100 & 179.9$_{-0.7}^{+0.8}$ & 180$_{-20}^{+20}$ & 100 & -- & -- & -- & \\ 
0229T03 & 2021 LV56 & 7:2 & Ss & 100 & 180.0$_{-0.5}^{+0.4}$ & 170$_{-10}^{+20}$ & 100 & -- & -- & -- & \\ 
1F32V14 & 2021 LS58 & 21:5 & Ss & 62 & 180$_{-9}^{+10}$ & 110$_{-80}^{+140}$ & 100 & -- & -- & -- & \\ 
1438U11 & 2021 LN58 & 13:2 & Ss & 100 & 179.2$_{-0.4}^{+0.7}$ & 25$_{-7}^{+13}$ & 100 & -- & -- & -- & \\ 
0W14B00 & 2020 VN40 & 10:1 & Ss & 11 & 90$_{-10}^{+20}$ & 110$_{-50}^{+150}$ & 90 & -- & -- & -- & \\ 
\hline
u0C06U06 & 2021 LM57 & 5:3 & Si & 93 & 180$_{-2}^{+1}$ & 170$_{-20}^{+30}$ & 100 & 190$_{-30}^{+30}$ & 290$_{-130}^{+70}$ & 35 & \\ 
u5429B02 & 2020 VR45 & 13:7 & Is & 29 & 180$_{-5}^{+5}$ & 170$_{-70}^{+90}$ & 100 & -- & -- & -- & \\ 
u0V28C07 & 2020 VS44 & 17:8:7:2 & Ss & 100 & 180$_{-10}^{+20}$ & 170$_{-40}^{+60}$ & 100 & -- & -- & -- & \\ 
u2036U11 & 2021 LX58 & 5:2 & Ss & 100 & 180$_{-7}^{+6}$ & 190$_{-20}^{+40}$ & 100 & 100$_{-10}^{+20}$ & 90$_{-40}^{+80}$ & 100 & \\ 
u0L15U09 & 2021 LA58 & 5:1 & Si & 68 & 180$_{-30}^{+30}$ & 300$_{-70}^{+50}$ & 98 & -- & -- & -- & \\ 
u1E26A04 & 2020 VD46 & 7:1 & Ss & 100 & 77$_{-6}^{+10}$ & 50$_{-30}^{+50}$ & 100 & -- & -- & -- & \\ 
\hline\hline
\end{longtable}
\footnotesize{
Additional information to the information presented in \autoref{tab:objects} for the resonant object. 
`Survey ID' is the internal designation that the \lido team gave the object.
`MPC designation' is the provisional designation assigned by the Minor Planet Center, as well as the permanent IAU designation in the cases where that is available.
`Resonance` is the objects class, as determined in \autoref{sec:classify}.
`S/I' indicates whether an object is Securely or Insecurely classified; the first (capital) letter refers to the primary classification, and the second (lower case) letter refers to whether the classification of Kozai or non-Kozai is secure or insecure, while an X indicates a lost object that could not be classified.
`Res Frac' is the percentage of the 201 clones that showed resonant behavious.
`RC' is the resonant center (the angle around which the resonant angle librates).
`RA' is the amplitude (peak to valley) of the resonant angle libration.
`Island Frac' is the percentage of the resonant clones that resonate within the same resonance island.
`KC' is the center value of $\omega$ for objects that feature Kozai resonance.
`KA' is the amplitude of Kozai resonance.
`KFrac' is the fraction of resonant clones that show Kozai resonant behavior for the entire 30~Myr integration. 
}
\end{longrotatetable}

\subsection{Classifying the Discoveries}\label{sec:classify}

We classified the TNOs using forward integrated n-body simulations, to determine how their orbits behave over time.
We use similar criteria to the \citep{Gladman2008} classification scheme, although we use 30~Myr instead of 10~Myr integrations, and 50-100~Myr simulations to confirm $\omega$ oscillation for resonant TNOs beyond $a>50$~au.
As in \citet{Gladman2008} we analyze the integrations and classify the objects into (1) Resonant TNOs, which show libration of a resonant angle $\phi$ for the majority of the integration (2) Scattering TNOs, which show changes to their semi-major axes $\Delta a\geq1.5$~au during the simulation (3) classical TNOs with $39.4<a<47.8$.
We have a few non-scattering objects with $a<39.4$~au, which are labeled as inner classical.
Beyond $a>47.8$, we classify objects as Detached ($e>0.24$) and Extended Hot Classical ($e<0.24$).
There is only one object which falls into this Extended Hot Classical group, and it is just sunward of the 5:2 and likely better classified as high-$q$ \citep{Lawler2019}.
There are no cold classical discoveries in this survey as all of the objects are $i>14\degr$.
For the Detached and Extended Hot Classical objects, these could also be sub-divided in other ways, for example based on their pericenters, $q>35$ or 40~au.

In order to properly classify the objects, we first quantified the uncertainties on the astrometric measurements.
The astrometry acquired for this program specifically has low uncertainties ($\sim0.1\arcsec$, see Section \ref{sec:design}).
A few objects have additional astrometry from other sources (see \autoref{sec:arc-extend}), and for those we used the Minor Planet Center (MPC) uncertainty model, which estimates the astrometric uncertainty for each telescope site by year and calibration catalog used\footnote{This is part of the MPC's internal OrbFit version, generously provided by Federica Spoto. This uncertainty model that will be published soon, along with a weighting scheme paper \citep{Dziadura_in_prep}}.
Note that the MPC uncertainty model for Maunakea observatories in recent years and the catalogs we use is also 0.1$\arcsec$, consistent with our assumed value for our own data.

We utilized the astrometry and uncertainty on each astrometric measurement to generate a cloud of clones consistent with our object's orbit.
We created a nominal orbit by fitting the astrometry using Project Pluto's \textit{Find\_Orb}\footnote{\url{https://www.projectpluto.com/}} \citep{Gray2022}, specifically the non-interactive locally-installed version named \textit{fo}.
We then ``fuzzed'' the astrometry by re-sampling each astrometric data point randomly using a Gaussian uncertainty and the 1$\sigma$ uncertainties from the MPC model.
We created 200 sets of fuzzed astrometry and a new orbital fit for each one, resulting in 201 orbits for each object, all consistent with the original astrometry and spanning the 3$\sigma$ orbital parameter space.
These clones preserve the $a-e$ dependence of the orbital fitting process and produce a representative set of orbits, spanning the same region described by the covariance matrix in \textit{Find\_Orb}.

We integrated all 201 clones of each object using an N-body integrator.
Our simulations used the Canadian Advanced Network for Astronomy Research (CANFAR) cluster\footnote{\url{https://www.opencadc.org/canfar/latest/}} \citep{Gaudet2009, Major2019} and Rebound MERCURIUS \citep{Rein2012,Rein2019}.
We added the mass and momentum of the terrestrial planets to the Sun and the center of mass was adjusted appropriately. 
The four giant planets were included as massive particles and were integrated for 30 Myr using a 0.5~year integration timestep and a write frequency of every 2000 years (15 thousand outputs).
Masses and starting positions were derived from the JPL Ephemeris DE441 via JPL Horizons\footnote{Giorgini, J. D. and JPL Solar System Dynamics Group, NASA/JPL Horizons On-Line Ephemeris System, \url{https://ssd.jpl.nasa.gov/horizons/}}.
These integrations were sufficiently well-sampled for resonant object classification, as minimal false positives were encountered due to beating.

\begin{figure}[tbp]
\centering
 	\includegraphics[width=0.7\textwidth]{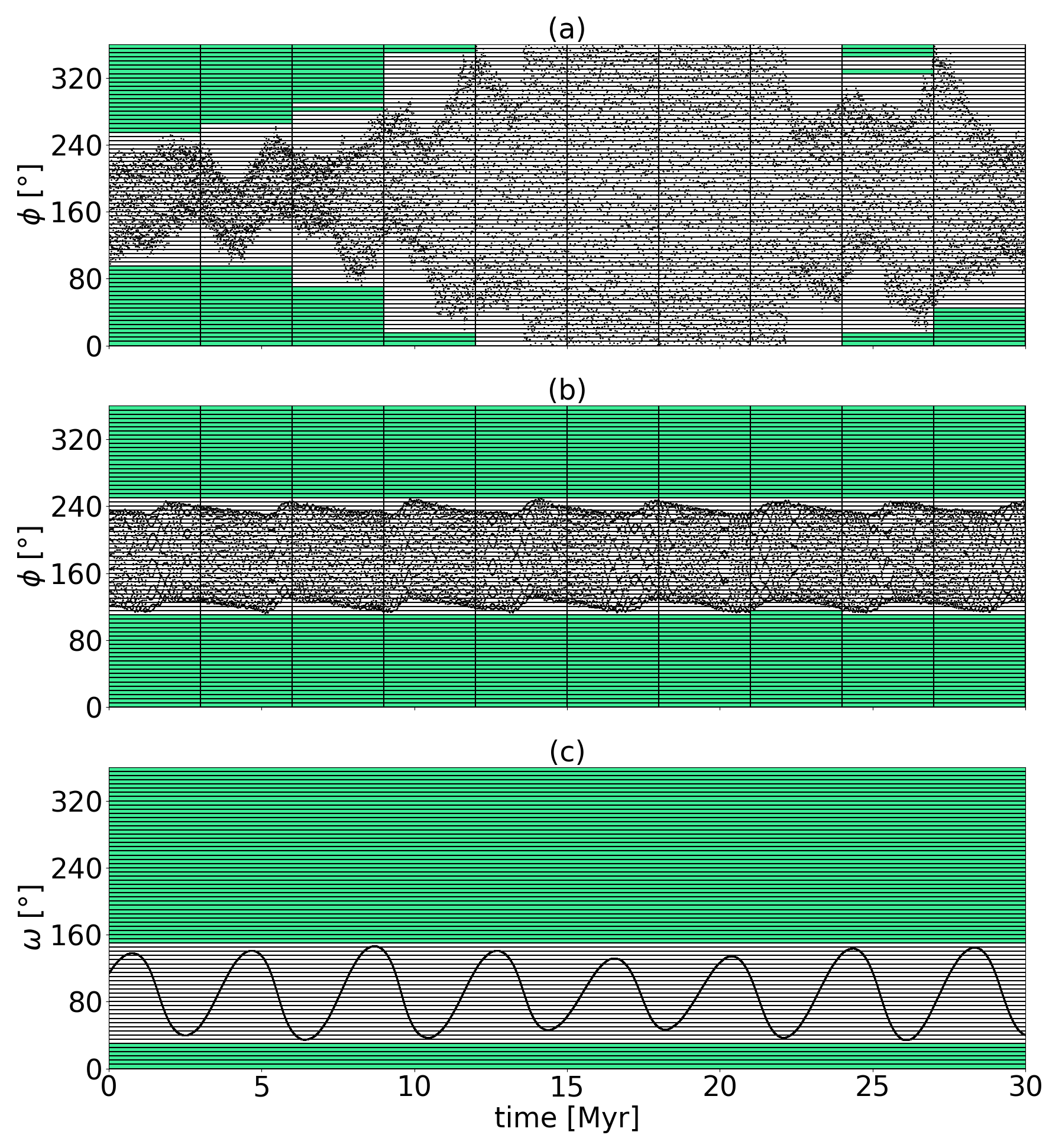}
	\caption{Demonstration of the classification method. The resonant angle vs time (panel (a) and (b)) is split into windows in time and bins in angle; in each time window, it is counted which angle bins are populated. If the object does not span all angles within a window, it is labeled as likely resonant in that window.  The TNO in panel (a) is resonant for some of the windows, and the TNO in panel (b) is resonant for all windows shown. Panel (c) shows the classification method for detecting Kozai resonance by analyzing the argument of pericenter vs. time. However in this case, the data is not windowed in the time direction, as the libration of $\omega$ occurs on a significantly longer timescale; windowing would result in many false positives.}
    \label{fig:window_classifier}
\end{figure}

We created a classification script to determine whether each clone was resonant.
This is identified by searching for confinement of the resonant angle over a series of windows.
Each 30~Myr simulation was divided into ten 3~Myr windows.
The classification script stepped through each window checking for resonant behavior in that window (see \autoref{fig:window_classifier} for a visual example).
We checked the resonant angle 
\begin{equation}
\phi = p\lambda - q\lambda_N - (p-q)\varpi
\end{equation}
for all resonances with $p\le25$, $q\le17$ and $q\le p$ within 2 AU of the average particle semi-major axis within each window, where $\omega$ is the argument of pericenter, $\Omega$ is the longitude of the ascending node, $M$ is the Mean Anomaly, $\lambda = \omega + \Omega + M$, and $\varpi=\omega+\Omega$.
Additional checks on the nominal clone were performed for objects labeled non-resonant, for all resonances $p$ up to 100 and $q$ up to 99.
For TNOs where the classifier identified intermittent periods of resonance and the semi-major axis appeared to show a consistent resonant evolution, mixed inclination-eccentricity mode resonances were also checked.
\begin{equation}
\phi = p\lambda - q\lambda_N - m\varpi - n\Omega
\end{equation}
where: $p=q+m+n$ and $n$ must be an even number.
If the script identified the TNO as resonant within a window, it also checked for Kozai behavior \citep{Kozai1962,Lidov1962}.
Kozai oscillations were diagnosed by checking for oscillation of the argument of pericenter, $\omega$, over the entire 30~Myr simulation as one window.  
The simulations performed were of an appropriate duration to diagnose circulation of $\omega$ for objects near the main belt, but not sufficiently long to record multiple oscillations for the more distant TNOs, $a\gtrsim75$.
The resonant center was determined from the average of the $\phi$ values in each window determined to be resonant, and the amplitude was calculated based on the 2$\sigma$ values (the resonant amplitude is peak to peak and includes 95\% of the points within that window and excludes the highest and lowest values).
If the TNO was identified as non-resonant, the script checks whether it meets the scattering criteria, of a change in semi-major axis of at least 1.5~au over the full simulation.
If the TNO is not scattering or resonant, the script labels it as a Centaur, hot Classical TNO (inner/main/extended), or Detached TNO, which are distinguished purely by $a$ and $e$ cuts.
The outputs of this resonant classification script were the classification of Kozai/non-Kozai, Kozai center, and Kozai amplitude for the full 30~Myrs, and for each window: Resonant or other classification, Resonant Center, Resonant amplitude.

\begin{figure}[tp]
\centering
    \includegraphics[width=0.7\textwidth]{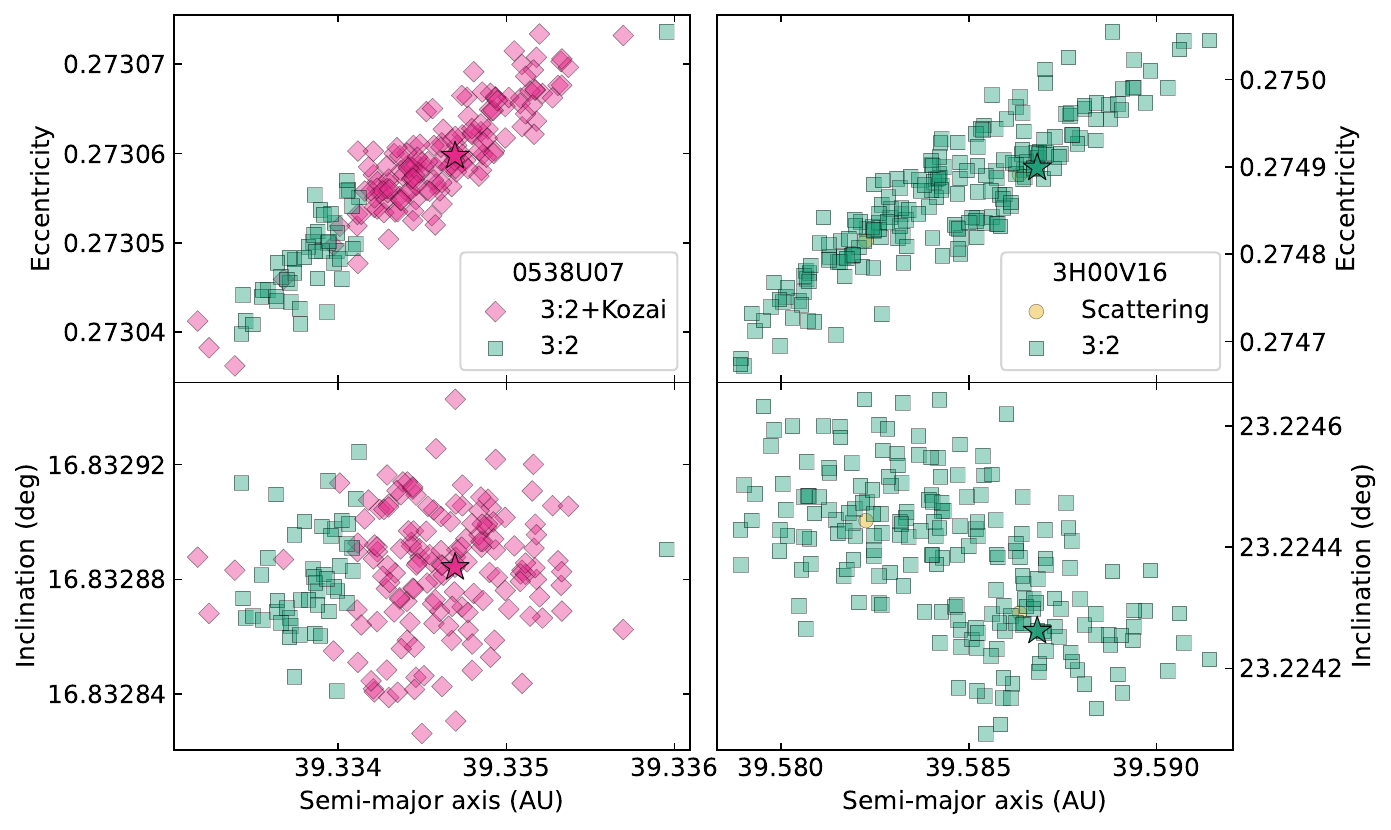}
 	\caption{Example results of classifying all 201 clones of two objects.
    Left: A securely classified \textit {3:2 resonant} object (0538U07, MPC designation 2021 LE57), but with insecure classification on whether it is also \textit{Kozai resonant}; in this case the two options appear to occupy separate phase space, so additional observations would, by reducing the orbital uncertainty, likely be able to reach a secure classification.
    Right: a securely classified \textit{3:2 resonant} objects (3H00V16, MPC designation 2021 LJ59); two clones show scattering behavior, but well over 90\% do not, so this is classified as secure. 
    For the vast majority of the 141 tracked \lido objects, all 201 clones show the same behavior as each other; the two examples shown above are a couple of the exceptions. 
    }
    \label{fig:clone_cloud}
\end{figure}

\begin{figure}[tbp]\label{fig:aei}
\centering
  \includegraphics[width=0.7\textwidth]{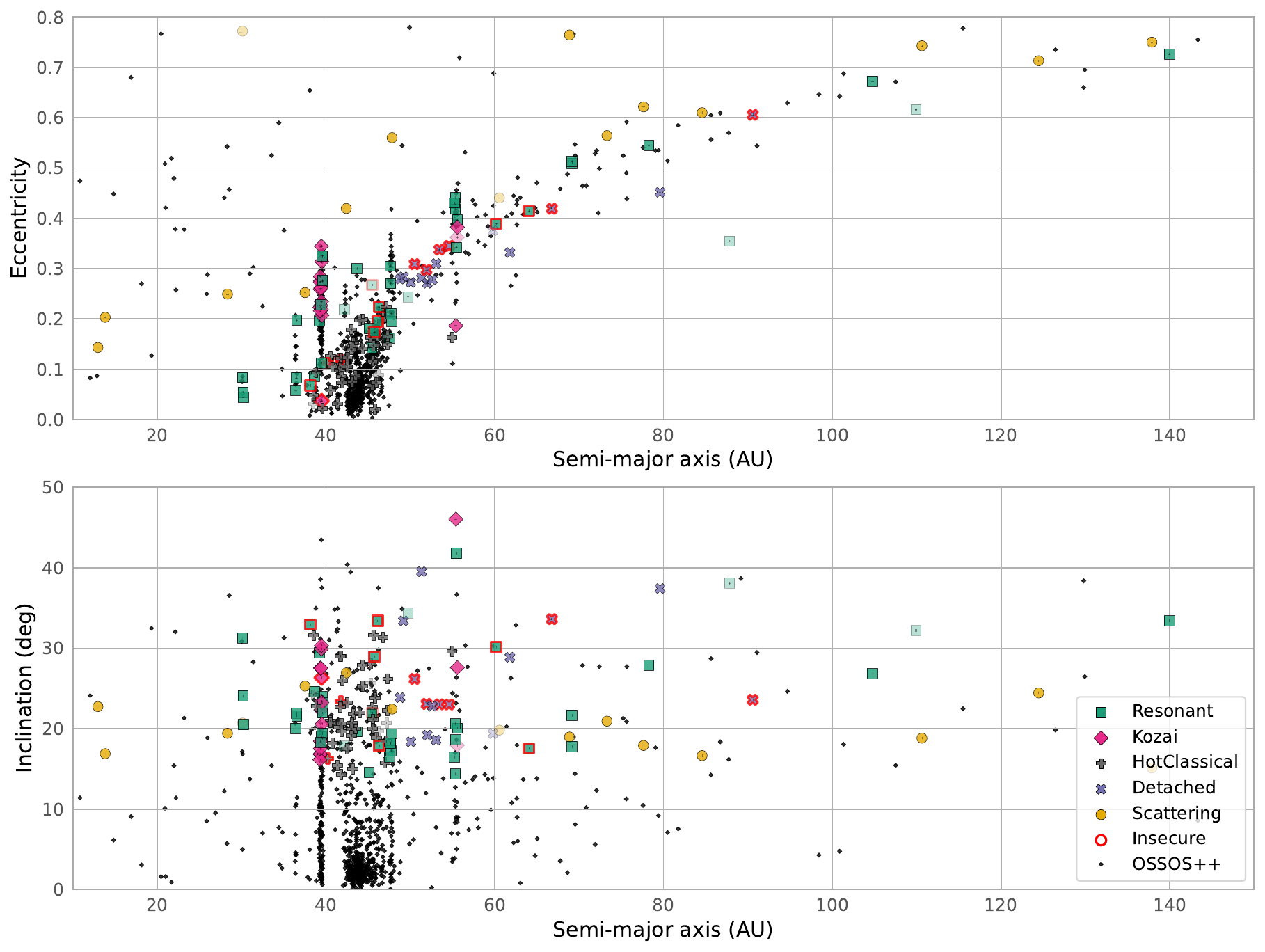}
  \caption{$a$, $e$, $i$ orbital elements and classification for all 141 tracked \lido objects.
  The seven lost objects are not shown, as their orbital parameter uncertainty are close to 100\%.
  The classification of the \lido objects is indicated by the point shape and color: Resonant (green square), Kozai resonant (magenta diamond), hot classical (grey plus), detached (purple cross), scattering (yellow circle); additionally, objects that are insecurely classified are highlighted with a red outline.
  The objects that are not within the characterized survey parameters (fainter than magnitude limit, or faster moving than the motion limit) have high transparency. 
  The OSSOS++ objects are shown with black points for comparison. 
  From this, it is clear that the OSSOS++ and \lido samples are comparable in number for $i>14\degr$.
  }
\end{figure}

The results from the classification scripts were processed in a few different ways.
Due to the high quality astrometry and reasonable arc lengths, the majority of objects were straightforward to classify.
Objects where more than 90\% of the clones were resonant more than 70\% of the time were automatically classified as resonant (see \autoref{fig:clone_cloud}).
This was 43 objects in a range of resonances.
There were a further 3 objects automatically identified as resonant for more than 60\% of the clones; for two we determined that the classifier was under-counting the resonant behavior and for one we determined it was an insecure classification.
We looked in detail at the objects which had some resonant behavior identified, but not $>60$\%, and found many for which the resonant behavior was under-counted, including the mixed mode resonances.
These were manually inspected to determine whether the classification code had identified beating/other non-resonant evolution as resonant or whether the objects were resonant and under-identified using the automated classifications.
Only a handful of objects had 1-2 instances of miss-identification as resonant when they were not resonant and were manually re-classified.
Several other objects were resonant TNOs which had been under-identified for a variety of reasons.
These included objects in an unstable resonant sticking configuration, where they entered and exited the resonance repeatedly throughout the simulation.
There were also objects with unusual resonant behavior, such as $n$:1 resonators with shifting libration centers and mixed mode libration.
When examined in the pure eccentricity resonance, the mixed mode librators found in our survey have a confined $\phi$ on short timescale, but that confined pattern cycles slowly through all angles.
With our classification script using only pure-$e$ resonances, this resulted in some windows being classified as resonant and some windows, where the $\phi$ straddles $0\degr$/$360\degr$, being classified as non-resonant.
We added the relevant mixed modes to our classification script.
We also did a random spot-check of the $a$-evolution of the objects identified as non-resonant and found a few high-order resonators; as a result we re-ran our classification code with an extended list of resonances to test.
The remaining non-resonant objects were classified as identified by the classification script as Centaurs, Scattering TNOs, Hot Classicals, and Detached objects.
Distant non-resonant objects can vary significantly in semi-major axis while being long-term stable, so we also manually inspected all $a>100$~au objects classified as scattering, confirming that they do in fact scatter significantly more than our 1.5 au threshold.
The classifications are listed in \autoref{tab:objects}, and are shown together with $a$, $e$, $i$ elements in \autoref{fig:aei}.

To confirm our classifications, we also used the classification software SBDynT \citep{Volk2025}, updated to identify specific librating resonant angles \citep{Volk:2025dda}.
We found nearly identical classifications, with a few differences due to the different classification timescales.
The SBDynT classifier uses 10~Myrs instead of 30~Myrs, and a few objects were securely classified on 10~Myr but not 30~Myr timescales.

Our survey was well-positioned to find Haumea Family Members, as those typically have inclinations $\sim25-30\degr$ \citep{Ragozzine2007}.
We examined the three objects in our discovery sample (0F36T02, 0K33T00, and 0L13B05) that had $40<a<45$~au, $0.08<e<0.15$, and $25\degr<i<30\degr$.
We calculated the $\delta$v in m~s$^{-1}$ of these objects compared to the Haumea family center from \citet{Ragozzine2007}.
We determined that these three objects have relatively low $\delta$v; 
\begin{itemize}
    \item 0F36T02 = 2021 LQ57, $H_{V}=8.70$: 157 m~s$^{-1}$  
    \item 0K33T00 = 2021 LY57, $H_{V}=8.78$: 151 m~s$^{-1}$  
    \item 0L13B05 = 2020 VQ44, $H_{V}=9.02$: 106 m~s$^{-1}$  
\end{itemize}
Here $H_{w}$ from \autoref{tab:objects} were converted to $H_V$ assuming typical Haumea family colors \citep[$g-r=0.55$, $B-V=0.70$,][]{Snodgrass2010, Schwamb2019} and standard band conversion equations \citep{Bilir2005,Jester2005, Jordi2006}, resulting in a $V-w\sim0.066$ band conversion.
These three objects all have rather large $H$-magnitudes, and assuming an albedo of $\rho_V = 0.48_{-0.18}^{+0.28}$, the median measured albedo of the family \citep{Vilenius2018}, they have diameters of 35, 34 and 30~km, respectively.
The size distribution of the Haumea family is more shallow than the bulk excited Kuiper belt \citep{Proudfoot2019, Pike2020}.
As demonstrated by \citet{Proudfoot2019}, the ratio of background interlopers to Haumea family members is approximately 1:1 at $H_V=7$ and will increase with increasing $H$-magnitude.
Background interlopers will outnumber family members at approximately 10:1 at $H_V\simeq8.5$, depending on the exact $\delta$v \citep{Proudfoot2019}.
Haumea family members have a range of $\delta v$, including many with $14.9\le\delta v<263$m~s$^{-1}$ and 14 objects with $\delta v>100$m~s$^{-1}$ \citep{Proudfoot2019} and even some confirmed members with $\delta v\sim$300m~s$^{-1}$ \citep{Proudfoot2024}.
As a result, even with the similar $\delta$v values for the \lido TNOs compared to the accepted family members, the \lido objects are more likely to be interlopers than family members.
These objects may still merit color or spectral followup, though, as determining the measured fraction of large $H_V$ Haumea family members will be an important test of the measured $H$-distribution of the family.

\subsection{Survey Characterization}
\label{sec:characterization}

In order to be useful for comparison with population models and other surveys, the survey must be characterized. 
In order for a survey to be characterized, parameters like the detection efficiency as a function of magnitude, detection efficiency as a function of rate of motion, survey footprint, false positive rate, tracking/loss fraction, rate of motion cuts, etc. must be well understood and quantified.
This knowledge allows the survey detections (which have survey biases and cuts built in) to be compared to population models by applying the same biases and cuts to those models. 
A survey simulator can thus simulate what the survey would have found if the population model was the true intrinsic distribution. 

The \lido survey was designed in a similar way as the previous TNO surveys from CFHT (CFEPS \citep{Petit2011,Petit2017}, the Alexandersen Survey \citep{Alexandersen2016} and OSSOS \citep{Bannister2018}), so that the discovery observations can be characterized using the same methodology and combined easily with the previous survey characterization files for the OSSOS survey simulator.
We used the same software as those surveys to measure the image point spread function (PSF), and used that PSF to generate 300-1300 synthetic objects per CCD that got implanted in time-scrambled copies of the discovery triplets; numerous copies are made, each with only about 40 synthetic sources per CCD, in order to avoid crowding the field.
The time scrambling is done such that we can also measure the false-positive rate; anything that is detected in the time scrambled data must either be a planted source or a false positive, as nothing real can move in that manner. 
The synthetic sources are generated with a magnitude in the range $22.0\leq m_w\leq 26.0$, with rates of motion from 0.5$\arcsec$/hr (ABCDE blocks) or 1$\arcsec$/hr (TUV blocks) to 10$\arcsec$/hr.
The detection efficiency at $<1\arcsec$/hr decreases dramatically and produced many false positives, so for the analysis of TUV objects we focused on the narrower rates of motion which included our targets of interest.
The images with implanted sources was subsequently treated identically to the real images, getting passed through the moving object pipeline at the same time in one large batch. 
Source candidates from the implanted and real images were subsequently vetted together (see \autoref{sec:discoveries}), with the human vetter unaware of whether they were looking at a real or implanted source, to ensure the implanted sources that are detected and accepted have the same biases as the real detections.

\subsubsection{Detection efficiency}
Once all the implanted sources have been vetted, we can measure the detection efficiency as a function of magnitude and rate of motion\footnote{In theory the detection efficiency could also be measured as a function of the angle of motion, but that has not been done by any of the previous surveys and it is likely not a strong dependency.}. 
The list of known implanted sources is compared with the list of detections from the time-scrambled images to identify which sources were detected and which were not (and also which detections were false positive, see below).
As can be seen in \autoref{fig:efficiency_vs_rate_vs_magnitude}, detection efficiency depends both on the rate of motion and on the magnitude of an object; this is because fast objects trail accross more pixels (resulting in a lower signal-to-noise ratio than a slow moving object) and very slow-moving faint objects can be difficult to distinguish from stationary sources.
In the OSSOS survey simulator, detection efficiency is parameterized as a parametric function of magnitude, where this function can then be different for different rate ranges.

\begin{figure}[tbp]\label{fig:efficiency_vs_rate_vs_magnitude}
\centering
  \includegraphics[width=0.42\textwidth]{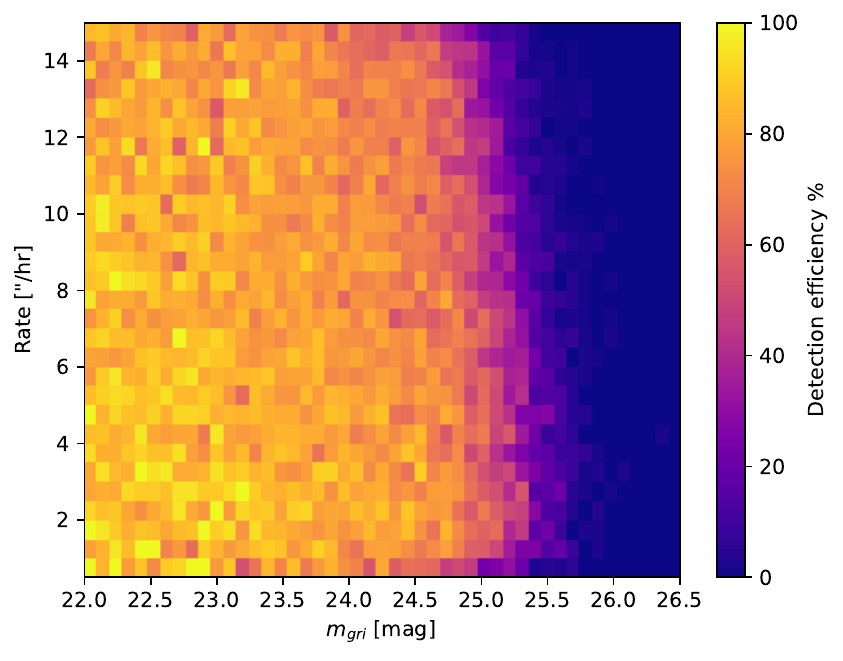}
  \includegraphics[width=0.42\textwidth]{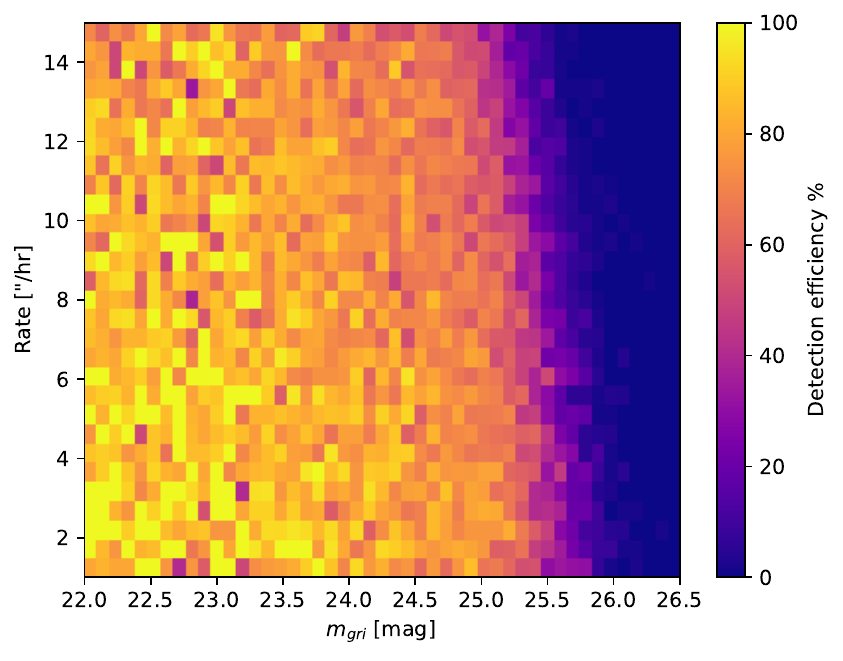}
  \caption{
  Rate of motion in $\arcsec$/hr vs magnitude $m_{gri}$, with color indicating the detection efficiency.
  The 2020 Nov block (ABCDE) is on the left, and the 2021 Jun block (TUV) is on the right.
  It is clear that the efficiency drops rapidly in the $0.5-1.0\arcsec$/hr bin (only present on the left plot), so these objects were characterized separately from the $1.0-2.5\arcsec$/hr bin.
  }
\end{figure}

\begin{figure}[tbp]\label{fig:efficiency_curves_tno_rates}
\centering
  \includegraphics[width=0.42\textwidth]{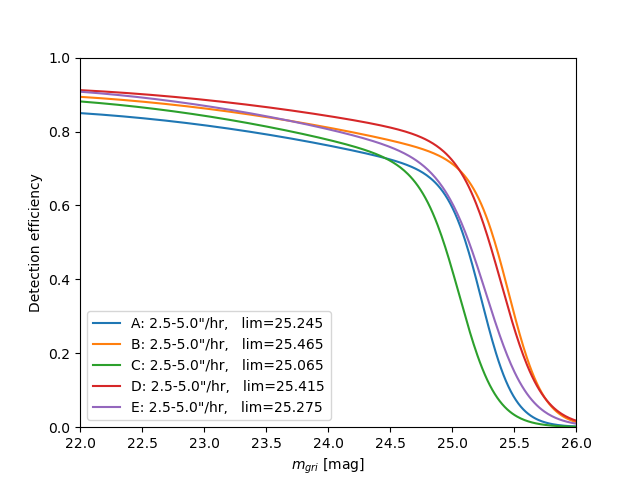}
  \includegraphics[width=0.42\textwidth]{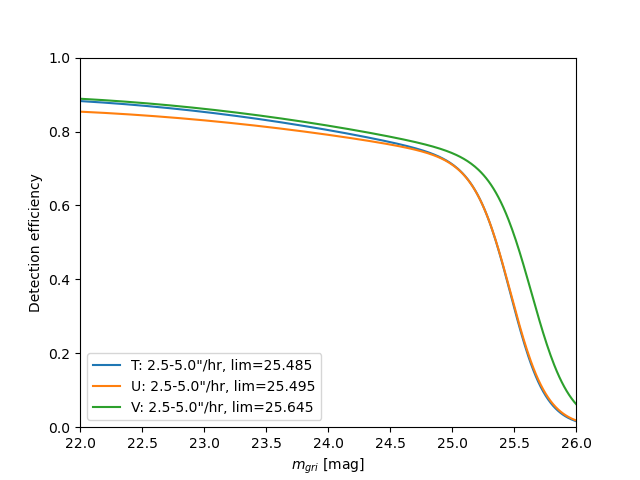}
  \caption{Efficiency curves for the rate bin 2.5-5.0\arcsec/hr (corresponding to most TNOs, \textasciitilde{}30-60~au) for each block of discovery fields.
  Autumn 2020 (left), Spring 2021 (right).
  These curves represent \autoref{eq:efficiency_function} fitted to the binned detection efficiency of implanted sources.
  The detection efficiency curves of every rate bin, as well as a representation of the data used to fit these curves, can  be seen in \autoref{fig:efficiency_curves}.
  }
\end{figure}

The implanted objects were divided into bins, typically 0.1 mag wide\footnote{Bins 0.2-0.5 mag wide were used for the smaller D-block and the narrow 0.5\arcsec-1.0\arcsec rate range where there were insufficient implanted sources to use 0.1 mag bins}, and the fraction of implanted objects that were successfully found was calculated in each bin.
A parametric function was fit to this data in order to have a parametric representation of the detection efficiency. 
This process was done separately for sources in different rate of motion bins. 
The efficiency function change as a function of rate is slight, so a few bins is sufficient.
For \lido we have used the following rate of motion bins: 0.5-1.0\arcsec/hr (ABCDE blocks only), 1.0-2.5\arcsec/hr, 2.5-5.0\arcsec/hr, 5.0-7.5\arcsec/hr and 7.5-10.0\arcsec/hr.
While we did plant synthetic sources at rates 10-15\arcsec/hr, we chose to not characterize those rates (corresponding to Centaurs at distances internal to about 15 au) as the survey tracking was not designed to track such fast moving objects. 
We have used the same functional form as OSSOS and \citet{Alexandersen2016} to fit the detection efficiency, originally described in \citet{Gladman2009}:
\begin{equation}\label{eq:efficiency_function}
\eta(m_r)=\frac{\eta_{21}-c(m_w-21)^2}{1+\exp\left(\frac{m_w-m_e}{w}\right)},
\end{equation}
where $\eta_{21}\simeq80-95\%$ is the maximum efficiency close to $m_w\simeq21$ (the detection efficiency is assumed to be constant for $m_w<21$), $c\simeq0.002-0.02$ is a measure for the strength of a quadratic drop (caused by crowding; objects being obscured by background stars is increasingly frequent for fainter objects), $m_e\simeq25.0-25.7$ is the characteristic magnitude where the function transitions to an exponential tail (as detection becomes noise limited) of width $w\simeq0.1-0.2$.
The detection efficiency, and thus these parameters, depend on the seeing conditions in the triplet images, both the absolute value but also the consistency between the three images. 
We therefore divided our observed triplets into eight blocks, such that only triplets observed on the same night are characterized together. 
The detection efficiency parameters for the eight blocks for the various rate bins are available in DOI\footnote{\href{https://doi.org/10.11570/26.0024}{doi:~10.11570/26.0024}} in files named $LiDO_?-block.eff$ where $?\in(A, B, C, D, E, T, U, V)$.
The detection efficiency of each block for typical TNO rates (2.5-5.0\arcsec/hr) can be seen in \autoref{fig:efficiency_curves_tno_rates} and other rates can be seen in \autoref{fig:efficiency_curves}.

\subsubsection{Magnitude limits}

As with the previous similar surveys, we have taken the limiting magnitude, the magnitude past which we no longer trust the characterization, to be where the detection efficiency has fallen to 40\% of its maximum value. 
The limiting magnitudes of each block and rate of motion is listed in the \textit{LiDO\_?-block.eff} files in the DOI\footnote{\href{https://doi.org/10.11570/26.0024}{doi:~10.11570/26.0024}}.
For reference, for the rate bin most relevant for TNOs (2.5-5.0\arcsec/hr, $\sim$60-30 au), the limiting magnitudes of our blocks are: A=25.245, B=25.465, C=25.065, D=25.415, E=25.275, T=25.485, U=25.495, V=25.645; the efficiency curve for each block in this rate bin can be seen in \autoref{fig:efficiency_curves_tno_rates}. 
It is clear that the spring-semester blocks (TUV) are much deeper than the fall semester blocks (ABCDE). 
This is not surprising, and is partly compensated for by the fact A/B and C/D blocks are repeats of each other, as discussed in \autoref{sec:triplets}.

\subsubsection{False positives}

As our synthetic sources are implanted in time-scrambled images, all sources detected in those images must either be implanted sources or false positives; no real object can move in a manner that it would be detected in these images. 
We can therefore compare the list of detected sources with the list of implanted sources and know that any detections that were not implanted are false positives. 
After human vetting (see \autoref{sec:discoveries}), we had seven known false positives from the time-scrambled images.
However, the brightest of these had mean magnitude $m_w=25.6$ and all seven false positives were well beyond the characterization limit of the block they were detected in. 
We thus conclude that while there may be a few false positives from the detections from the real data images, those false positives are likely all beyond the characterization limit of the survey.
The fact that we achieved 100\% tracking for the characterized sample supports that there were no bright false positives.

\subsubsection{Survey footprint}

\begin{figure}[tbp] \label{fig:survey_footprintABCDE}
\centering
 	\includegraphics[width=0.35\textwidth]{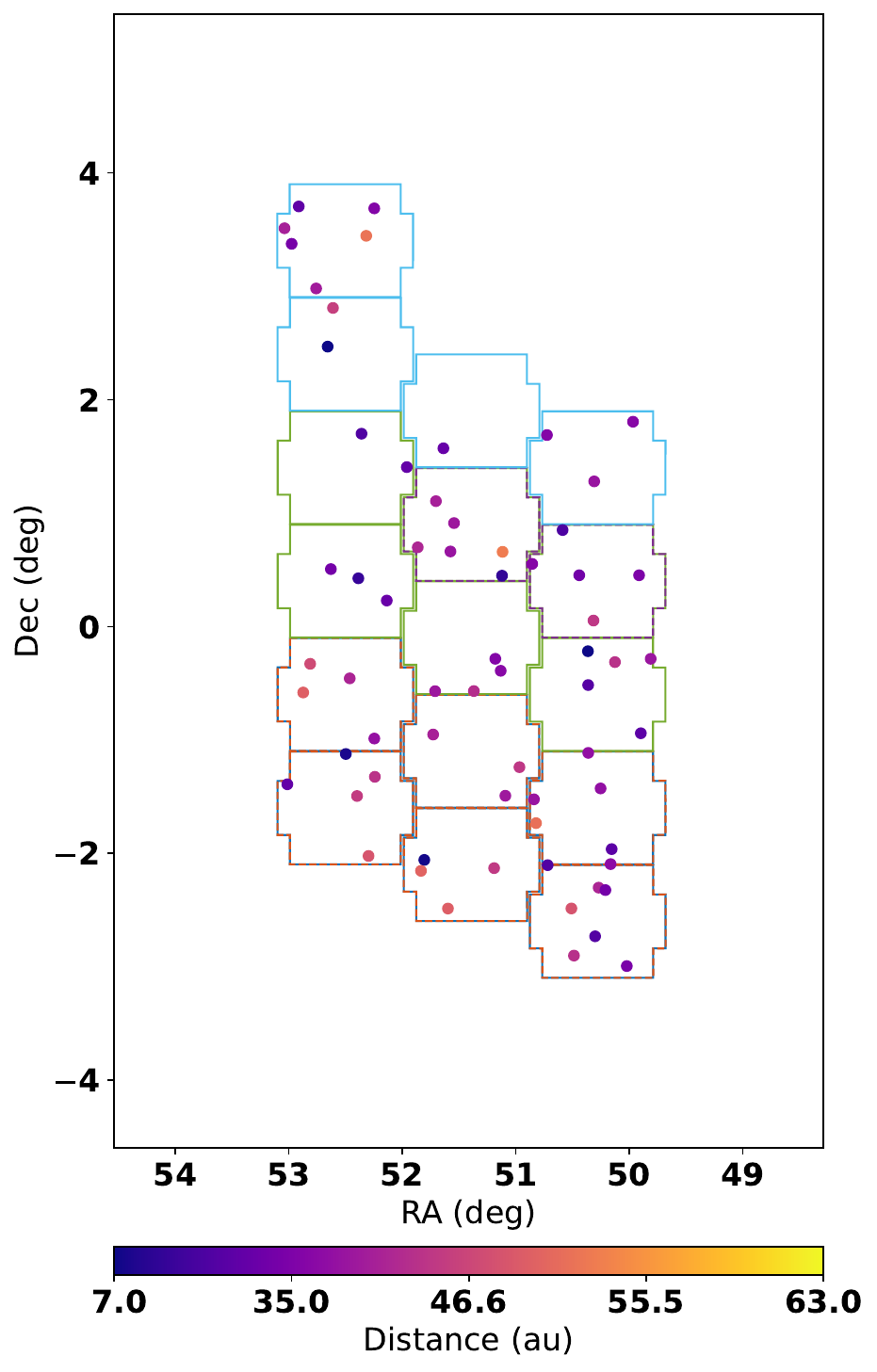}
	\caption{
    Survey footprint of the fall 2020 blocks, with each field and object plotted on the night of discovery.
    Each outlined polygon shows one CFHT MegaCam field, while the color of the field shows which block that field belongs to; blue (A-block), dashed orange (B-block), green (C-block), dashed purple (D-block) and cyan (E-block).
    Note that A and B block were the exact same pointings on two sequential nights (the same goes for D and two of the fields from C block).
    In the fall semester, the fields were held fixed for the whole discovery lunation, so the tesselation looks perfect.
    The online HTML version of this publication contains a video lasting 3 minutes and 21 seconds long and consuming 7.1 MB. 
    This video is an animated version of this plot, showing all the observed fields and the motion of all the objects from 2020-01-01 to 2025-06-22.
    In the animations, each frame corresponds to one day.
    Objects are circled in black on days that they were measured, and all exposures that resulted in at least one object being measured have their outline plotted (i.e. unused poor-seeing/cloud-hampered exposures were not plotted). 
    The uncertainty in the location of each object on each night, based on the final orbit published in this work, is also plotted as an ellipse and can, in the animation, be seen growing rapidly for the lost objects as time diverges from the discovery night, while for most other objects it is hardly visible at all.
    The figure in the PDF version represent a snapshot of each block on the discovery night.
    A similar plot for the spring 2021 blocks can be seen in \autoref{fig:survey_footprintTUV}.
    }
\end{figure}

\begin{figure}[tbp] \label{fig:survey_footprintTUV}
\centering
    \includegraphics[width=0.35\textwidth]{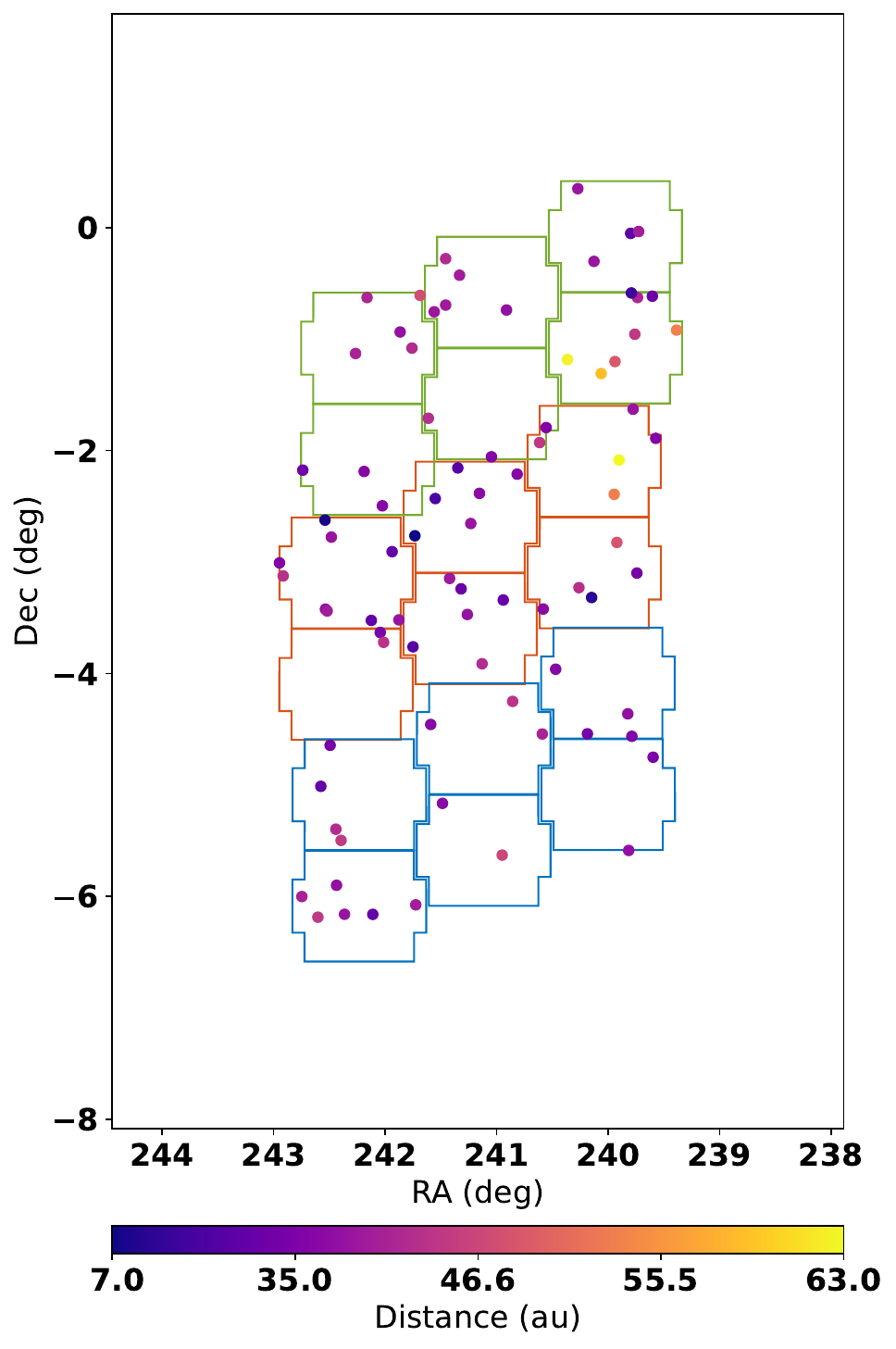}\\
	\caption{
    Survey footprint, as in \autoref{fig:survey_footprintABCDE}, but for the spring 2021 blocks, with each field and object plotted on the night of discovery.
    Each outlined polygon shows one CFHT MegaCam field, while the color of the field shows which block that field belongs; blue (T-block), orange (U-block) and green (V-block).
    In the spring semester, the fields were shifted at the average TNO rate to maximize the usefulness of the 2020 observations; the U block was observed first, T block 5 days later and V block an additional 3 nights later. 
    This spread of the observing nights and the shifting fields causes apparent gaps and overlaps in the tesselation.
    However, in terms of not having gaps/overlap in the observed TNOs this layout is in fact optimal.
    For example, the two objects in the apparent overlap between the green V-block and orange U-block were not seen in U-block because they were off the edge and only moved into that part of the sky by the time V-block was observed 8 nights later. 
    The online HTML version of this publication contains a video lasting 3 minutes and 21 seconds and consuming 5.1 MB. 
    That video is an animated version of this plot, showing all the observed fields and the motion of all the objects from 2020-01-01 to 2025-06-22.
    In the animation, each frame corresponds to one day.
    Objects are circled in black on days that they were measured, and all exposures that resulted in at least one object being measured have their outline plotted (i.e. unused poor-seeing/cloud-hampered exposures were not plotted). 
    The uncertainty in the location of each object on each night, based on the final orbit published in this work, is also plotted as an ellipse and can, in the animation, be seen growing rapidly for the lost objects as time diverges from the discovery night, while for most other objects it is hardly visible at all.
    The figure in the PDF version represents a snapshot of each block on the discovery night.
    }
\end{figure}

In order to accurately simulate the survey, it is important to know exactly where the survey pointed and on what dates, as that has very large impact on which objects are detectable.
\autoref{fig:survey_footprintABCDE} and \autoref{fig:survey_footprintTUV} show the footprint of \lido. 
Previous surveys have lumped multiple fields (usually all fields that are contiguous and observed on the same night) together into a polygon and calculated the fill factor (fraction of the polygon filled with ccd pixels) for each block of fields. 
We have instead calculated the fill factor for a single MegaCam image, as well as the shape of a single image relative to the center, and then used the same shape and fill factor for 42 identical polygons corresponding to each telescope pointing. 
This results in slightly slower survey simulations, as there are many more "survey footprints" to check whether any given object was within, but will also result in a more accurate simulation of the actual survey. 
\autoref{tab:polygon} shows the shape of every field relative to the field centers listed in \autoref{tab:fields}.
We also measured the shape (and thus area) of each of the 40 CCD chips, allowing us to calculate a fill factor of $93.15\%$ (the remaining being gaps between chips).

\begin{table}[tbp]
\setlength{\tabcolsep}{2.5pt}
\caption{Shape of MegaPrime field of view, with offsets in degrees along the RA and Dec axis, relative to the pointing coordinate (loosely referred to as the ``center'').}
\begin{center}
\begin{tabular}{c c} \label{tab:polygon}
$\Delta_{RA}$ & $\Delta_{Dec}$ \\ \hline
0.49220295 & 0.24865226 \\
0.59738348 & 0.24870259 \\
0.60126822 & -0.22868328 \\
0.49589672 & -0.23003771 \\
0.49237541 & -0.48772592 \\
-0.49606355 & -0.49342289 \\
-0.50266439 & -0.23590507 \\
-0.60801834 & -0.23574556 \\
-0.61018697 & 0.24149839 \\
-0.50502853 & 0.24265469 \\
-0.50048081 & 0.50044066 \\
0.48433452 & 0.50631853 \\
\hline
\end{tabular}
\end{center}
\end{table}

\begin{table}[tbp]
\setlength{\tabcolsep}{2.5pt}
\caption{Field centers, observed date (center time of first exposure of triplet), which block they belong to for characterization purposes.}
\begin{center}
\begin{tabular}{c c c c} \label{tab:fields}
RA          & Dec         & Date            & Block\#\# \\ \hline
03:30:01.31 & -01:36:30.0 & 2459161.8467031517 & A00 \\
03:25:34.46 & -02:06:30.0 & 2459161.851372452  & A01 \\
03:21:07.62 & -02:36:30.0 & 2459161.8557066517 & A02 \\
03:30:01.18 & -00:36:30.0 & 2459161.8340165517 & A03 \\
03:25:34.46 & -01:06:30.0 & 2459161.838202852  & A04 \\
03:21:07.75 & -01:36:30.0 & 2459161.842388352  & A05 \\
03:30:01.31 & -01:36:30.0 & 2459162.829743552  & B00 \\
03:25:34.46 & -02:06:30.0 & 2459162.833928852  & B01 \\
03:21:07.62 & -02:36:30.0 & 2459162.838120552  & B02 \\
03:30:01.18 & -00:36:30.0 & 2459162.842308952  & B03 \\
03:25:34.46 & -01:06:30.0 & 2459162.8466951516 & B04 \\
03:21:07.75 & -01:36:30.0 & 2459162.850881952  & B05 \\
03:30:01.13 & +00:23:30.0 & 2459162.9063057518 & C06 \\
03:25:34.46 & -00:06:30.0 & 2459162.911997352  & C07 \\
03:21:07.80 & -00:36:30.0 & 2459162.9161842517 & C08 \\
03:30:01.16 & +01:23:30.0 & 2459162.920477252  & C09 \\
03:25:34.46 & +00:53:30.0 & 2459162.924664552  & C10 \\
03:21:07.77 & +00:23:30.0 & 2459162.928851952  & C11 \\
03:25:34.46 & +00:53:30.0 & 2459163.812470052  & D10 \\
03:21:07.77 & +00:23:30.0 & 2459163.816657352  & D11 \\
03:30:01.28 & +02:23:30.0 & 2459163.872671952  & E12 \\
03:25:34.46 & +01:53:30.0 & 2459163.876862852  & E13 \\
03:21:07.65 & +01:23:30.0 & 2459163.881050152  & E14 \\
03:30:01.47 & +03:23:30.0 & 2459163.885236852  & E15 \\
\hline
16:00:00.64 & -05:05:49.5 & 2459373.837067252  & T00 \\
16:04:28.58 & -05:35:49.5 & 2459373.8412817516 & T01 \\
16:08:56.53 & -06:05:49.5 & 2459373.845498952  & T02 \\
16:00:01.05 & -04:05:49.5 & 2459373.849716852  & T03 \\
16:04:28.58 & -04:35:49.5 & 2459373.8539318517 & T04 \\
16:08:56.11 & -05:05:49.5 & 2459373.858160052  & T05 \\
16:00:30.48 & -03:06:34.7 & 2459368.821320452  & U06 \\
16:04:57.68 & -03:36:34.7 & 2459368.825546452  & U07 \\
16:09:24.88 & -04:06:34.7 & 2459368.829763852  & U08 \\
16:00:30.74 & -02:06:34.7 & 2459368.833983252  & U09 \\
16:04:57.68 & -02:36:34.7 & 2459368.838201052  & U10 \\
16:09:24.62 & -03:06:34.7 & 2459368.842419352  & U11 \\
16:08:38.18 & -01:05:30.2 & 2459376.833452452  & V12 \\
15:59:44.73 & -01:05:30.2 & 2459376.837670152  & V13 \\
16:04:11.50 & -01:35:30.2 & 2459376.841890052  & V14 \\
16:08:38.27 & -02:05:30.2 & 2459376.846107352  & V15 \\
15:59:44.82 & -00:05:30.2 & 2459376.8504609517 & V16 \\
16:04:11.50 & -00:35:30.2 & 2459376.854927652  & V17 \\
\hline
\end{tabular}
\end{center}
\end{table}

\subsubsection{Tracking fraction}

It is necessary to know the fraction of objects that were actually tracked sufficiently for the object to be included in any analysis work; if some fraction of objects were lost and can't be included, the same fraction of simulated model detections should be ``lost'' and not included in the analysis. 
Fortunately the \lido survey achieved 100\% tracking of objects within the characterized sample. 
Only objects that were fainter than the magnitude limit of the block they were detected in, and objects moving faster than our 10\arcsec/hr limit, were lost. 
We have attempted to parameterize the tracking loss rate for the uncharacterized objects in the \mbox{\textit{lido\_?-block.eff}} files in the DOI\footnote{\href{https://doi.org/10.11570/26.0024}{doi:~10.11570/26.0024}}, but since these parameters do not affect the characterized sample, they should not affect the results and will not be discussed further here.

\subsubsection{Magnitude error parameterization} \label{sec:trombone}

In order for a survey simulator to simulate the apparent magnitudes in a way consistent with the real objects, it must know the error distribution as a function of magnitude, as well as whether one, two or three measurements went into the average.
Most real objects will have three measurements, but some might have two or one if their photometry would be polluted in some of the images, for example by a nearby star, chip defect or cosmic ray.
The \textit{LiDO\_?.eff} files in the DOI\footnote{\href{https://doi.org/10.11570/26.0024}{doi:~10.11570/26.0024}} contain a \textit{phot\_frac} parameter, which encompasses the fraction of objects in that block that has one, two or three photometric measurements included in their averages discovery magnitude. 
The \textit{LiDO\_?.eff} files also contain the six \textit{mag\_error} parameters, which describe the growth of the magnitude uncertainty towards fainter magnitudes as well as how asymmetric the error distribution is, as described in Appendix A of \citep{Bannister2016}. 
We calculated these parameters using the same software as \citet{Bannister2016}.
A graphical representation of the \textit{mag\_error} functions and parameters, and the data they were fit to, can be seen in \autoref{fig:trombone_plots}.

\section{Constraining the orbital distribution of Hot Classical TNOs}
\label{sec:science}

\lido has 125 characterized mid to high inclination objects, which provides a strong constraint on the inclination distribution of many distant TNO sub-populations.
OSSOS has 185 objects with $i>14\degr$; a comparable number to \lido, while the full OSSOS++ sample has 292 objects with $i>14\degr$.
The addition of \lido to the OSSOS++ sample provides a significant increase in the sample of $i>14\degr$ objects.
We investigate the Hot Classical TNOs in OSSOS++ and \lido, focusing in particular on the features of the $i>14\degr$ portion of the sample.
The OSSOS++ sample has 86 characterized hot classical TNOs with $i>14\degr$ and between 40-48~au, and \lido has 38.
The \lido detections provide useful constraints on the orbital distribution of the hot classical TNOs.

\subsection{The \lido Hot Classical TNOs}
Here we consider the characterized non-resonant and non-scattering objects with semi-major axes between 40-48~au as the hot classical TNO sample.
The \lido survey discovered 44 TNOs that meet these orbital element criteria, 38 of which are in the characterized sample (the rest are fainter than the magnitude limit).
We note that there is an extended outer classical object in \lido, near the 5:2 resonance, which is more likely to be dynamically linked to the 5:2 than the main classical belt, as well as three `inner' hot classicals.
These objects are not considered in this analysis.
The sample of hot classical \lido objects show significant sub-structure in their semi-major axis distribution and eccentricity distribution, see the cumulative distribution of real objects in Figure \ref{fig:lidoCumul}.
Much of the semi-major axis distribution is clearly related to the proximity of some of Neptune's mean motion resonances, and is also apparent in the OSSOS++ sample.
The \lido hot classical objects specifically sample the poorly constrained $i>14\degr$ population and provide important constraints on the orbital distribution of the population when combined with the previous OSSOS++ known hot classicals.

\subsection{The OSSOS++ Classical TNOs}
The OSSOS++ detections have been utilized in a variety of works to create population models, \citep[e.g.][]{Lawler2013,Volk2016,Petit2017}.
While the orbital and absolute magnitude distribution ($H$) of most TNO subpopulations detected by OSSOS are well-described by parametric distributions \citep[e.g.,][]{Shankman2017,Crompvoets2022}, the OSSOS model of the hot classical TNO orbital and $H$-distribution is calculated by directly debiasing the real discovered TNOs \citep{Petit2023}. 
This is done by dividing the hot classical phase space into individual cells in $a$, $q$, $\sin i$, and $H_r$. 
The detection bias for each cell with a real TNO in it is then inverted to create a debiased distribution, down to a limiting $H_r$ magnitude of 8.3. 
The debiased distribution is then ``smeared'' using a kernel density estimator in order to avoid the unrealistic fluctuations imposed by a small number of real TNOs in very small multidimensional bins.
The resulting $H$-magnitude distribution is the same for the hot population \citep{Petit2023} and the cold population \citep{Kavelaars2021}, and is strikingly similar to the size distribution obtained by \citet{Napier2024}.
This $H$-distribution is described as an exponentially tapered power law size distribution, see \citet{Kavelaars2021} for additional details.
Once debiased, the complex orbital distribution of the classical TNOs produced a good model of the cold and low-$i$ classical belt because the large number of TNO discoveries in that phase space combined with the relatively constrained range of inclinations resulted in a sufficient density to map the region.
At larger inclination, the density is lower due to both the reduced number of detections and the greater volume available.
As a result, the OSSOS++ debiased hot classical model has visually apparent gaps at mid to high inclinations, which may be either intrinsic features or sampling effects.
The debiased OSSOS++ model unsurprisingly provides a statistically acceptable match for the OSSOS++ hot classical population \citep{Petit2017}, and testing the model with new independent surveys is critical.

\subsection{Modeling the Hot Classical Population}

\begin{figure}
    \centering
    \includegraphics[width=0.7\linewidth]{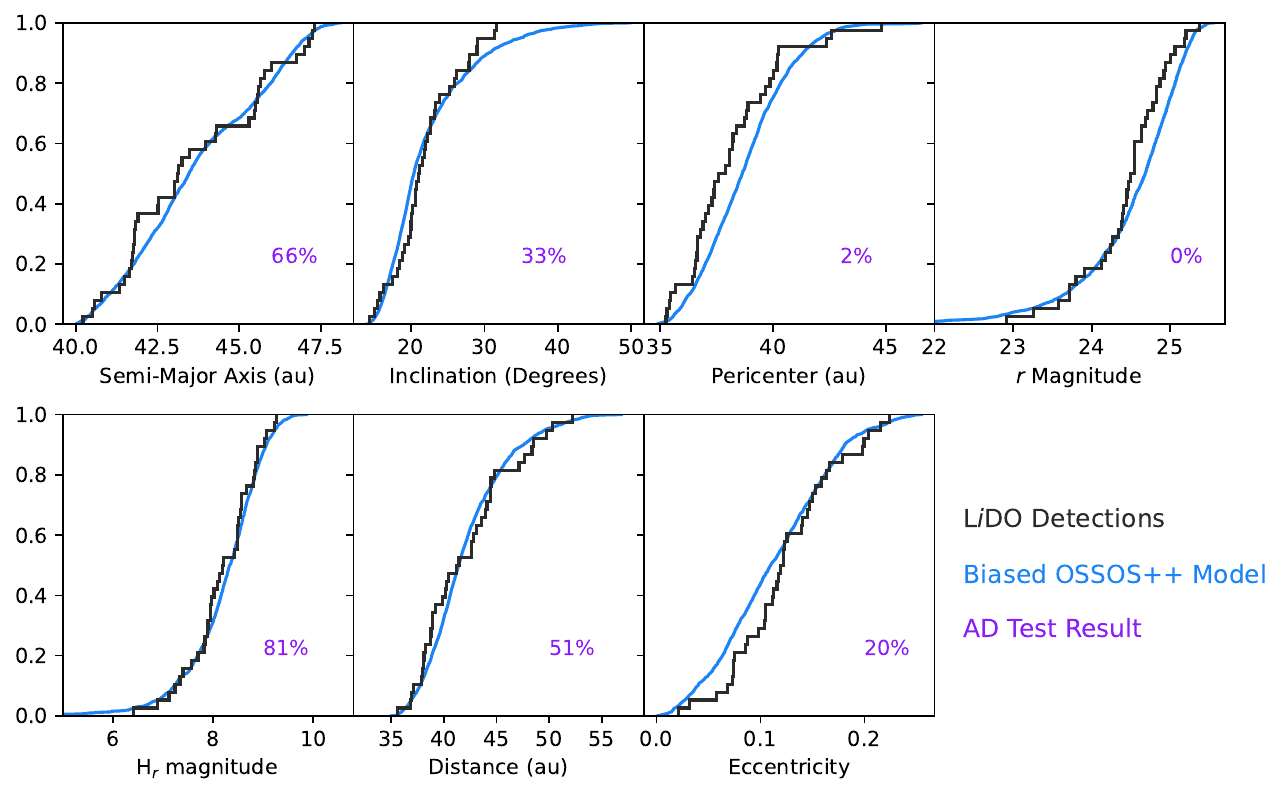}
    \caption{The cumulative distribution of orbital elements for the OSSOS++ model, biased to match \lido, compared to the \lido detections.  The pericenter and $r$ magnitude distributions are statistically rejectable. While it would likely be possible to introduce an $e$-based weighting to adjust the OSOSS++ model, or to add additional model objects derived from the \lido objects, issues resulting from the under-sampled objects used to create the model would reappear each time the real discovery list is expanded. As a result of how the model was generated, the $a$-distribution of the model is very smoothed, and does not reproduce several real features in the real object distribution found in both \lido and OSSOS++ $i>14\degr$ objects.}
    \label{fig:ossospp_cumul}
\end{figure}

We first tested the OSSOS++ model of the hot classical TNOs against the \lido and combined OSSOS++ and \lido samples.
We found that the OSSOS++ model was statistically rejectable in pericenter and apparent magnitude distributions, see \autoref{fig:ossospp_cumul} for details.
Additionally, the real objects show considerable structure in semi-major axis (gaps with no detections) that is not reflected in the OSSOS++ model.
We attribute this discrepency to the comparatively small number of large-$i$ detections compared to the sample at low inclinations, and the resulting low density of detections in this phase space.
Even with smearing, the OSSOS++ model does not adequately fill the phase space at $i>14\degr$.
We were unable to make simple changes to this model to extend it to fit the \lido data, and we decided not to simply repeat the direct debiasing for the \lido objects, as addition of the \lido data is not sufficiently large to solve the under-sampling issue at large-$i$; the same issue would be expected to recur as soon as another sample was added.
As a result, we used a different approach to create our model of the $a-e-i$ distribution for the hot classical population.

We observed that the majority of the visually apparent structure in the semi-major axis distribution appeared to correlate with regions of the phase space that we do not expect to be long-term stable.
The apparent gaps are visible in both the \lido and OSSOS++ data, and many are clearly correlated with mean motion resonances which fall within the main belt region.
Additional unstable regions are expected due to secular modes, which are especially difficult to calculate analytically.
To understand the dynamical stability of the hot classical region, we used N-body simulations to explore the effects of long-term steady-state evolution on a smooth initial model of the hot classical belt.

\begin{figure}
    \centering
    \includegraphics[width=0.7\linewidth]{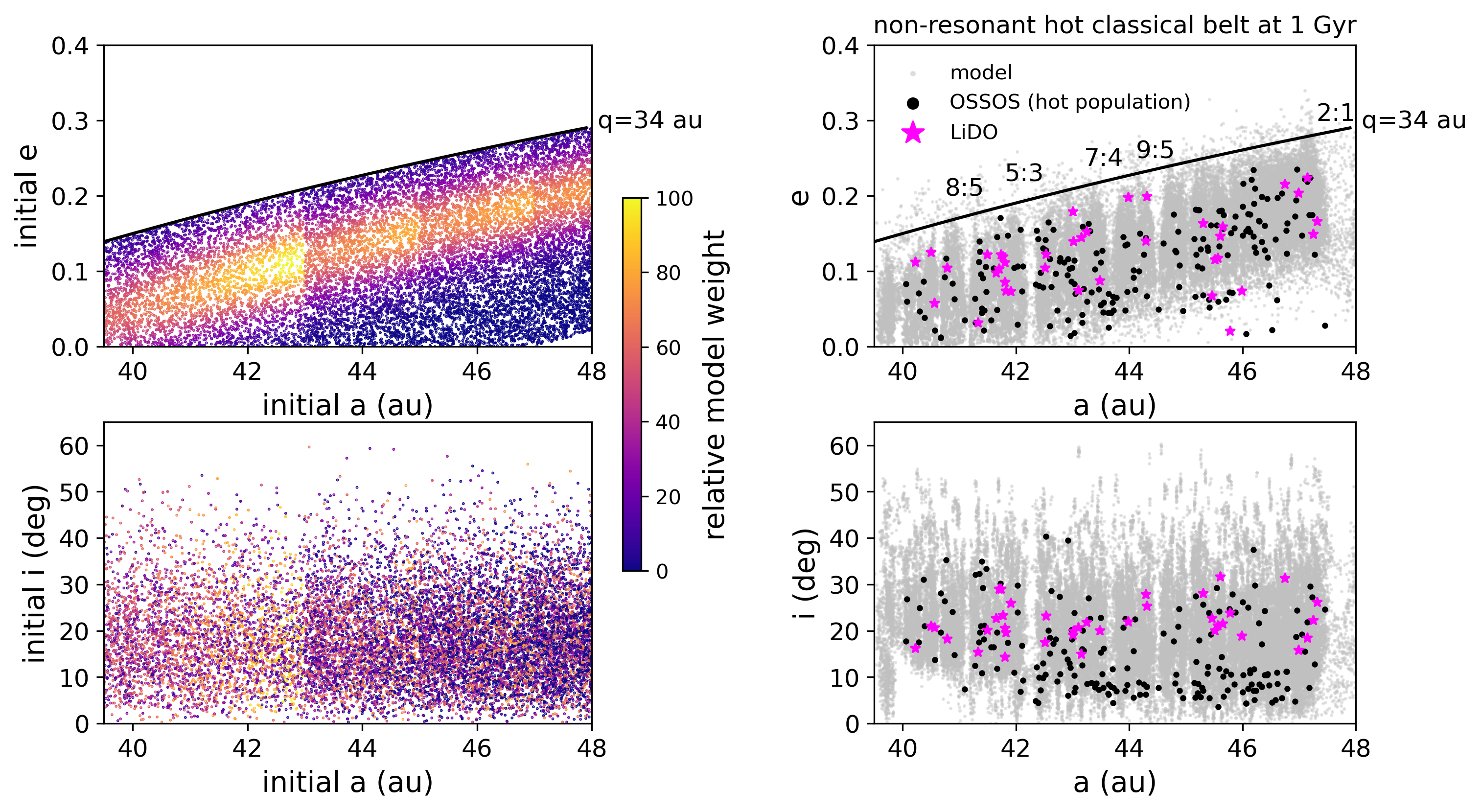}
    \caption{Left panels: hot classical region test particle initial conditions in eccentricity vs semi-major axis (top) and inclination vs semi-major axis (bottom). Each test particle is color-coded by its weight in our parameterized model for the hot classical region's initial filling; note that the abrupt changes in weighting factors at distinct semi-major axis values is simply a correction factor that smooths the semi-major axis distribution relative to how we selected the boundaries of initial bins for simulation initialization purposes and not abrupt changes in the final model. 
    Right panels: stability-sculpted hot classical population after 1 Gyr of evolution with surviving resonant and scattering particles removed; surviving particles were cloned according to their model weight.
    The hot classical detections from OSSOS (black circles) and \lido (pink stars) are shown for comparison.
    We label some of the significant classical belt resonances that have clearly associated gaps in the top right panel.
    }
    \label{fig:hc_model}
\end{figure}

We built a hot classical model based on stability sculpting from a smooth initial distribution of test particles.
As a baseline, we ran a 1 Gyr simulation of 13100 test particles in the main classical belt region using rebound's MERCURIUS integrator \citep{Rein2012,Rein2019}.
Similar to our classification integrations in \autoref{sec:classify}, these integrations added the mass/momentum of the terrestrial planets to the Sun and only included the four giant planets as massive perturbing particles.
Masses and starting positions were derived from the JPL Ephemeris DE441 via JPL Horizons\footnote{Giorgini, J. D. and JPL Solar System Dynamics Group, NASA/JPL Horizons On-Line Ephemeris System, \url{https://ssd.jpl.nasa.gov/horizons/}}.
The massless TNO particles were initialized on barycentric osculating orbits with $a=39.5$--$48$~au and pericenter distances in the range $q=34$--$min(a,47)$~au, as non-resonant TNOs with $q<34$ are not expected to be stable. 
The initial ecliptic inclinations of the particles were drawn from sin($i$) times a Gaussian with a width of $\sigma_i=15\degr$; a width consistent with the present-day hot population as constrained by OSSOS \citep{VanLaerhoven2019}.
All orbital angles (argument of pericenter, longitude of ascending node, and mean anomaly) were set randomly from 0--$2\pi$.
These initial particles are shown in the left panels of \autoref{fig:hc_model}.
Note that the initial, raw distribution of test particles has some discontinuities in particle density as a function of semimajor axis, but in all analyses below, we use weighting factors to resample the particles into our desired smooth distributions in $a$ and $q$/$e$. 
The particles were integrated with a timestep of 0.4 yr and the four giant planets in their current configuration for 1~Gyr.
The outputs of the simulation from $t=990$--1000 Myr were used to classify all remaining particles according to the \cite{Gladman2008} scheme, using the Small Body Dynamics Tool (SBDynT) TNO classification software \citep{Volk2025,Spencer2026,Volk2026Zenodo}.
We used these classifications to remove all resonant and scattering particles from the main belt region, resulting in a set of stability-sculpted hot classical particles.

Once we had a set of stability-sculpted particles, we used a re-sampling procedure to effectively re-weight our final hot classical particles according to different initial assumed $a$ and $q$ distributions.
This is done by `cloning' each final particle a different number of times by randomly sampling its final 10 Myr of evolution a prescribed number of times determined by the weighting factor of its initial orbit; for particles that are removed or are scattering/resonant at the end of the simulation, they are not cloned, regardless of their initial orbit's weighting factor.
Once we had the full list of particles with orbital parameters, we assigned each one an $H_r$ magnitude. 
We used the OSSOS++ $H_r$ magnitude distribution directly from the OSSOS++ model \citep{Petit2023}.
Then the full model was ready to be analyzed with a survey simulator and compared to the real detections.

We utilized the OSSOS survey simulator \citep{Petit2011, Lawler2018}, the OSSOS++ characterizations and detections \citep{Bannister2018}, and the \lido characterization and detections to test the hot classical models.
The input test particle model is provided to the survey simulator, and the survey simulator replicates the biases of the surveys used and outputs a list of biased detections.
These detections can be directly compared to the real detections.
We ran a variety of survey simulations, including OSSOS++, \lido, and the combination of the two surveys.
We assumed an average color offset of $g-r=0.7$ and $r-w=0.15$, which is necessary as some of CFEPS was conducted in $g$-band, \lido was conducted in $w$-band, and the rest of the survey pointings were in $r$-band.
We note that the bulk of the $g$-band data is shallower and on the ecliptic plane, and thus resulted in few $i>14\degr$ detections.
The effect of the small $w$ band offset from $r$ band was noticeable in the biased apparent magnitude distribution if uncorrected, thus we are confident in these color transformations.
The survey simulator was run until several thousand biased model objects were ``detected'' so that a statistical comparison could be made between the biased model and the real detections.

The Anderson-Darling (AD) statistical test \citep{ad1954} was used to compare the biased model objects with the real TNOs.
This test quantifies the distance between the real objects and the biased model.
In order to determine whether this number is significant, we use a bootstrapping methodology, where we draw randomly from the model until we have the same number of drawn model objects as our real object sample, then compute the AD value for this fake sample.
We repeat this process 500 times, determining a list of AD values, and then compute where our real objects fall within that distribution as a percentage.
If the percentage is 5\% or greater, we find that the model is not rejectable at $2\sigma$ significance.
We used this statistical analysis on a variety of models created using our stability-sculpted particles.

\begin{figure}[tbp]
    \centering
    \includegraphics[width=0.5\linewidth]{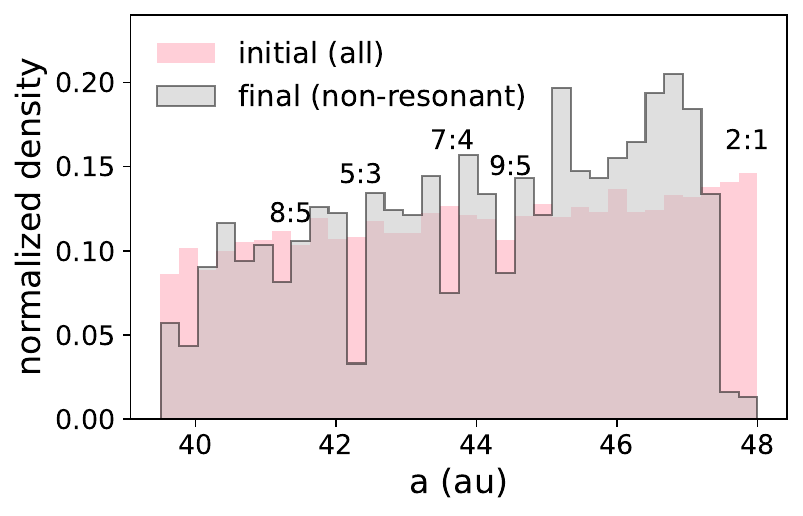}
    \caption{Initial (pink) vs 1 Gyr stability sculpted (outlined gray) non-resonant hot belt semimajor axis distribution. The major resonances sculpting this distribution are indicated. }
    \label{fig:hc_model_adist}
\end{figure}

We first tested our stability sculpted model by confirming that uniformly filling all available classical belt $a$-$q$ space (weighting our initial particles to yield initially equal numbers of particles per square-au within that space, down to our minimum pericenter distance) and running the stability sculpting simulation does not result in a final hot classical population that is consistent with the \lido or OSSOS detections.
This uniform filling was statistically rejectable at 2$\sigma$ significance for the $r$ magnitude, pericenter, and eccentricity distributions.
We experimented with different boundaries in the $q$ distribution while maintaining uniformity in $a$-$q$, but did not find acceptable matches to the observations.
To find an acceptable match to the observations, we ended up prescribing weighting factors to yield an initial $a$-$q$ distribution such that the pericenter distance distribution at any given $a$ value is set by a gaussian with a peak at $q=38$~au and a width of $\sigma_q=1.85$~au (truncated as necessary at an upper bound corresponding to $e=0$ and an initial lower bound of $q=34$~au).
We also added a semi-major axis-dependent weighting factor proportional to $a^{-1}$ (based on a particle's initial semi-major axis) to model an initial population with an $a$-dependent density per square-au rather than uniform density.
We iteratively experimented over a coarsely sampled range of values for the semi-major axis exponent (from -2 to 2) and of the center and width of the $q$ distribution ($q=34\pm3$, $37\pm3.5$, $39\pm1.75$, and $38\pm(1.75-2)$) to find a satisfactory model.
The $q$-based weighting is required to produce statistically acceptable $e$ and $q$ distributions, while the $a$-based weighting adjustment results in an improved semi-major axis distribution match that is not strictly statistically required.
As a result, we do not reject different $a$-based weighting schemes, only identify that weighting method as a useful adjustment which is likely to be relevant for future work with larger sample populations.
While we save a more thorough parameter search for future work, we have selected a preferred non-rejectable model for the remainder of our analysis based on both the statistical results and the qualitative match to the $a$-distribution.
These relative weighting factors for our preferred non-rejectable model are shown with colors in the left panels of Figure~\ref{fig:hc_model}, where our selected initial $q$ distribution appears clearly as the brighter curve in $a$-$e$ space. 
As mentioned above, the initial set of test particles is not entirely smooth in $a$, so the semi-major axis weighting is less clear; Figure~\ref{fig:hc_model_adist} shows a histogram of our initial weighed $a$ distribution for clarity, as well as the resulting stability-sculpted $a$ distribution.
These weighting factors were used to clone the final particles, and those clones are shown as a scatter-plot in the right side of Figure~\ref{fig:hc_model}.
As discussed below in Section~\ref{ss:hc_survey_sims}, this model provides an acceptable match to the \lido detections and the combined high-inclination OSSOS and \lido detections.

\subsection{Hot Classical Model Results}\label{ss:hc_survey_sims}

\begin{figure}[ptb]
\centering
  \includegraphics[width=0.7\textwidth]{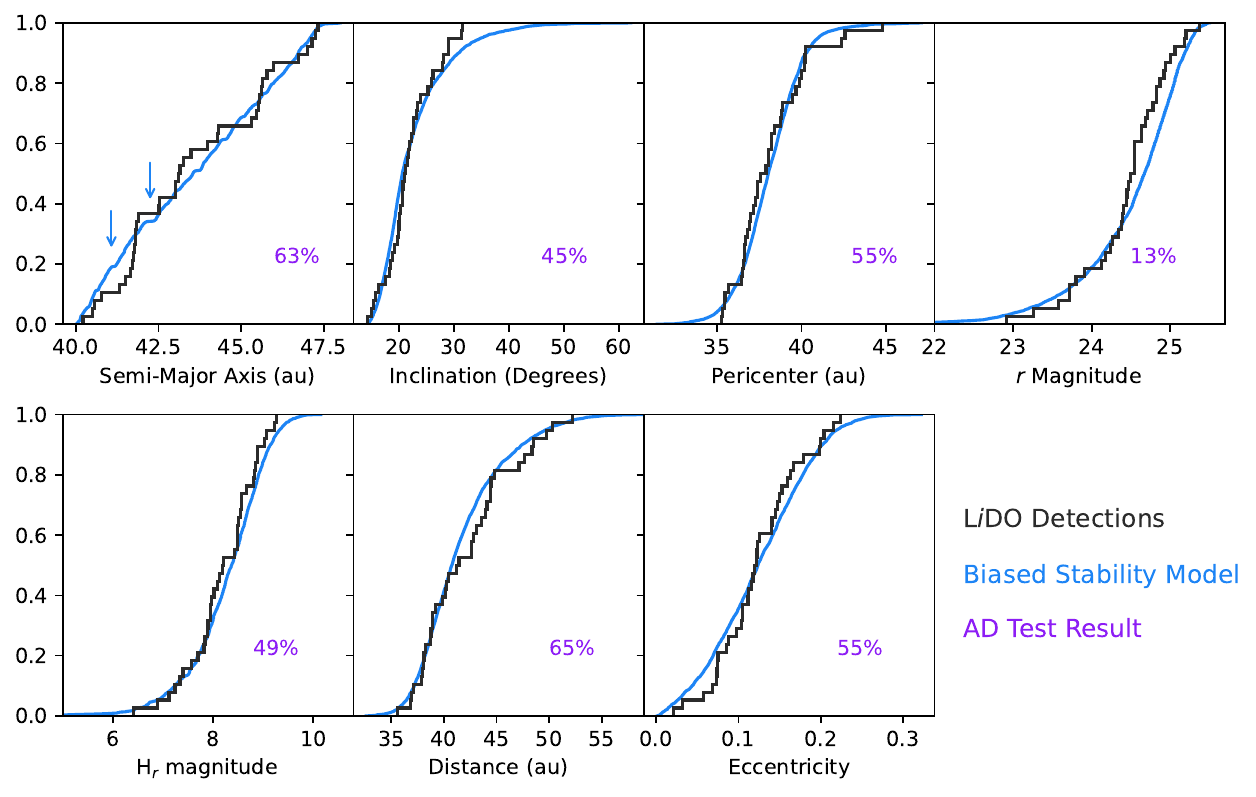}
  \caption{
    Cumulative distributions of \lido TNOs and survey simulator biased model objects from our stability-sculpted hot classical TNO model, in seven orbital element/magnitude parameters: semi-major axis $a$, inclination $i$, pericenter distance $q$, magnitude in $r$-band $m_r$, absolute magnitude $H_r$, distance at detection $d$, and eccentricity $e$.
    Several gaps in $a$ are apparent in the real detections (black).
    The biased model (blue) produces a noticeable gap in $a$ at $\sim42.5$~au, and a narrow gap at $\sim41$~au, indicated by blue arrows in the upper left panel.
    These gap positions match the real object gaps, but the width is considerably narrower.
    Overall, the $a$-distribution of the stability sculpted model has some clear structural similarities as well as noticeable differences compared to the real \lido tno $a$-distribution.
    This includes structures in the real TNO $e$-distribution that are not all reflected in the biased model.
    The AD statistic percentages for each parameter are shown in the relevant plots, and all are $>5$\% and therefore non-rejectable.
    } \label{fig:lidoCumul}
\end{figure}

\begin{figure}[ptb]
\centering
  \includegraphics[width=0.7\textwidth]{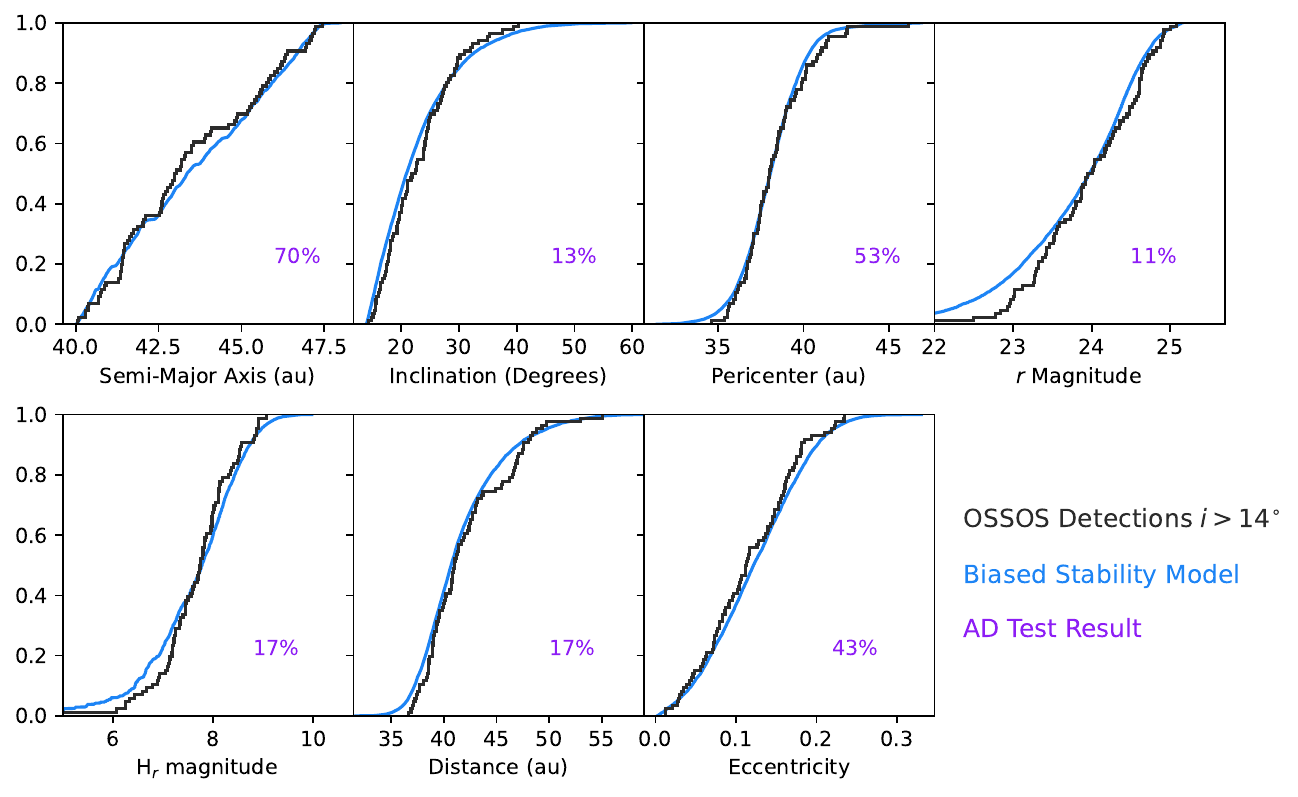}
  \caption{
    Cumulative distributions as in \autoref{fig:lidoCumul}, but for real TNOs with $i>14\degr$ from OSSOS++.
    Similar structure as in \autoref{fig:lidoCumul} is apparent in the real OSSOS++ $a$-distribution, and our biased model has a slightly higher AD statistic when compared to OSSOS++.
    The distance distribution shows a stronger version of the dip seen in the \lido objects at approximately 45~au, which is not seen in the biased model.
    The biased model is non-rejectable based on the AD statistics in all 7 parameters.
    } \label{fig:ossosCumul}
\end{figure}

\begin{figure}[tbp]
\centering
  \includegraphics[width=0.7\textwidth]{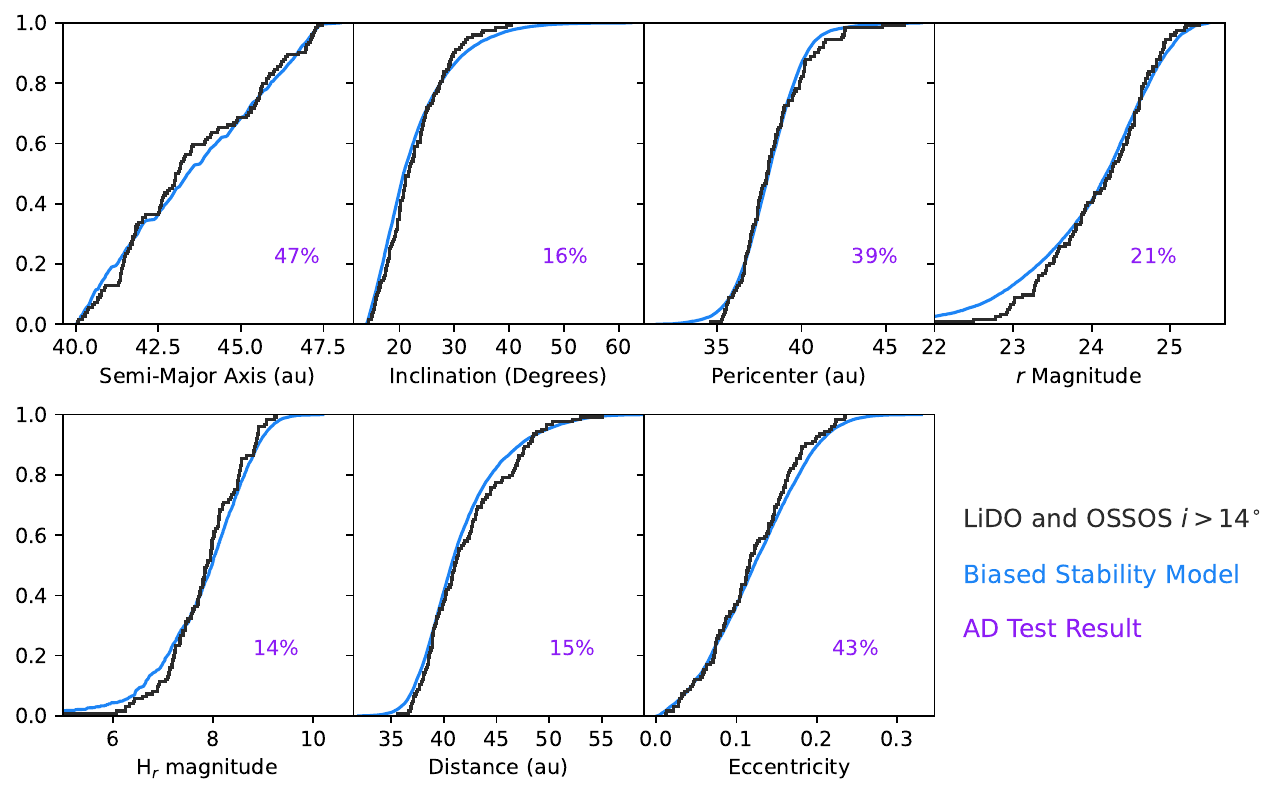}
  \caption{
    Cumulative distributions as in Figures \ref{fig:lidoCumul} and \ref{fig:ossosCumul}, but with the two surveys combined.
    Using the real detections and survey characterization of both surveys simultaneously is also non-rejectable according to the AD test. 
    As in both individual surveys, the stability model shows some promising features, but does not reproduce all of the sub-structure in semi-major axis.
    } \label{fig:bothCumul}
\end{figure}

We tested a variety of weighted models of stability-sculpted hot classical particles using the survey simulator.
We present the results of one acceptable model here, but we note that we were also able to identify other slightly different weighted distributions that were also non-rejectable. 
This model is intended to be a proof of concept for generating an $a$-$e$-$i$ distribution using a smooth input distribution, sculpted by the current Solar System architecture to model the mid to high inclination classical object distribution.

We analyzed the weighted stability-sculpted hot classical model with the OSSOS++ size distribution with the survey simulator. 
We did separate survey simulations to determine which model objects were ``detected'' for OSSOS++, \lido, and both surveys combined.
Our analysis focuses on the $i>14\degr$ objects, as this is the inclination range where \lido is sensitive.
For OSSOS++, we ran the survey simulation on the full model, but only compared the model detections that were $i>14\degr$ to the real OSSOS++ objects that were $i>14\degr$.
The results of these simulations are shown in Figures \ref{fig:lidoCumul}, \ref{fig:ossosCumul}, and \ref{fig:bothCumul}.
Comparing Figures \ref{fig:lidoCumul} and \ref{fig:ossosCumul}, many of the same gaps are visually apparent.
The stability-sculpted hot classical model does reproduce some of the semi-major axis distribution gaps seen in the real objects, although for the real objects these features are stronger.
There are likely some improvements that could be made with the input inclination distribution for the model (before it is stability sculpted); running the stability sculpting for a full 4.5 Gyr would likely reduce the $i>40\degr$ tail in the model, which we do not see in the real detections.
There are some features in the eccentricity and distance distribution of the real objects (gaps at certain values, the tails of the e-distribution being under-represented compared to the model, etc.) that are likely real and appear in both \lido and OSSOS++, but are not well reproduced in the stability sculpted model.
The semi-major axis distribution is very promising as a starting point for a different way of producing a model $a$-distribution.
The biased model is non-rejectable for \lido, OSSOS++, and both surveys combined.
Further improvements to this model, including 4.5 Gyr of sculpting, will be the subject of future work.

Using our non-rejectable model, we computed a population estimate for the hot classical TNOs.
We determined the intrinsic number of hot classical objects needed to produce the number of detections found in the \lido survey for three $H_r$ cuts: 8.3, 8.9, and 9.6.
The first two $H_r$ limits were selected for direct comparison to OSSOS++, and the third was selected to encompass the full number of \lido detections.
The \lido survey included 20, 34, and 38 characterized hot classical objects brighter than $H_r$ of 8.3, 8.9, and 9.6 respectively.
We ran the survey simulator 1000 times for models limited to each of these magnitudes, and recorded the number of model objects required to produce the expected number of detections, and computed the mean population and 2$\sigma$ limits.
For $H_r$<8.9, the intrinsic population required by \lido is $41,000^{+14,000}_{-12,000}$.
In \citep{Petit2023}, the hot classical population size using the OSSOS++ debiasing model requires an intrinsic population of $46,000\pm7,000$.
The results for the other limits are reported in Table \ref{tab:popEst}.
This is in excellent agreement with the results from our model, where the orbital distribution was generated in an entirely different way.

\begin{table}[tbp]
\newcommand{\noteOSSOS}{\citet{Petit2023}}
\newcommand{\noteDESA}{\citet{Bernardinelli2025}}
\newcommand{\noteDESB}{Derived from \citet{Bernardinelli2022}, which states that 695 of the 26031 synthetic objects in the CFEPS L7 model \citep{Petit2011} are detectable in their survey simulations, and that there were 251 real detections of the same population. This thus translates to $26031/695*251=9400$. Uncertainties quoted here come from assuming Poisson statistics on the 251 real detections.}
\newcommand{\noteLiDO}{Using the stability-sculpted hot classical model presented in this work.}
\centering
\begin{tabular}{c c c c c c}
\hline
                                       & $H_r < 7.8$                                & $5.5 \leq H_r \leq 8.2$                      & $H_r<$ 8.3                  & $H_r<$ 8.9                   & $H_r<$ 9.6  \\
\hline
OSSOS++\footnote{\noteOSSOS}           & --                                         & --                                           & $20,000\pm3,000$             & $46,000\pm7,000$             & --          \\
DES                            & $9,400\pm1,200$ \footnote{\noteDESB} & $11,487^{+1,775}_{-1,627}$ \footnote{\noteDESA} & --                           & --                           & --          \\
\hline
\lido \footnote{\noteLiDO}             & $8,800^{+5,100}_{-3,900}$                                      & $19,000^{+8,000}_{-6,000}$                                        & $20,000^{+8,000}_{-7,000}$   & $41,000^{+14,000}_{-12,000}$ & $90,000^{+30,000}_{-27,000}$ \\
\lido and OSSOS++ \footnote{\noteLiDO} & $9,200^{+1,700}_{-1,500}$                                      & $19,000^{+3,000}_{-3,000}$                                        & $21,000^{+3,000}_{-3,000}$   & $45,000^{+6,000}_{-5,000}$   &$106,000^{+14,000}_{-12,000}$ \\
\hline
\end{tabular}
\caption{Population estimates for different absolute magnitude ranges and from different surveys. 
The \lido population estimates from \lido use the stability-sculpted hot classical model presented in this work, while the OSSOS++ population estimates use the OSSOS++ hot classical model from \citep{Petit2023}. 
Uncertainties quoted are $2\sigma$.
The combined ``\lido and OSSOS++'' population estimate uses the stability-sculpted hot classical model from this work; combining the two surveys together reduces the population estimate uncertainty.
The \lido and OSOSS++ models produce nearly identical population estimates, which are consistent with \citet{Bernardinelli2022}, but are significantly larger than the estimate in \citet{Bernardinelli2025}, see text for details.}
\label{tab:popEst}
\end{table}

We tested the population estimate agreement by determining the population for \lido, OSSOS++, and the combined surveys using only the $i>14\degr$ portion of the model.
This provides confirmation that the portion of the model we are most focused on modeling gives consistent results for the three different configurations.
We selected only the $i>14\degr$ model objects, and ran them through the survey simulator in the same way.
For \lido, we found the intrinsic population of $i>14\degr$ and $H_r$<8.9 was $31,500^{+10,300}_{-8,900}$, for OSSOS++ it was $26,700^{+5,800}_{-4,700}$, and for the combination it was $27,600^{+5,100}_{-4,700}$ (2$\sigma$ uncertainties).
The \lido estimate is 16\% larger than the OSSOS++ estimate, but it does have larger uncertainties due to the smaller number of objects in \lido that meet the selection criteria.
This is also apparent because the combination of the two models increases the population estimate slightly compared to OSSOS++ alone.
The three sets of population estimates are all within 1$\sigma$ of each other, and confirm that our models produce compatible populations with $i>14\degr$.

From the DES, \citet{Bernardinelli2025} produced a hot classical population estimate for $5.5\leq H_r\leq 8.2$ of $11,487^{1,775}_{-1,627}$.
For the same magnitude range, we get a value of $19,000^{+3,000}_{-3,000}$ ($2\sigma$ uncertainties) based on \lido + OSSOS++ simulations. 
This is in noticeable tension with the \citet{Bernardinelli2025} result, at the $>4\sigma$ level.
The \citet{Bernardinelli2022} DES work does not explicitly state a population estimate, but states that 695 of the 26,031 synthetic objects with $H_r < 7.8$ in the CFEPS L7 model \citep{Petit2011} are detectable in their survey simulations, and that there were 251 real detections of the same population. We can thus derive a population estimate of $26,031/695\times251=9,400\pm1200$, where uncertainties come from assuming Poisson statistics on the 251 real detections. 
For this same magnitude cut, we get a value of $9,200^{+1,700}_{-1,500}$, in perfect agreement with the estimate derived from \citet{Bernardinelli2022}.
The \lido and OSSOS++ estimates are internally consistent and the estimates from \lido and OSSOS++ combined agree well with \citet{Bernardinelli2022}.
The discrepancy between the \citet{Bernardinelli2025} result and the OSSOS++ hot and cold population estimates \citep{Kavelaars2021,Petit2023} and the likelihood that a joint analysis is required to reconcile these results is noted in \citet{Bernardinelli2025}.
While the DES survey is characterized and published together with a DES survey simulator, the large differences in how the surveys were executed and characterized makes running combined survey simulations challenging and beyond the scope of this work.

As our resulting population estimates are consistent with published OSSOS++ estimates, and the estimates at $i>14\degr$ using our stability-sculpted hot classical model are consistent, we calculate a combined population estimate for \lido and OSSOS++.
We note that our model has been extensively tested at $i>14\degr$ and less rigorously tested at lower inclinations, as these objects were not found by the \lido survey where our analysis is focused.  
This population estimate is provided for comparison purposes to the \lido estimate, but some model refinement for the OSSOS++ $i<14\degr$ hot classicals would likely be warranted in future work.
This combined population estimate includes all hot objects (not limited to $i>14\degr$), and uses 155, 184, and 189 OSSOS++ objects in addition to the 20, 34, and 38 \lido objects for the three $H_r$ cuts of 8.3, 8.9, and 9.6.
The results for all three $H_r$ cuts (as well as two other selected for direct comparison with other work) are given in Table \ref{tab:popEst}, and for $H_r$<8.9 the population is $45,000^{+6,000}_{-5,000}$.
This is consistent with the original OSSOS++ population estimates as well as the \lido estimates with our new orbital distribution model.
The version of our stability-sculpted hot classical model used in these population estimates and Figures \ref{fig:lidoCumul} to \ref{fig:bothCumul} is included in the \textit{lido\_hot\_classical\_model.dat} file in the DOI\footnote{\href{https://doi.org/10.11570/26.0024}{doi:~10.11570/26.0024}}.

\section{Discussion}
\label{sec:Discussion}

\lido has explored a higher inclination portion of the outer Solar System than most previous surveys, and discovered 125 characterized objects.
These TNOs and Centaurs all have $i>14\degr$, which makes the discoveries a significant addition to the previous characterized survey efforts.
The \lido survey tracked all characterized discoveries in order to determine robust orbital classifications.
For the near-resonant TNOs, this required very high orbital precision of $\sigma_a/a\lesssim 10^{-3}$.
The classified discoveries include a number of previously unidentified and very interesting populations, which are preferentially large-inclination and thus much more discoverable in an off-ecliptic survey such as \lido. 

\lido discovered quite a few unprecedented resonant TNOs.
These include the first known 10:1 resonator (0W14B00, MPC designation 2020 VN40), which showed the previously undetected libration around 0$\degr$ for some argument of pericenter angles \cite{Pike2025}. The uncharacterized object u4A12E12 (MPC designation 2020 VJ46) is securely in the 5:1 resonance and exhibits the same libration around $0\degr$.  
Additionally, u4A12E12 has one of the highest known pericenter distances, with $q=56.65$ au and an $e=0.35$, the lowest $e$ of the known 5:1 resonators. 
The 7:1 resonator u1E26A04 (MPC designation 2020 VD46) also exibits some of the libration around 0$\degr$, but to a lesser extent than the other distant \lido $n$:1s.
These objects illustrate the ubiquity of unusual resonant behavior and orbits at large inclination.

\lido also identified several mixed inclination-eccentricity mode resonant TNOs, some of which were securely classified in the mixed mode and some of which displayed repeated transient behavior.
The structure of these mixed mode resonances is quite complex, and will be explored in further work \citep{DraperInPrep}.
There are 6 securely classified mixed mode resonators, and 6 that show at least some transient sticking with multiple clones in these mixed modes for varying timescales.
Temporary or long-term capture in these mixed eccentricity-inclination resonances can result in significant evolution of a TNO's eccentricity and inclination, so the long-term effect of these resonances may actually shape the orbital distribution of non-resonant nearby populations, such as the hot classical TNOs, even when the objects are not securely resonant.

These unusual resonant TNOs present a challenge for classification.
Mixed-mode resonators are challenging to identify with a fixed list of equations for different resonant angles, with a classification methodology similar to the one applied here.
If all possible mixed mode angle combinations are included, the list of resonances to compute $\phi$ for to do a classification search can become unreasonably long. 
In the majority of cases, a mixed inclination-eccentricity resonant object will show some confinement of $\phi$ for the pure eccentricity resonance.
This typically manifests as constrained oscillations of $\phi$ centered around a moving center, which slowly circulates.
Depending on the timescale of the $\phi$ structure, these clones are often flagged as having some resonant behavior.
We were able to identify these objects in \lido by doing a visual inspection of our clones which showed inconsistent classification (such as a low resonant fraction for many clones), and then repeating the $\phi$ calculation and window analysis specifically for the relevant mixed eccentricity inclination modes for that object.
This method is effective, but not recommended for much larger samples.
We also tested the SBDynT classification software \citep{Volk2025,Spencer2026} on our full sample.
This classifier uses bulk properties of the integration of a TNO's orbit (for example, the histogram of the $a$-distribution) to identify resonant TNOs without calculating $\phi$; a second classifier then searches for and identifies the specific librating resonant angle (including a very wide range of mixed mode angles).
SBDynT was trained on a large data set \cite[described in][]{Volk2025}, however, example training data for mixed mode resonators and unusual resonant behavior at high-inclination (such as the 10:1 oscillation around $\phi=0$) is less well represented in the training set than the more common, eccentricity-type resonators.
Fortunately, the SBDynT classifier still did an excellent job of identifying the resonant TNOs in our sample.
The classifier flagged all of the TNOs we identified as resonant as either resonant or possibly resonant in the same resonance mode identified in our procedure.
We are confident in the ability of this classifier to provide robust dynamical classifications even at large inclinations.

\lido discovered three possible Haumea family members, which all have relatively low $\delta v$.
However, these objects are quite small, and in this $H$ range a large amount of contamination from background TNOs is expected.
Multi-band photometry or spectroscopy would be required to confirm Haumea family membership.
If any of these objects are determined to be Haumea family members from followup surface characterization, the detection here would provide a valuable lever for constraining the small-end size distribution of the Haumea family.

\lido discovered 38 characterized hot classical TNOs, which provide a significant addition to the 86 OSSOS++ hot classicals with $i>14\degr$.
These additional discoveries are inconsistent with the OSSOS++ model of the hot classical TNOs, which was created by debiasing the real objects and fuzzing their orbital parameters.
At low-inclination, the large sample of OSSOS++ classical TNOs probes the phase space sufficiently well for this debiasing approach to produce a statistically acceptable model.
At larger inclinations, the volume increases and the sample size decreases dramatically.
This results in a debaised model which does not sufficiently sample the $a-e-i$ distribution, although we note that qualitatively the individual distributions of the \lido and OSSOS++ samples appear similar.
We see similar sub-structure in the $a$-distribution, for example, which is a good indication that this sub-structure is a real feature of the high-$i$ classical TNOs.
As many of the gaps in the $a$-distributions for \lido and OSSOS++ clearly correlated with mean motion resonances, we decided to test whether the orbital distribution in $a$, $e$, and $i$ was primarily sculpted by steady-state effects.
We created a variety of initial distributions for hot classical TNOs with smooth inclination and pericenter distributions.
We then allowed these distributions to evolve with the current planetary architecture for 1~Gyr using REBOUND simulations.
After removing the resonant TNOs, we found that we re-created many of the $a$-gaps.
We also produce hotter $e$ and $i$ distributions than the input particles, particularly near the mean-motion resonance boundaries.
Importantly, we retain an interdependence between $a$, $e$, and $i$ which is not smoothed and has clear links to nearby mean motion resonances.
The structure of this Gyr-evolved model provides a good qualitative match to the observations as well as a statistically non-rejectable orbital distribution model.
Some of the resonance gaps are not sufficiently large compared to those in the real TNOs. 
This may be only because our simulations were 1~Gyr instead of 4.5~Gyrs, or it may be related to some specific dynamical effects of planetary migration (which we did not attempt to model).
Longer simulations would also be useful for determining whether other features in the large-inclination population are signatures of implantation or driven by longer-term evolution.
We note that the model presented here is a statistically acceptable model, but other inclination distributions and weighting schemes also produce statistically acceptable models.
Future work is needed to test a larger range of input smooth models and a longer evolution time-scale.
Our proof-of-concept analysis demonstrates the significant role that long-term steady-state evolution plays in sculpting the high-inclination Kuiper belt region, and the importance of identifying the features related to this steady-state evolution when searching for implantation signatures.

Our model provides a statistically acceptable fit for both \lido and OSSOS, and produces a similar but more complicated distribution to other modeling work.
The DES inclination distribution used for hot classical TNOs \citep{Bernardinelli2025} has a similar characteristic width to our model,  and that inclination distribution would likely be statistically acceptable for \lido.
However, our model maintains the inter-dependence of the various orbital parameters (visible in the bottom right panel of \autoref{fig:hc_model}, particularly in the lack of $i<10\degr$ objects with $a=$~40~-~41~au). 
The \lido model has a less smoothed distribution, which provides a qualitatively better match to the real observations (compare our \autoref{fig:bothCumul} with Figure 14 of \citet{Bernardinelli2025}).
All of the models (OSSOS++, \lido, and DES) have the extended large-$i$ tail that over-predicts the $i>40\degr$ objects, and should be a subject of future work.
Comparing our \autoref{fig:bothCumul} with Figure 9 of \citet{Bernardinelli2025}, it is clear that both models share the problem of slightly over-predicting the amount of objects in the $H=$~6~-~7 range.
Our stability-based approach has many significant improvements over both a parametric approach and a debiasing approach.

The \lido survey detected 125 characterized high-inclination TNOs and highlights the unusual objects found off the ecliptic plane.
As the number of mid- to large-inclination TNOs increases, with future off-ecliptic surveys such as Vera C. Rubin Observatory's Legacy Survey of Space and Time \citep[LSST,][]{Schwamb2023} and the ROMAN Space Telescope's High Latitude Wide Area Survey \citep[HLWAS,][]{Spergel2015, Holler2018}, the discoveries of TNOs with unusual orbits will increase significantly.
It is critical to utilize an orbit classification system that is robust to identifying mixed inclination-eccentricity resonators as well as resonators with unexpected libration centers.
The role that these large-inclination resonant configurations play in sculpting the Kuiper belt region also requires further consideration.

\section{Conclusions}
\label{sec:conclusions}

The \lido survey was designed to detect high-inclination TNOs, and found 125 characterized objects.
As with other surveys in the OSSOS++ well-characterized surveys, \lido has a carefully measured characterization so that it is possible to extract the intrinsic orbital distributions and populations within different subpopulations.
The high-inclination region hosts a variety of interesting and unusual orbital classes, including many resonant TNOs.
\lido discovered 55 securely resonant characterized TNOs, including the first securely classified 10:1 resonator \citep{Pike2025}.
The large number of mixed inclination-eccentricity resonators, 6 secure and 6 short-term sticking, was unpredicted.
The way in which the many relevant high order and mixed-mode resonances sculpt the high-inclination kuiper belt region requires further study.
The \lido survey detected 21 Plutinos, 13 of which are also in Kozai resonance.
The Kozai Plutinos come to pericenter close to their maximum separation from the ecliptic plane, and are therefore most detectable off the ecliptic plane.
Analysis of the 3:2 resonant population is ongoing, and includes \citet{Lawler2025} which presents a filled-parameter-space model of the Plutino stable orbital space, including the Kozai resonance.
The majority of the structure in the orbital distribution of hot classical TNOs with inclinations $i>14\degr$ appears to be created by long-term steady-state sculpting by mean motion and secular resonances, with a few features plausibly resulting from primordial emplacement.
We have explored this with a smoothly implanted model of the hot classical TNOs integrated with the current Solar System configuration for 1~Gyr which produces a non-rejectable $a-e-i$ distribution.
More work is needed to robustly identify which features of the orbital distribution of hot classical TNOs are indications the specifics of implantation and which are dominated by long-term evolution.
As future surveys such as LSST discover mid to high inclination TNOs, it will be important to carefully classify these objects, keeping in mind the high prevalence of high order and mixed mode resonant objects in addition to the many resonators in major resonances such as the 3:2.
The 125 TNOs \lido discovered at high inclinations provide critical insight into the orbital distribution of many TNO populations, and explore interesting dynamics previously unprobed by observational surveys.

\begin{acknowledgements}

Based on observations obtained with MegaPrime/MegaCam, a joint project of CFHT and CEA/DAPNIA, at the Canada-France-Hawaii Telescope (CFHT) which is operated by the National Research Council (NRC) of Canada, the Institut National des Science de l'Univers of the Centre National de la Recherche Scientifique (CNRS) of France, and the University of Hawaii. The observations at the Canada-France-Hawaii Telescope were performed with care and respect from the summit of Maunakea which is a significant cultural and historic site.
Based in part on observations obtained at the international Gemini Observatory, a program of NSF NOIRLab, which is managed by the Association of Universities for Research in Astronomy (AURA) under a cooperative agreement with the U.S. National Science Foundation on behalf of the Gemini Observatory partnership: the U.S. National Science Foundation (United States), National Research Council (Canada), Agencia Nacional de Investigaci\'{o}n y Desarrollo (Chile), Ministerio de Ciencia, Tecnolog\'{i}a e Innovaci\'{o}n (Argentina), Minist\'{e}rio da Ci\^{e}ncia, Tecnologia, Inova\c{c}\~{o}es e Comunica\c{c}\~{o}es (Brazil), and Korea Astronomy and Space Science Institute (Republic of Korea), acquired through the Gemini Observatory Archive at NSF NOIRLab and processed using the Gemini IRAF package.
This paper includes data gathered with the 6.5 meter Magellan Telescopes located at Las Campanas Observatory, Chile.
This research is based in part on data collected at the Subaru Telescope, which is operated by the National Astronomical Observatory of Japan.
This research used the Canadian Advanced Network For Astronomy Research (CANFAR) operated in partnership by the  Canadian Astronomy Data Centre and The Digital Research Alliance of Canada with support from the National Research Council of Canada, the Canadian Space Agency, CANARIE, and the Canadian Foundation for Innovation

MA, REP, CM, and CC acknowledge funding from NASA Solar System Observations grant 80NSSC21K0289. 
MA and REP acknowledge support from the NASA Minor Planet Center Award 80NSSC22M0024. 
REP, KV, and JP acknowledge support from NASA Solar System Observations grant 80NSSC23K0680.
REP and KV acknowledge support from NASA Emerging Worlds grant 80NSSC21K0376.
KV acknowledges additional support from NASA grants 80NSSC23K0886 and 80NSSC23K1169.
SL acknowledges support from NSERC Discover Grant RGPIN-2020-04111.
CC was supported in part by a Massachusetts Space Grant Consortium (MASGC) Award.
JP was supported by an internal Smithsonian Institution grant.

Data Access: Data supporting this study are included within the article and in observatory archives.  The CFHT data can be accessed through the CADC archive (\url{https://www.cadc-ccda.hia-iha.nrc-cnrc.gc.ca/en/}). The Subaru data was obtained through the same CADC archive and can be accessed there. The Magellan data is archived by CfA and can be accessed on request to the corresponding author (\url{http://tdc-www.harvard.edu/computing/magellan/}).
The final astrometry and previously published astrometry can be accessed through the Minor Planet Center (MPC, \url{https://www.minorplanetcenter.net/}).

\end{acknowledgements}


\facilities{CFHT, CANFAR, Magellan, Subaru, MPC} 
\software{This research was made possible by the open-source projects 
\texttt{matplotlib} \citep{Hunter2007,Droettboom2016},
\texttt{MegaPipe} \citep{Gwyn2008}
\texttt{REBOUND} \citep{Rein2012},
\texttt{Jupyter} \citep{Kluyver2016},
\texttt{MERCURIUS} \citep{Rein2019},
\texttt{SBDynT} \citep{Volk2026Zenodo, Spencer2026},
\texttt{numpy} \citep{Harris2020},
\texttt{Find\_Orb} \cite{Gray2022},
the updated version of the CFEPS Moving Object detection Pipeline \citep{Petit2004} used by OSSOS \citep{Bannister2016},
}

\appendix

\section{Appendix Survey Characterization plots}

\begin{figure}[tbp]\label{fig:efficiency_curves}
\centering
  \includegraphics[scale=0.42]{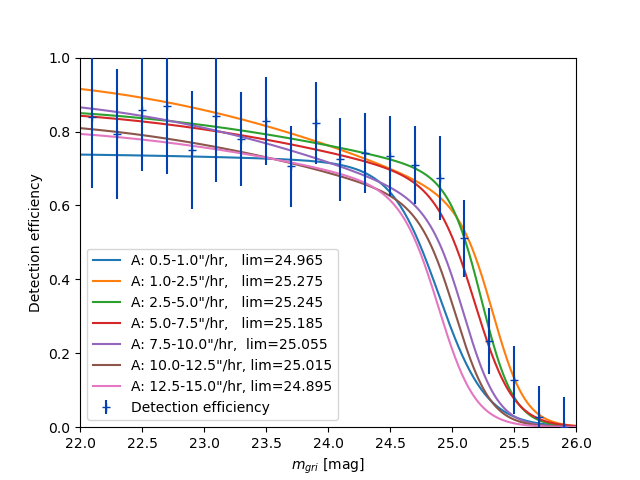}
  \includegraphics[scale=0.42]{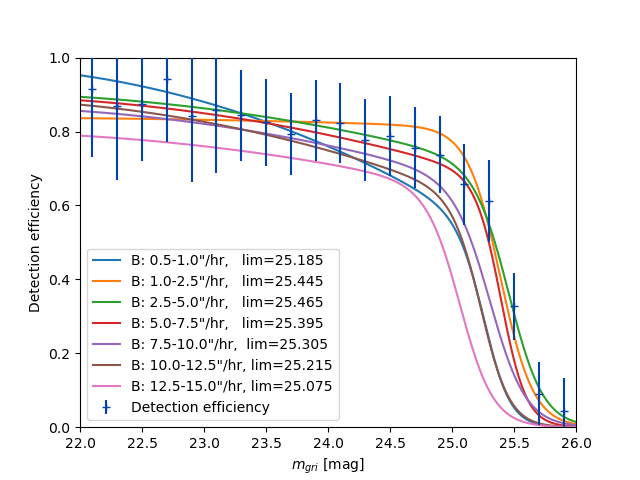}
  \includegraphics[scale=0.42]{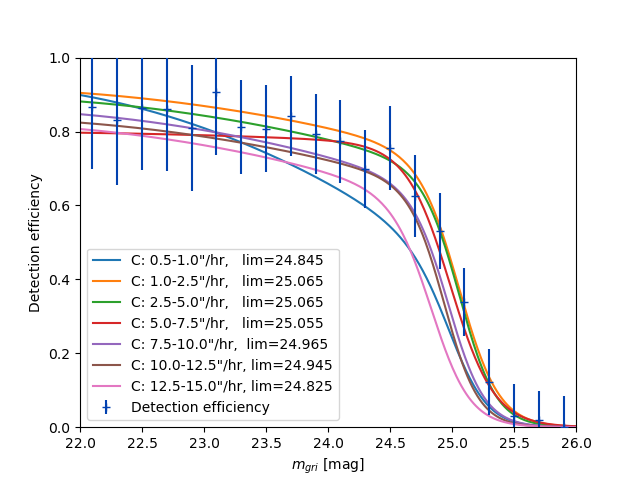}
  \includegraphics[scale=0.42]{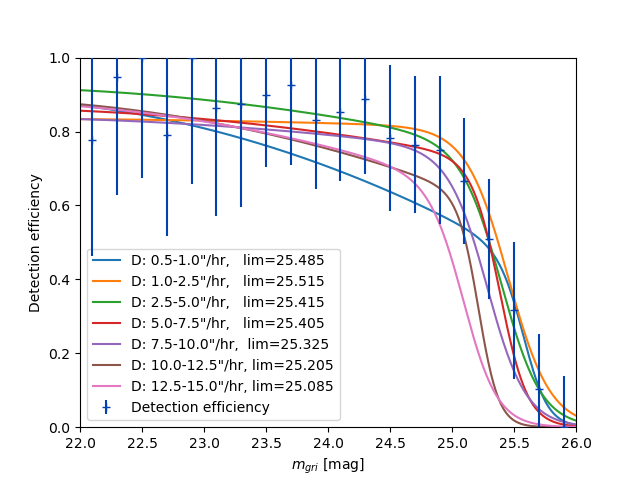}
  \includegraphics[scale=0.42]{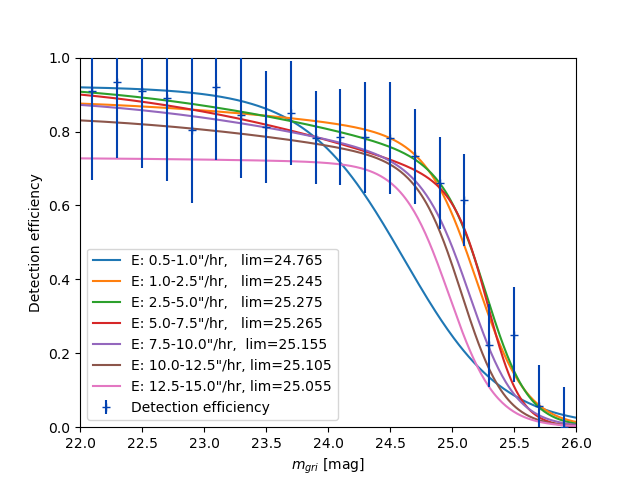}
  \includegraphics[scale=0.42]{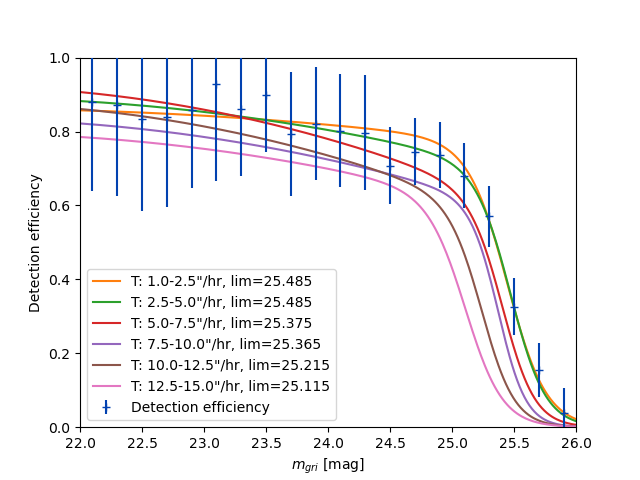}
  \includegraphics[scale=0.42]{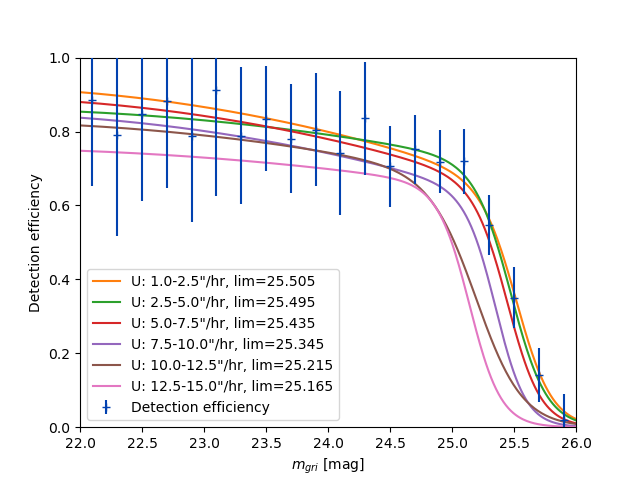}
  \includegraphics[scale=0.42]{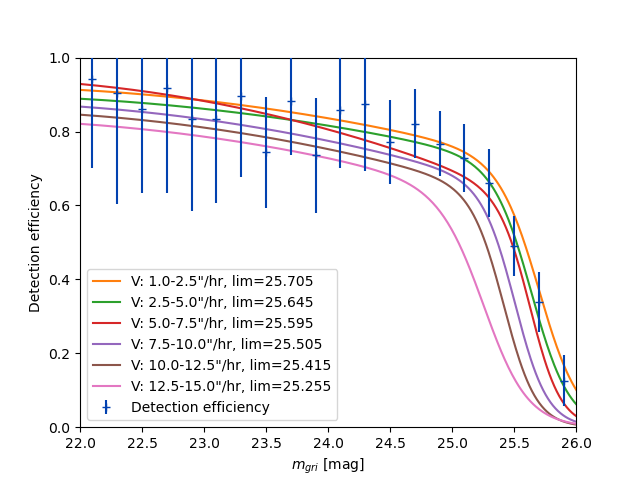}
  \caption{
  Efficiency curves for all rates and blocks.
  Each subplot shows efficiency for a single sub-block.
  In each plot, the points and error bars show the measured detection efficiency in the $2.5-5.0\arcsec$/hr rate range (binned in 0.2~mag intervals), corresponding to the green curve. 
  The colored curves are best fit curves to the detection efficiencies in different rate intervals.
  Efficiency curves represent \autoref{eq:efficiency_function} fit to the measured detection efficiencies in 0.1 mag bins, except where that produced spurious results ($0.5-1.0\arcsec$/hr interval and for the smaller D-block), in which case larger bins (0.2 to 0.5 mag) were used.
  While these plots show efficiency curves for rates $10-15\arcsec$/hr, these rates were eventually abandoned from the characterization efforts as the survey observing cadence was not designed to track such fast moving objects. 
  }
\end{figure}

\begin{figure}[tbp]\label{fig:trombone_plots}
\centering
  \includegraphics[width=0.33\textwidth]{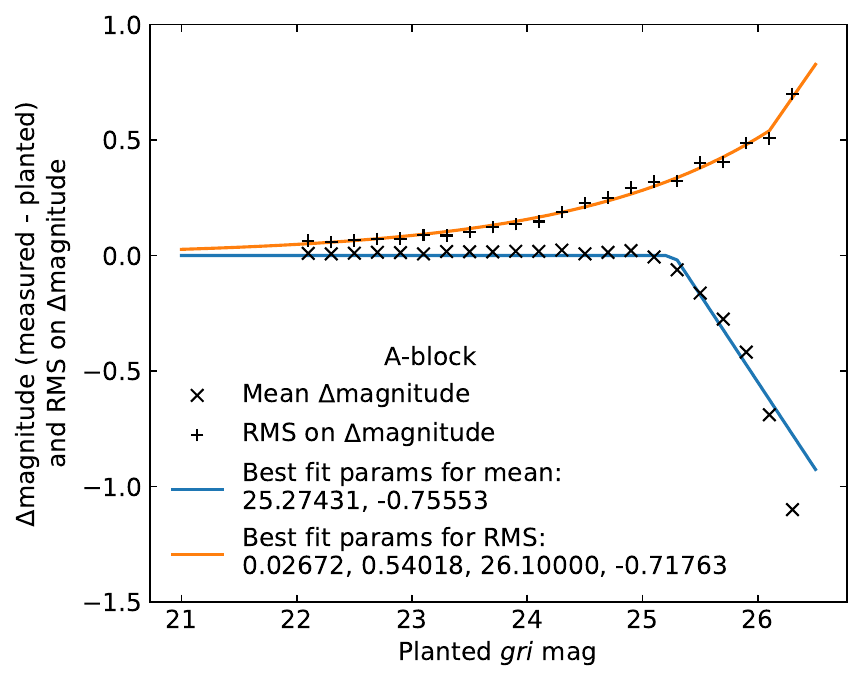}
  \includegraphics[width=0.33\textwidth]{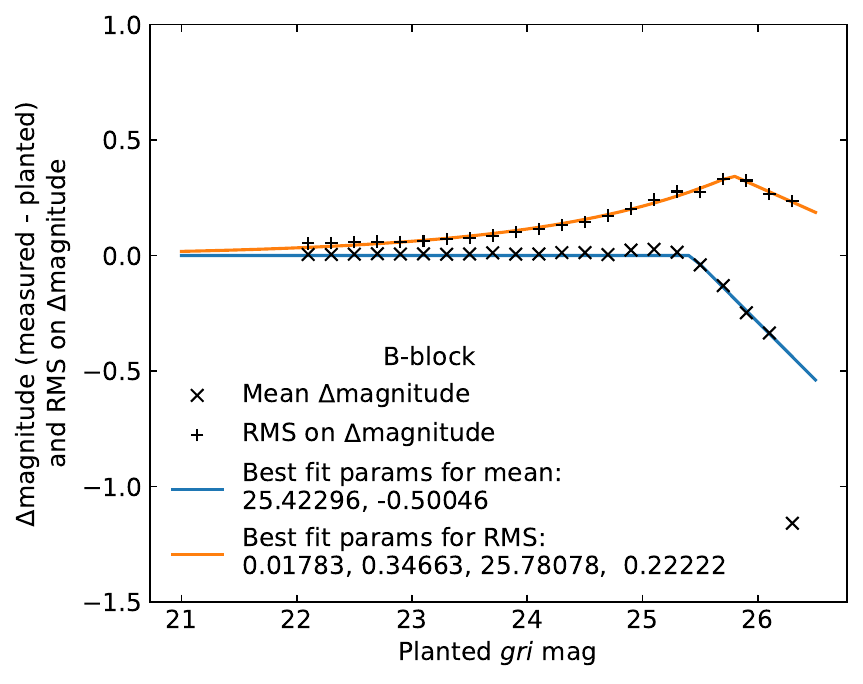}
  \includegraphics[width=0.33\textwidth]{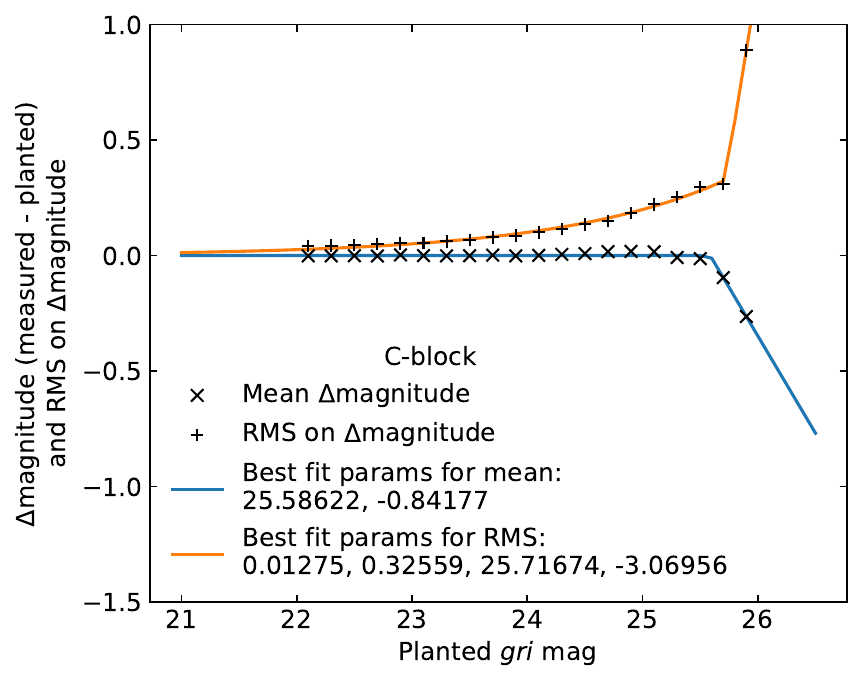}
  \includegraphics[width=0.33\textwidth]{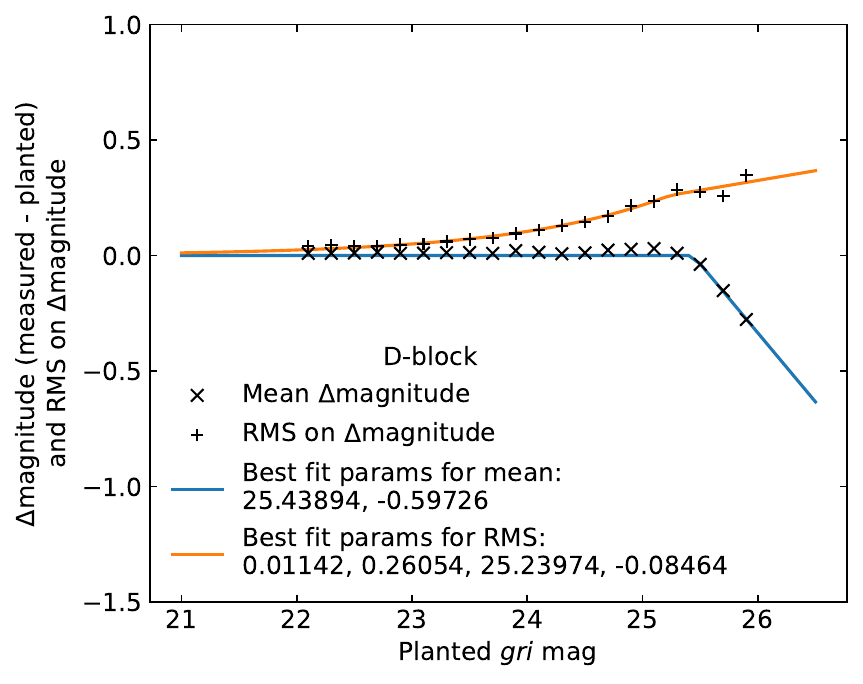}
  \includegraphics[width=0.33\textwidth]{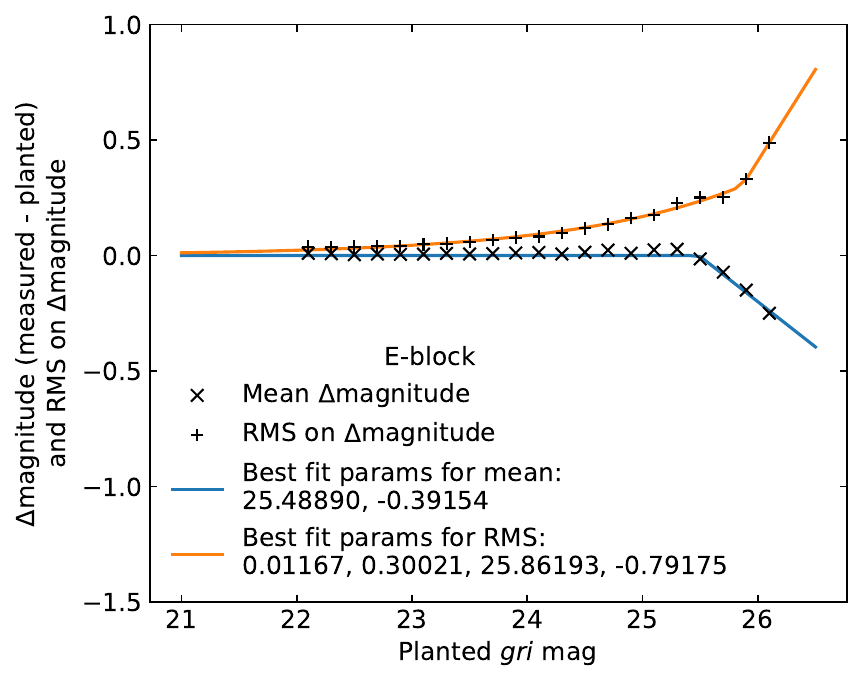}\\
  \includegraphics[width=0.33\textwidth]{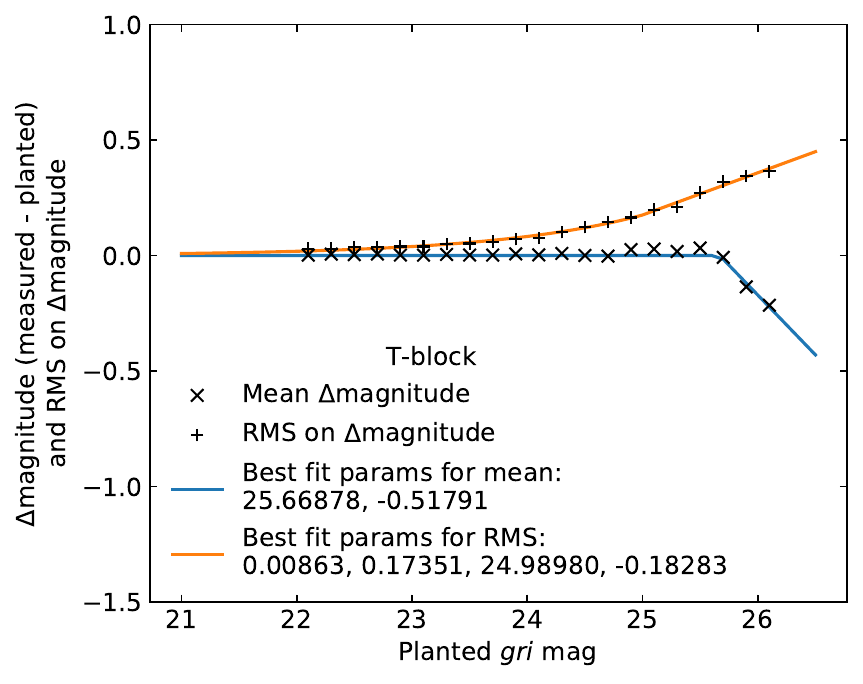}
  \includegraphics[width=0.33\textwidth]{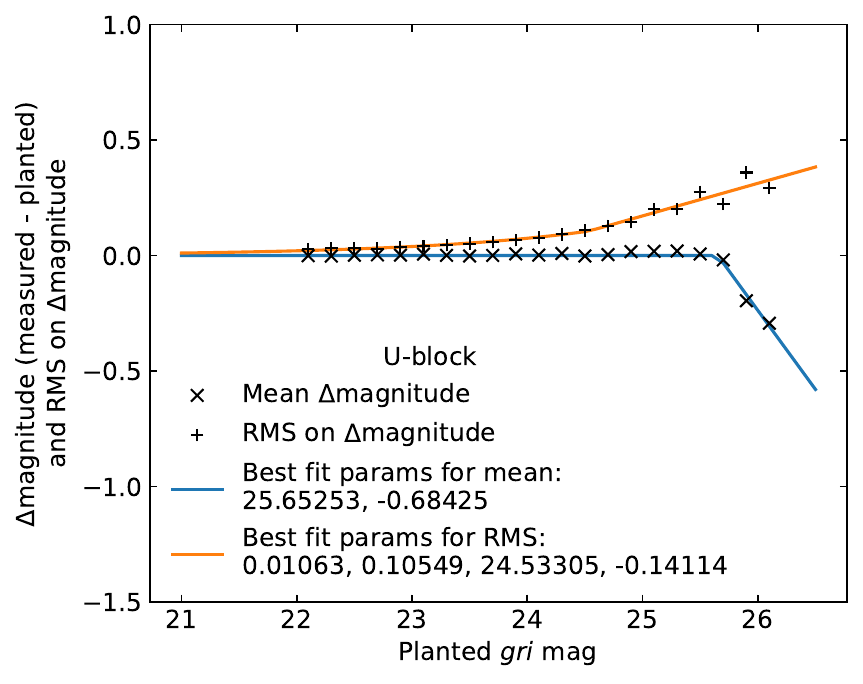}
  \includegraphics[width=0.33\textwidth]{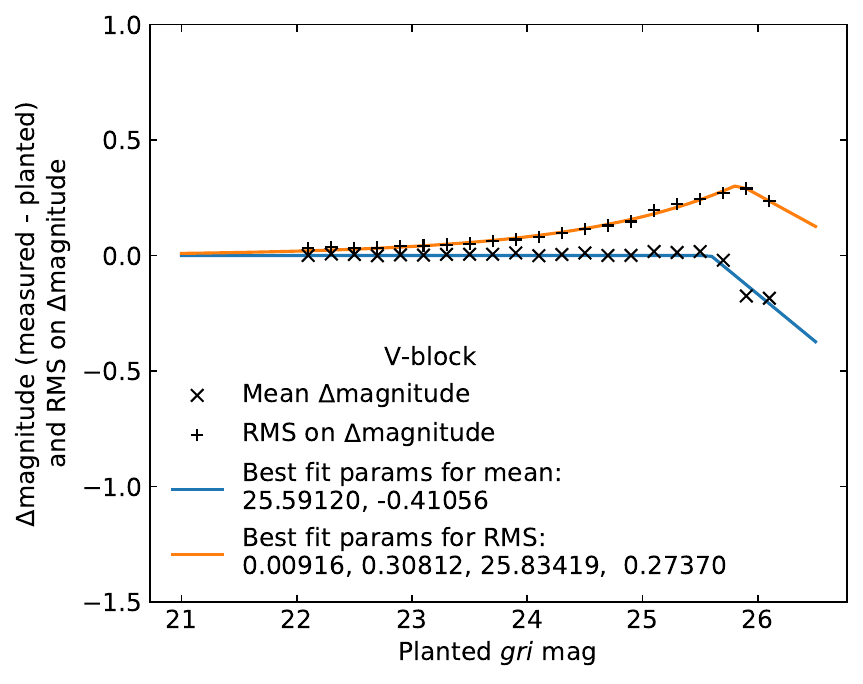}
  \caption{
    Visual representation of the data and parameters described in \autoref{sec:trombone}.
    Each panel (one for each survey sub-block) shows the mean magnitude difference $\Delta$magnitude (measured - planted) for implanted synthetic sources as well as the RMS on $\Delta$magnitude. 
    Crosses shows the mean difference between the measured and planted magnitudes, while the blue curve shows the two-parameter function fitted to that data, as described in Appendix A of \citep{Bannister2016}.
    Plusses show the RMS of the measured$-$planted magnitude difference, while the blue curve shows the four-parameter function fitted to the dashed line, as described in Appendix A of \citep{Bannister2016}.
    Any bins with less than 3 measurements (occasionally happens in the mag>26 range) were ignored in the fit, as these bins disrupted the function in unrealistic ways. 
    In cases where the last bin has <10 measurements, a human decided whether or not to include it in the fit (included if it does not change the fit significantly compared to if it's ignored).
    }
\end{figure}

\bibliographystyle{aasjournal}
\bibliography{lido2026,manual_citations}

@INPROCEEDINGS{Lawler2026,
       author = {{Lawler}, Samantha M. and {Pike}, Rosemary E.},
        title = "{Small bodies in the distant Solar System}",
    booktitle = {Encyclopedia of Astrophysics, Volume 1},
         year = 2026,
       volume = {1},
        month = jan,
        pages = {106-121},
          doi = {10.1016/B978-0-443-21439-4.00023-7},
archivePrefix = {arXiv},
       eprint = {2410.04338},
 primaryClass = {astro-ph.EP},
       adsurl = {https://ui.adsabs.harvard.edu/abs/2026enap....1..106L}
}

@ARTICLE{Collyer2025,
       author = {{Collyer}, Cameron and {Pike}, Rosemary E. and {Chen}, Ying-Tung and {Alexandersen}, Mike and {Comte}, Mark and {Lawler}, Samantha M. and {Holler}, Bryan J. and {Kavelaars}, J.~J. and {Peltier}, Lowell},
        title = "{The Surfaces of Small to Midsize Plutinos: Evidence of an Association between Inclination and Surface Type}",
      journal = {\psj},
         year = 2025,
        month = sep,
       volume = {6},
       number = {9},
          eid = {211},
        pages = {211},
          doi = {10.3847/PSJ/adef36},
archivePrefix = {arXiv},
       eprint = {2507.10766},
 primaryClass = {astro-ph.EP},
       adsurl = {https://ui.adsabs.harvard.edu/abs/2025PSJ.....6..211C}
}

@ARTICLE{Pike2025,
       author = {{Pike}, Rosemary E. and {Murray-Clay}, Ruth and {Volk}, Kathryn and {Alexandersen}, Mike and {Comte}, Mark and {Lawler}, Samantha M. and {Chen}, Ying-Tung and {Hermosillo Ruiz}, Arcelia and {Semenchuck}, Cameron and {Collyer}, Cameron and {Kavelaars}, J.~J. and {Peltier}, Lowell},
        title = "{LiDO: Discovery of a 10:1 Resonator with a Novel Libration State}",
      journal = {\psj},
         year = 2025,
        month = jul,
       volume = {6},
       number = {7},
          eid = {156},
        pages = {156},
          doi = {10.3847/PSJ/addd22},
archivePrefix = {arXiv},
       eprint = {2508.05627},
 primaryClass = {astro-ph.EP},
       adsurl = {https://ui.adsabs.harvard.edu/abs/2025PSJ.....6..156P}
}

@ARTICLE{Lawler2025,
       author = {{Lawler}, Samantha M. and {Comte}, Mark and {Pike}, Rosemary E. and {Alexandersen}, Mike and {Chen}, Ying-Tung and {Collyer}, Cameron and {Holman}, Matthew and {Kavelaars}, J.~J. and {Peltier}, Lowell and {Semenchuck}, Cameron},
        title = "{LiDO: Exploring the Stable Plutino Parameter Space}",
      journal = {\psj},
         year = 2025,
        month = apr,
       volume = {6},
       number = {4},
          eid = {100},
        pages = {100},
          doi = {10.3847/PSJ/adc10c},
archivePrefix = {arXiv},
       eprint = {2503.10847},
 primaryClass = {astro-ph.EP},
       adsurl = {https://ui.adsabs.harvard.edu/abs/2025PSJ.....6..100L}
}

@INCOLLECTION{Volk2025,
       author = {{Volk}, Kathryn and {Malhotra}, Renu},
        title = "{Machine learning-assisted dynamical classification of trans-Neptunian objects}",
    booktitle = {Machine Learning for Small Bodies in the Solar System. Edited by V. Carruba},
         year = 2025,
       editor = {{Carruba}, Valerio and {Smirnov}, Evgeny and {Oszkiewicz}, Dagmara},
        pages = {173-207},
          doi = {10.1016/B978-0-44-324770-5.00012-X},
       adsurl = {https://ui.adsabs.harvard.edu/abs/2025mlsm.book..173V}
}

@ARTICLE{Proudfoot2024,
       author = {{Proudfoot}, Benjamin C.~N. and {Fern{\'a}ndez-Valenzuela}, Estela and {Stansberry}, John A. and {Kiss}, Csaba and {Ragozzine}, Darin and {Fraser}, Wesley C. and {Pike}, Rosemary E. and {Pinilla-Alonso}, Noemi},
        title = "{A Near-infrared Survey of Candidate Haumea Family Members}",
      journal = {\aj},
         year = 2024,
        month = dec,
       volume = {168},
       number = {6},
          eid = {269},
        pages = {269},
          doi = {10.3847/1538-3881/ad864e},
       adsurl = {https://ui.adsabs.harvard.edu/abs/2024AJ....168..269P}
}

@ARTICLE{Smotherman2024,
       author = {{Smotherman}, Hayden and {Bernardinelli}, Pedro H. and {Portillo}, Stephen K.~N. and {Connolly}, Andrew J. and {Kalmbach}, J. Bryce and {Stetzler}, Steven and {Juri{\'c}}, Mario and {Bekte{\v{s}}evi{\'c}}, Dino and {Langford}, Zachary and {Adams}, Fred C. and {Oldroyd}, William J. and {Holman}, Matthew J. and {Chandler}, Colin Orion and {Fuentes}, Cesar and {Gerdes}, David W. and {Lin}, Hsing Wen and {Markwardt}, Larissa and {McNeill}, Andrew and {Mommert}, Michael and {Napier}, Kevin J. and {Payne}, Matthew J. and {Ragozzine}, Darin and {Rivkin}, Andrew S. and {Schlichting}, Hilke and {Sheppard}, Scott S. and {Strauss}, Ryder and {Trilling}, David E. and {Trujillo}, Chadwick A.},
        title = "{The DECam Ecliptic Exploration Project (DEEP). VI. First Multiyear Observations of Trans-Neptunian Objects}",
      journal = {\aj},
         year = 2024,
        month = mar,
       volume = {167},
       number = {3},
          eid = {136},
        pages = {136},
          doi = {10.3847/1538-3881/ad1524},
archivePrefix = {arXiv},
       eprint = {2310.03678},
 primaryClass = {astro-ph.EP},
       adsurl = {https://ui.adsabs.harvard.edu/abs/2024AJ....167..136S}
}

@ARTICLE{Trujillo2024,
       author = {{Trujillo}, Chadwick A. and {Fuentes}, Cesar and {Gerdes}, David W. and {Markwardt}, Larissa and {Sheppard}, Scott S. and {Strauss}, Ryder and {Chandler}, Colin Orion and {Oldroyd}, William J. and {Trilling}, David E. and {Lin}, Hsing Wen and {Adams}, Fred C. and {Bernardinelli}, Pedro H. and {Holman}, Matthew J. and {Juri{\'c}}, Mario and {McNeill}, Andrew and {Mommert}, Michael and {Napier}, Kevin J. and {Payne}, Matthew J. and {Ragozzine}, Darin and {Rivkin}, Andrew S. and {Schlichting}, Hilke and {Smotherman}, Hayden},
        title = "{The DECam Ecliptic Exploration Project (DEEP). II. Observational Strategy and Design}",
      journal = {\aj},
         year = 2024,
        month = mar,
       volume = {167},
       number = {3},
          eid = {133},
        pages = {133},
          doi = {10.3847/1538-3881/ad1523},
archivePrefix = {arXiv},
       eprint = {2310.19864},
 primaryClass = {astro-ph.EP},
       adsurl = {https://ui.adsabs.harvard.edu/abs/2024AJ....167..133T}
}

@ARTICLE{Bernardinelli2024,
       author = {{Bernardinelli}, Pedro H. and {Smotherman}, Hayden and {Langford}, Zachary and {Portillo}, Stephen K.~N. and {Connolly}, Andrew J. and {Kalmbach}, J. Bryce and {Stetzler}, Steven and {Juri{\'c}}, Mario and {Oldroyd}, William J. and {Lin}, Hsing Wen and {Adams}, Fred C. and {Chandler}, Colin Orion and {Fuentes}, Cesar and {Gerdes}, David W. and {Holman}, Matthew J. and {Markwardt}, Larissa and {McNeill}, Andrew and {Mommert}, Michael and {Napier}, Kevin J. and {Payne}, Matthew J. and {Ragozzine}, Darin and {Rivkin}, Andrew S. and {Schlichting}, Hilke and {Sheppard}, Scott S. and {Strauss}, Ryder and {Trilling}, David E. and {Trujillo}, Chadwick A.},
        title = "{The DECam Ecliptic Exploration Project (DEEP). III. Survey Characterization and Simulation Methods}",
      journal = {\aj},
         year = 2024,
        month = mar,
       volume = {167},
       number = {3},
          eid = {134},
        pages = {134},
          doi = {10.3847/1538-3881/ad1527},
archivePrefix = {arXiv},
       eprint = {2310.03671},
 primaryClass = {astro-ph.EP},
       adsurl = {https://ui.adsabs.harvard.edu/abs/2024AJ....167..134B}
}

@ARTICLE{Napier2024,
       author = {{Napier}, Kevin J. and {Lin}, Hsing Wen and {Gerdes}, David W. and {Adams}, Fred C. and {Simpson}, Anna M. and {Porter}, Matthew W. and {Weber}, Katherine G. and {Markwardt}, Larissa and {Gowman}, Gabriel and {Smotherman}, Hayden and {Bernardinelli}, Pedro H. and {Juri{\'c}}, Mario and {Connolly}, Andrew J. and {Kalmbach}, J. Bryce and {Portillo}, Stephen K.~N. and {Trilling}, David E. and {Strauss}, Ryder and {Oldroyd}, William J. and {Trujillo}, Chadwick A. and {Chandler}, Colin Orion and {Holman}, Matthew J. and {Schlichting}, Hilke E. and {McNeill}, Andrew},
        title = "{The DECam Ecliptic Exploration Project (DEEP). V. The Absolute Magnitude Distribution of the Cold Classical Kuiper Belt}",
      journal = {\psj},
         year = 2024,
        month = feb,
       volume = {5},
       number = {2},
          eid = {50},
        pages = {50},
          doi = {10.3847/PSJ/ad1528},
archivePrefix = {arXiv},
       eprint = {2309.09478},
 primaryClass = {astro-ph.EP},
       adsurl = {https://ui.adsabs.harvard.edu/abs/2024PSJ.....5...50N}
}

@INPROCEEDINGS{Malhotra2023,
       author = {{Malhotra}, Renu and {Ito}, Takashi},
        title = "{The doubly resonant Plutinos}",
    booktitle = {55th Annual Meeting of the Division for Planetary Sciences},
         year = 2023,
       series = {AAS/Division for Planetary Sciences Meeting Abstracts},
       volume = {55},
        month = oct,
          eid = {202.08},
        pages = {202.08},
       adsurl = {https://ui.adsabs.harvard.edu/abs/2023DPS....5520208M}
}

@ARTICLE{Balaji2023,
       author = {{Balaji}, S. and {Zaveri}, N. and {Hayashi}, N. and {Hermosillo Ruiz}, A. and {Barnes}, J. and {Murray-Clay}, R. and {Volk}, K. and {Gerhardt}, J. and {Syed}, Z.},
        title = "{Can the orbital distribution of Neptune's 3:2 mean-motion resonance result from stability sculpting?}",
      journal = {\mnras},
         year = 2023,
        month = sep,
       volume = {524},
       number = {2},
        pages = {3039-3051},
          doi = {10.1093/mnras/stad2026},
archivePrefix = {arXiv},
       eprint = {2307.06280},
 primaryClass = {astro-ph.EP},
       adsurl = {https://ui.adsabs.harvard.edu/abs/2023MNRAS.524.3039B}
}

@ARTICLE{Schwamb2023,
       author = {{Schwamb}, Megan E. and {Jones}, R. Lynne and {Yoachim}, Peter and {Volk}, Kathryn and {Dorsey}, Rosemary C. and {Opitom}, Cyrielle and {Greenstreet}, Sarah and {Lister}, Tim and {Snodgrass}, Colin and {Bolin}, Bryce T. and {Inno}, Laura and {Bannister}, Michele T. and {Eggl}, Siegfried and {Solontoi}, Michael and {Kelley}, Michael S.~P. and {Juri{\'c}}, Mario and {Lin}, Hsing Wen and {Ragozzine}, Darin and {Bernardinelli}, Pedro H. and {Chesley}, Steven R. and {Daylan}, Tansu and {{\v{D}}urech}, Josef and {Fraser}, Wesley C. and {Granvik}, Mikael and {Knight}, Matthew M. and {Lisse}, Carey M. and {Malhotra}, Renu and {Oldroyd}, William J. and {Thirouin}, Audrey and {Ye}, Quanzhi},
        title = "{Tuning the Legacy Survey of Space and Time (LSST) Observing Strategy for Solar System Science}",
      journal = {\apjs},
         year = 2023,
        month = jun,
       volume = {266},
       number = {2},
          eid = {22},
        pages = {22},
          doi = {10.3847/1538-4365/acc173},
archivePrefix = {arXiv},
       eprint = {2303.02355},
 primaryClass = {astro-ph.EP},
       adsurl = {https://ui.adsabs.harvard.edu/abs/2023ApJS..266...22S}
}

@ARTICLE{GaiaCollaboration2018,
       author = {{Gaia Collaboration} and {Brown}, A.~G.~A. and {Vallenari}, A. and {Prusti}, T. and {de Bruijne}, J.~H.~J. and {Babusiaux}, C. and {Bailer-Jones}, C.~A.~L. and {Biermann}, M. and {Evans}, D.~W. and {Eyer}, L. and {Jansen}, F. and {Jordi}, C. and {Klioner}, S.~A. and {Lammers}, U. and {Lindegren}, L. and {Luri}, X. and {Mignard}, F. and {Panem}, C. and {Pourbaix}, D. and {Randich}, S. and {Sartoretti}, P. and {Siddiqui}, H.~I. and {Soubiran}, C. and {van Leeuwen}, F. and {Walton}, N.~A. and {Arenou}, F. and {Bastian}, U. and {Cropper}, M. and {Drimmel}, R. and {Katz}, D. and {Lattanzi}, M.~G. and {Bakker}, J. and {Cacciari}, C. and {Casta{\~n}eda}, J. and {Chaoul}, L. and {Cheek}, N. and {De Angeli}, F. and {Fabricius}, C. and {Guerra}, R. and {Holl}, B. and {Masana}, E. and {Messineo}, R. and {Mowlavi}, N. and {Nienartowicz}, K. and {Panuzzo}, P. and {Portell}, J. and {Riello}, M. and {Seabroke}, G.~M. and {Tanga}, P. and {Th{\'e}venin}, F. and {Gracia-Abril}, G. and {Comoretto}, G. and {Garcia-Reinaldos}, M. and {Teyssier}, D. and {Altmann}, M. and {Andrae}, R. and {Audard}, M. and {Bellas-Velidis}, I. and {Benson}, K. and {Berthier}, J. and {Blomme}, R. and {Burgess}, P. and {Busso}, G. and {Carry}, B. and {Cellino}, A. and {Clementini}, G. and {Clotet}, M. and {Creevey}, O. and {Davidson}, M. and {De Ridder}, J. and {Delchambre}, L. and {Dell'Oro}, A. and {Ducourant}, C. and {Fern{\'a}ndez-Hern{\'a}ndez}, J. and {Fouesneau}, M. and {Fr{\'e}mat}, Y. and {Galluccio}, L. and {Garc{\'\i}a-Torres}, M. and {Gonz{\'a}lez-N{\'u}{\~n}ez}, J. and {Gonz{\'a}lez-Vidal}, J.~J. and {Gosset}, E. and {Guy}, L.~P. and {Halbwachs}, J.-L. and {Hambly}, N.~C. and {Harrison}, D.~L. and {Hern{\'a}ndez}, J. and {Hestroffer}, D. and {Hodgkin}, S.~T. and {Hutton}, A. and {Jasniewicz}, G. and {Jean-Antoine-Piccolo}, A. and {Jordan}, S. and {Korn}, A.~J. and {Krone-Martins}, A. and {Lanzafame}, A.~C. and {Lebzelter}, T. and {L{\"o}ffler}, W. and {Manteiga}, M. and {Marrese}, P.~M. and {Mart{\'\i}n-Fleitas}, J.~M. and et al.},
        title = "{Gaia Data Release 2. Summary of the contents and survey properties}",
      journal = {\aap},
         year = 2018,
        month = aug,
       volume = {616},
          eid = {A1},
        pages = {A1},
          doi = {10.1051/0004-6361/201833051},
archivePrefix = {arXiv},
       eprint = {1804.09365},
 primaryClass = {astro-ph.GA},
       adsurl = {https://ui.adsabs.harvard.edu/abs/2018A&A...616A...1G}
}

@ARTICLE{Petit2023,
       author = {{Petit}, Jean-Marc and {Gladman}, Brett and {Kavelaars}, J.~J. and {Bannister}, Michele T. and {Alexandersen}, Mike and {Volk}, Kathryn and {Chen}, Ying-Tung},
        title = "{The Hot Main Kuiper Belt Size Distribution from OSSOS}",
      journal = {\apjl},
         year = 2023,
        month = apr,
       volume = {947},
       number = {1},
          eid = {L4},
        pages = {L4},
          doi = {10.3847/2041-8213/acc525},
archivePrefix = {arXiv},
       eprint = {2309.01530},
 primaryClass = {astro-ph.EP},
       adsurl = {https://ui.adsabs.harvard.edu/abs/2023ApJ...947L...4P}
}

@ARTICLE{Lei2022,
       author = {{Lei}, Hanlun and {Li}, Jian and {Huang}, Xiumin and {Li}, Muzi},
        title = "{The Von Zeipel-Lidov-Kozai Effect inside Mean Motion Resonances with Applications to Trans-Neptunian Objects}",
      journal = {\aj},
         year = 2022,
        month = sep,
       volume = {164},
       number = {3},
          eid = {74},
        pages = {74},
          doi = {10.3847/1538-3881/ac7c6a},
archivePrefix = {arXiv},
       eprint = {2207.12954},
 primaryClass = {astro-ph.EP},
       adsurl = {https://ui.adsabs.harvard.edu/abs/2022AJ....164...74L}
}

@ARTICLE{Crompvoets2022,
       author = {{Crompvoets}, B.~L. and {Lawler}, S.~M. and {Volk}, K. and {Chen}, Y.-T. and {Gladman}, B. and {Peltier}, L. and {Alexandersen}, M. and {Bannister}, M.~T. and {Gwyn}, S. and {Kavelaars}, J.~J. and {Petit}, J.-M.},
        title = "{OSSOS XXV: Large Populations and Scattering-Sticking in the Distant Trans-Neptunian Resonances}",
      journal = {\psj},
         year = 2022,
        month = may,
       volume = {3},
       number = {5},
          eid = {113},
        pages = {113},
          doi = {10.3847/PSJ/ac67e0},
archivePrefix = {arXiv},
       eprint = {2204.09139},
 primaryClass = {astro-ph.EP},
       adsurl = {https://ui.adsabs.harvard.edu/abs/2022PSJ.....3..113C}
}

@software{Gray2022,
       author = {{Gray}, Bill},
        title = "{Find\_Orb: Orbit determination from observations}",
 howpublished = {Astrophysics Source Code Library, record ascl:2202.016},
         year = 2022,
        month = feb,
          eid = {ascl:2202.016},
archivePrefix = {ascl},
       eprint = {2202.016},
       adsurl = {https://ui.adsabs.harvard.edu/abs/2022ascl.soft02016G}
}

@ARTICLE{Bernardinelli2022,
       author = {{Bernardinelli}, Pedro H. and {Bernstein}, Gary M. and {Sako}, Masao and {Yanny}, Brian and {Aguena}, M. and {Allam}, S. and {Andrade-Oliveira}, F. and {Bertin}, E. and {Brooks}, D. and {Buckley-Geer}, E. and {Burke}, D.~L. and {Carnero Rosell}, A. and {Carrasco Kind}, M. and {Carretero}, J. and {Conselice}, C. and {Costanzi}, M. and {da Costa}, L.~N. and {De Vicente}, J. and {Desai}, S. and {Diehl}, H.~T. and {Dietrich}, J.~P. and {Doel}, P. and {Eckert}, K. and {Everett}, S. and {Ferrero}, I. and {Flaugher}, B. and {Fosalba}, P. and {Frieman}, J. and {Garc{\'\i}a-Bellido}, J. and {Gerdes}, D.~W. and {Gruen}, D. and {Gruendl}, R.~A. and {Gschwend}, J. and {Hinton}, S.~R. and {Hollowood}, D.~L. and {Honscheid}, K. and {James}, D.~J. and {Kent}, S. and {Kuehn}, K. and {Kuropatkin}, N. and {Lahav}, O. and {Maia}, M.~A.~G. and {March}, M. and {Menanteau}, F. and {Miquel}, R. and {Morgan}, R. and {Myles}, J. and {Ogando}, R.~L.~C. and {Palmese}, A. and {Paz-Chinch{\'o}n}, F. and {Pieres}, A. and {Plazas Malag{\'o}n}, A.~A. and {Romer}, A.~K. and {Roodman}, A. and {Sanchez}, E. and {Scarpine}, V. and {Schubnell}, M. and {Serrano}, S. and {Sevilla-Noarbe}, I. and {Smith}, M. and {Soares-Santos}, M. and {Suchyta}, E. and {Swanson}, M.~E.~C. and {Tarle}, G. and {To}, C. and {Varga}, T.~N. and {Walker}, A.~R.},
        title = "{A Search of the Full Six Years of the Dark Energy Survey for Outer Solar System Objects}",
      journal = {\apjs},
         year = 2022,
        month = feb,
       volume = {258},
       number = {2},
          eid = {41},
        pages = {41},
          doi = {10.3847/1538-4365/ac3914},
archivePrefix = {arXiv},
       eprint = {2109.03758},
 primaryClass = {astro-ph.EP},
       adsurl = {https://ui.adsabs.harvard.edu/abs/2022ApJS..258...41B}
}

@ARTICLE{Kavelaars2021,
       author = {{Kavelaars}, J.~J. and {Petit}, Jean-Marc and {Gladman}, Brett and {Bannister}, Michele T. and {Alexandersen}, Mike and {Chen}, Ying-Tung and {Gwyn}, Stephen D.~J. and {Volk}, Kathryn},
        title = "{OSSOS Finds an Exponential Cutoff in the Size Distribution of the Cold Classical Kuiper Belt}",
      journal = {\apjl},
         year = 2021,
        month = oct,
       volume = {920},
       number = {2},
          eid = {L28},
        pages = {L28},
          doi = {10.3847/2041-8213/ac2c72},
archivePrefix = {arXiv},
       eprint = {2107.06120},
 primaryClass = {astro-ph.EP},
       adsurl = {https://ui.adsabs.harvard.edu/abs/2021ApJ...920L..28K}
}

@ARTICLE{Gladman2021,
       author = {{Gladman}, Brett and {Volk}, Kathryn},
        title = "{Transneptunian Space}",
      journal = {\araa},
         year = 2021,
        month = sep,
       volume = {59},
        pages = {203-246},
          doi = {10.1146/annurev-astro-120920-010005},
       adsurl = {https://ui.adsabs.harvard.edu/abs/2021ARA&A..59..203G}
}

@ARTICLE{Magnier2020,
       author = {{Magnier}, Eugene A. and {Chambers}, K.~C. and {Flewelling}, H.~A. and {Hoblitt}, J.~C. and {Huber}, M.~E. and {Price}, P.~A. and {Sweeney}, W.~E. and {Waters}, C.~Z. and {Denneau}, L. and {Draper}, P.~W. and {Hodapp}, K.~W. and {Jedicke}, R. and {Kaiser}, N. and {Kudritzki}, R.-P. and {Metcalfe}, N. and {Stubbs}, C.~W. and {Wainscoat}, R.~J.},
        title = "{The Pan-STARRS Data-processing System}",
      journal = {\apjs},
         year = 2020,
        month = nov,
       volume = {251},
       number = {1},
          eid = {3},
        pages = {3},
          doi = {10.3847/1538-4365/abb829},
archivePrefix = {arXiv},
       eprint = {1612.05240},
 primaryClass = {astro-ph.IM},
       adsurl = {https://ui.adsabs.harvard.edu/abs/2020ApJS..251....3M}
}

@ARTICLE{Harris2020,
       author = {{Harris}, Charles R. and {Millman}, K. Jarrod and {van der Walt}, St{\'e}fan J. and {Gommers}, Ralf and {Virtanen}, Pauli and {Cournapeau}, David and {Wieser}, Eric and {Taylor}, Julian and {Berg}, Sebastian and {Smith}, Nathaniel J. and {Kern}, Robert and {Picus}, Matti and {Hoyer}, Stephan and {van Kerkwijk}, Marten H. and {Brett}, Matthew and {Haldane}, Allan and {del R{\'\i}o}, Jaime Fern{\'a}ndez and {Wiebe}, Mark and {Peterson}, Pearu and {G{\'e}rard-Marchant}, Pierre and {Sheppard}, Kevin and {Reddy}, Tyler and {Weckesser}, Warren and {Abbasi}, Hameer and {Gohlke}, Christoph and {Oliphant}, Travis E.},
        title = "{Array programming with NumPy}",
      journal = {\nat},
         year = 2020,
        month = sep,
       volume = {585},
       number = {7825},
        pages = {357-362},
          doi = {10.1038/s41586-020-2649-2},
archivePrefix = {arXiv},
       eprint = {2006.10256},
 primaryClass = {cs.MS},
       adsurl = {https://ui.adsabs.harvard.edu/abs/2020Natur.585..357H}
}

@ARTICLE{Bernardinelli2020,
       author = {{Bernardinelli}, Pedro H. and {Bernstein}, Gary M. and {Sako}, Masao and {Liu}, Tongtian and {Saunders}, William R. and {Khain}, Tali and {Lin}, Hsing Wen and {Gerdes}, David W. and {Brout}, Dillon and {Adams}, Fred C. and {Belyakov}, Matthew and {Somasundaram}, Aditya Inada and {Sharma}, Lakshay and {Locke}, Jennifer and {Franson}, Kyle and {Becker}, Juliette C. and {Napier}, Kevin and {Markwardt}, Larissa and {Annis}, James and {Abbott}, T.~M.~C. and {Avila}, S. and {Brooks}, D. and {Burke}, D.~L. and {Carnero Rosell}, A. and {Carrasco Kind}, M. and {Castander}, F.~J. and {da Costa}, L.~N. and {De Vicente}, J. and {Desai}, S. and {Diehl}, H.~T. and {Doel}, P. and {Everett}, S. and {Flaugher}, B. and {Garc{\'\i}a-Bellido}, J. and {Gruen}, D. and {Gruendl}, R.~A. and {Gschwend}, J. and {Gutierrez}, G. and {Hollowood}, D.~L. and {James}, D.~J. and {Johnson}, M.~W.~G. and {Johnson}, M.~D. and {Krause}, E. and {Kuropatkin}, N. and {Maia}, M.~A.~G. and {March}, M. and {Miquel}, R. and {Paz-Chinch{\'o}n}, F. and {Plazas}, A.~A. and {Romer}, A.~K. and {Rykoff}, E.~S. and {S{\'a}nchez}, C. and {Sanchez}, E. and {Scarpine}, V. and {Serrano}, S. and {Sevilla-Noarbe}, I. and {Smith}, M. and {Sobreira}, F. and {Suchyta}, E. and {Swanson}, M.~E.~C. and {Tarle}, G. and {Walker}, A.~R. and {Wester}, W. and {Zhang}, Y. and {DES Collaboration}},
        title = "{Trans-Neptunian Objects Found in the First Four Years of the Dark Energy Survey}",
      journal = {\apjs},
         year = 2020,
        month = mar,
       volume = {247},
       number = {1},
          eid = {32},
        pages = {32},
          doi = {10.3847/1538-4365/ab6bd8},
archivePrefix = {arXiv},
       eprint = {1909.01478},
 primaryClass = {astro-ph.EP},
       adsurl = {https://ui.adsabs.harvard.edu/abs/2020ApJS..247...32B}
}

@ARTICLE{Pike2020,
       author = {{Pike}, Rosemary E. and {Proudfoot}, Benjamin C.~N. and {Ragozzine}, Darin and {Alexandersen}, Mike and {Maggard}, Steven and {Bannister}, Michele T. and {Chen}, Ying-Tung and {Gladman}, Brett J. and {Kavelaars}, J.~J. and {Gwyn}, Stephen and {Volk}, Kathryn},
        title = "{A dearth of small members in the Haumea family revealed by OSSOS}",
      journal = {Nature Astronomy},
         year = 2020,
        month = jan,
       volume = {4},
        pages = {89-96},
          doi = {10.1038/s41550-019-0867-z},
archivePrefix = {arXiv},
       eprint = {1908.10286},
 primaryClass = {astro-ph.EP},
       adsurl = {https://ui.adsabs.harvard.edu/abs/2020NatAs...4...89P}
}

@ARTICLE{Ito2019,
       author = {{Ito}, Takashi and {Ohtsuka}, Katsuhito},
        title = "{The Lidov-Kozai Oscillation and Hugo von Zeipel}",
      journal = {Monographs on Environment, Earth and Planets},
         year = 2019,
        month = nov,
       volume = {7},
       number = {1},
        pages = {1-113},
          doi = {10.5047/meep.2019.00701.0001},
archivePrefix = {arXiv},
       eprint = {1911.03984},
 primaryClass = {astro-ph.EP},
       adsurl = {https://ui.adsabs.harvard.edu/abs/2019MEEP....7....1I}
}

@INPROCEEDINGS{Major2019,
       author = {{Major}, Brian and {Kavelaars}, JJ and {Fabbro}, Sebastien and {Durand}, Daniel and {Jeeves}, Helena},
        title = "{Arcade: An Interactive Science Platform in CANFAR}",
    booktitle = {Astronomical Data Analysis Software and Systems XXVII},
         year = 2019,
       editor = {{Teuben}, Peter J. and {Pound}, Marc W. and {Thomas}, Brian A. and {Warner}, Elizabeth M.},
       series = {Astronomical Society of the Pacific Conference Series},
       volume = {523},
        month = oct,
        pages = {277},
       adsurl = {https://ui.adsabs.harvard.edu/abs/2019ASPC..523..277M}
}

@ARTICLE{VanLaerhoven2019,
       author = {{Van Laerhoven}, Christa and {Gladman}, Brett and {Volk}, Kathryn and {Kavelaars}, J.~J. and {Petit}, Jean-Marc and {Bannister}, Michele T. and {Alexandersen}, Mike and {Chen}, Ying-Tung and {Gwyn}, Stephen D.~J.},
        title = "{OSSOS. XIV. The Plane of the Kuiper Belt}",
      journal = {\aj},
         year = 2019,
        month = jul,
       volume = {158},
       number = {1},
          eid = {49},
        pages = {49},
          doi = {10.3847/1538-3881/ab24e1},
       adsurl = {https://ui.adsabs.harvard.edu/abs/2019AJ....158...49V}
}

@ARTICLE{Schwamb2019,
       author = {{Schwamb}, Megan E. and {Fraser}, Wesley C. and {Bannister}, Michele T. and {Marsset}, Micha{\"e}l and {Pike}, Rosemary E. and {Kavelaars}, J.~J. and {Benecchi}, Susan D. and {Lehner}, Matthew J. and {Wang}, Shiang-Yu and {Thirouin}, Audrey and {Delsanti}, Audrey and {Peixinho}, Nuno and {Volk}, Kathryn and {Alexandersen}, Mike and {Chen}, Ying-Tung and {Gladman}, Brett and {Gwyn}, Stephen D.~J. and {Petit}, Jean-Marc},
        title = "{Col-OSSOS: The Colors of the Outer Solar System Origins Survey}",
      journal = {\apjs},
         year = 2019,
        month = jul,
       volume = {243},
       number = {1},
          eid = {12},
        pages = {12},
          doi = {10.3847/1538-4365/ab2194},
archivePrefix = {arXiv},
       eprint = {1809.08501},
 primaryClass = {astro-ph.EP},
       adsurl = {https://ui.adsabs.harvard.edu/abs/2019ApJS..243...12S}
}

@ARTICLE{Lawler2019,
       author = {{Lawler}, S.~M. and {Pike}, R.~E. and {Kaib}, N. and {Alexandersen}, M. and {Bannister}, M.~T. and {Chen}, Y.-T. and {Gladman}, B. and {Gwyn}, S. and {Kavelaars}, J.~J. and {Petit}, J.-M. and {Volk}, K.},
        title = "{OSSOS. XIII. Fossilized Resonant Dropouts Tentatively Confirm Neptune{\textquoteright}s Migration Was Grainy and Slow}",
      journal = {\aj},
         year = 2019,
        month = jun,
       volume = {157},
       number = {6},
          eid = {253},
        pages = {253},
          doi = {10.3847/1538-3881/ab1c4c},
archivePrefix = {arXiv},
       eprint = {1808.02618},
 primaryClass = {astro-ph.EP},
       adsurl = {https://ui.adsabs.harvard.edu/abs/2019AJ....157..253L}
}

@ARTICLE{Rein2019,
       author = {{Rein}, Hanno and {Hernandez}, David M. and {Tamayo}, Daniel and {Brown}, Garett and {Eckels}, Emily and {Holmes}, Emma and {Lau}, Michelle and {Leblanc}, R{\'e}jean and {Silburt}, Ari},
        title = "{Hybrid symplectic integrators for planetary dynamics}",
      journal = {\mnras},
         year = 2019,
        month = jun,
       volume = {485},
       number = {4},
        pages = {5490-5497},
          doi = {10.1093/mnras/stz769},
archivePrefix = {arXiv},
       eprint = {1903.04972},
 primaryClass = {astro-ph.EP},
       adsurl = {https://ui.adsabs.harvard.edu/abs/2019MNRAS.485.5490R}
}

@ARTICLE{Proudfoot2019,
       author = {{Proudfoot}, Benjamin C.~N. and {Ragozzine}, Darin},
        title = "{Modeling the Formation of the Family of the Dwarf Planet Haumea}",
      journal = {\aj},
         year = 2019,
        month = jun,
       volume = {157},
       number = {6},
          eid = {230},
        pages = {230},
          doi = {10.3847/1538-3881/ab19c4},
archivePrefix = {arXiv},
       eprint = {1904.00038},
 primaryClass = {astro-ph.EP},
       adsurl = {https://ui.adsabs.harvard.edu/abs/2019AJ....157..230P}
}

@ARTICLE{Sheppard2019,
       author = {{Sheppard}, Scott S. and {Trujillo}, Chadwick A. and {Tholen}, David J. and {Kaib}, Nathan},
        title = "{A New High Perihelion Trans-Plutonian Inner Oort Cloud Object: 2015 TG387}",
      journal = {\aj},
         year = 2019,
        month = apr,
       volume = {157},
       number = {4},
          eid = {139},
        pages = {139},
          doi = {10.3847/1538-3881/ab0895},
archivePrefix = {arXiv},
       eprint = {1810.00013},
 primaryClass = {astro-ph.EP},
       adsurl = {https://ui.adsabs.harvard.edu/abs/2019AJ....157..139S}
}

@ARTICLE{Vilenius2018,
       author = {{Vilenius}, E. and {Stansberry}, J. and {M{\"u}ller}, T. and {Mueller}, M. and {Kiss}, C. and {Santos-Sanz}, P. and {Mommert}, M. and {P{\'a}l}, A. and {Lellouch}, E. and {Ortiz}, J.~L. and {Peixinho}, N. and {Thirouin}, A. and {Lykawka}, P.~S. and {Horner}, J. and {Duffard}, R. and {Fornasier}, S. and {Delsanti}, A.},
        title = "{``TNOs are Cool'': A survey of the trans-Neptunian region. XIV. Size/albedo characterization of the Haumea family observed with Herschel and Spitzer}",
      journal = {\aap},
         year = 2018,
        month = oct,
       volume = {618},
          eid = {A136},
        pages = {A136},
          doi = {10.1051/0004-6361/201732564},
archivePrefix = {arXiv},
       eprint = {1904.06333},
 primaryClass = {astro-ph.EP},
       adsurl = {https://ui.adsabs.harvard.edu/abs/2018A&A...618A.136V}
}

@ARTICLE{Becker2018,
       author = {{Becker}, J.~C. and {Khain}, T. and {Hamilton}, S.~J. and {Adams}, F.~C. and {Gerdes}, D.~W. and {Zullo}, L. and {Franson}, K. and {Millholland}, S. and {Bernstein}, G.~M. and {Sako}, M. and {Bernardinelli}, P. and {Napier}, K. and {Markwardt}, L. and {Lin}, Hsing Wen and {Wester}, W. and {Abdalla}, F.~B. and {Allam}, S. and {Annis}, J. and {Avila}, S. and {Bertin}, E. and {Brooks}, D. and {Carnero Rosell}, A. and {Carrasco Kind}, M. and {Carretero}, J. and {Cunha}, C.~E. and {D'Andrea}, C.~B. and {da Costa}, L.~N. and {Davis}, C. and {De Vicente}, J. and {Diehl}, H.~T. and {Doel}, P. and {Eifler}, T.~F. and {Flaugher}, B. and {Fosalba}, P. and {Frieman}, J. and {Garc{\'\i}a-Bellido}, J. and {Gaztanaga}, E. and {Gruen}, D. and {Gruendl}, R.~A. and {Gschwend}, J. and {Gutierrez}, G. and {Hartley}, W.~G. and {Hollowood}, D.~L. and {Honscheid}, K. and {James}, D.~J. and {Kuehn}, K. and {Kuropatkin}, N. and {Maia}, M.~A.~G. and {March}, M. and {Marshall}, J.~L. and {Menanteau}, F. and {Miquel}, R. and {Ogando}, R.~L.~C. and {Plazas}, A.~A. and {Sanchez}, E. and {Scarpine}, V. and {Schindler}, R. and {Sevilla-Noarbe}, I. and {Smith}, M. and {Smith}, R.~C. and {Soares-Santos}, M. and {Sobreira}, F. and {Suchyta}, E. and {Swanson}, M.~E.~C. and {Walker}, A.~R. and {DES Collaboration}},
        title = "{Discovery and Dynamical Analysis of an Extreme Trans-Neptunian Object with a High Orbital Inclination}",
      journal = {\aj},
         year = 2018,
        month = aug,
       volume = {156},
       number = {2},
          eid = {81},
        pages = {81},
          doi = {10.3847/1538-3881/aad042},
archivePrefix = {arXiv},
       eprint = {1805.05355},
 primaryClass = {astro-ph.EP},
       adsurl = {https://ui.adsabs.harvard.edu/abs/2018AJ....156...81B}
}

@ARTICLE{Holler2018,
       author = {{Holler}, Bryan J. and {Milam}, Stefanie N. and {Bauer}, James M. and {Alcock}, Charles and {Bannister}, Michele T. and {Bjoraker}, Gordon L. and {Bodewits}, Dennis and {Bosh}, Amanda S. and {Buie}, Marc W. and {Farnham}, Tony L. and {Haghighipour}, Nader and {Hardersen}, Paul S. and {Harris}, Alan W. and {Hirata}, Christopher M. and {Hsieh}, Henry H. and {Kelley}, Michael S.~P. and {Knight}, Matthew M. and {Kramer}, Emily A. and {Longobardo}, Andrea and {Nixon}, Conor A. and {Palomba}, Ernesto and {Protopapa}, Silvia and {Quick}, Lynnae C. and {Ragozzine}, Darin and {Reddy}, Vishnu and {Rhodes}, Jason D. and {Rivkin}, Andy S. and {Sarid}, Gal and {Sickafoose}, Amanda A. and {Simon}, Amy A. and {Thomas}, Cristina A. and {Trilling}, David E. and {West}, Robert A.},
        title = "{Solar system science with the Wide-Field Infrared Survey Telescope}",
      journal = {Journal of Astronomical Telescopes, Instruments, and Systems},
         year = 2018,
        month = jul,
       volume = {4},
          eid = {034003},
        pages = {034003},
          doi = {10.1117/1.JATIS.4.3.034003},
archivePrefix = {arXiv},
       eprint = {1709.02763},
 primaryClass = {astro-ph.EP},
       adsurl = {https://ui.adsabs.harvard.edu/abs/2018JATIS...4c4003H}
}

@ARTICLE{Yu2018,
       author = {{Yu}, Tze Yeung Mathew and {Murray-Clay}, Ruth and {Volk}, Kathryn},
        title = "{Trans-Neptunian Objects Transiently Stuck in Neptune{\textquoteright}s Mean-motion Resonances: Numerical Simulations of the Current Population}",
      journal = {\aj},
         year = 2018,
        month = jul,
       volume = {156},
       number = {1},
          eid = {33},
        pages = {33},
          doi = {10.3847/1538-3881/aac6cd},
archivePrefix = {arXiv},
       eprint = {1805.08228},
 primaryClass = {astro-ph.EP},
       adsurl = {https://ui.adsabs.harvard.edu/abs/2018AJ....156...33Y}
}

@ARTICLE{Lawler2018,
       author = {{Lawler}, Samantha M. and {Kavelaars}, J.~J. and {Alexandersen}, Mike and {Bannister}, Michele T. and {Gladman}, Brett and {Petit}, Jean-Marc and {Shankman}, Cory},
        title = "{OSSOS: X. How to use a Survey Simulator: Statistical Testing of Dynamical Models Against the Real Kuiper Belt}",
      journal = {Frontiers in Astronomy and Space Sciences},
         year = 2018,
        month = may,
       volume = {5},
          eid = {14},
        pages = {14},
          doi = {10.3389/fspas.2018.00014},
archivePrefix = {arXiv},
       eprint = {1802.00460},
 primaryClass = {astro-ph.EP},
       adsurl = {https://ui.adsabs.harvard.edu/abs/2018FrASS...5...14L}
}

@ARTICLE{Bannister2018,
       author = {{Bannister}, Michele T. and {Gladman}, Brett J. and {Kavelaars}, J.~J. and {Petit}, Jean-Marc and {Volk}, Kathryn and {Chen}, Ying-Tung and {Alexandersen}, Mike and {Gwyn}, Stephen D.~J. and {Schwamb}, Megan E. and {Ashton}, Edward and {Benecchi}, Susan D. and {Cabral}, Nahuel and {Dawson}, Rebekah I. and {Delsanti}, Audrey and {Fraser}, Wesley C. and {Granvik}, Mikael and {Greenstreet}, Sarah and {Guilbert-Lepoutre}, Aur{\'e}lie and {Ip}, Wing-Huen and {Jakubik}, Marian and {Jones}, R. Lynne and {Kaib}, Nathan A. and {Lacerda}, Pedro and {Van Laerhoven}, Christa and {Lawler}, Samantha and {Lehner}, Matthew J. and {Lin}, Hsing Wen and {Lykawka}, Patryk Sofia and {Marsset}, Micha{\"e}l and {Murray-Clay}, Ruth and {Pike}, Rosemary E. and {Rousselot}, Philippe and {Shankman}, Cory and {Thirouin}, Audrey and {Vernazza}, Pierre and {Wang}, Shiang-Yu},
        title = "{OSSOS. VII. 800+ Trans-Neptunian Objects{\textemdash}The Complete Data Release}",
      journal = {\apjs},
         year = 2018,
        month = may,
       volume = {236},
       number = {1},
          eid = {18},
        pages = {18},
          doi = {10.3847/1538-4365/aab77a},
archivePrefix = {arXiv},
       eprint = {1805.11740},
 primaryClass = {astro-ph.EP},
       adsurl = {https://ui.adsabs.harvard.edu/abs/2018ApJS..236...18B}
}

@ARTICLE{Bosch2018,
       author = {{Bosch}, James and {Armstrong}, Robert and {Bickerton}, Steven and {Furusawa}, Hisanori and {Ikeda}, Hiroyuki and {Koike}, Michitaro and {Lupton}, Robert and {Mineo}, Sogo and {Price}, Paul and {Takata}, Tadafumi and {Tanaka}, Masayuki and {Yasuda}, Naoki and {AlSayyad}, Yusra and {Becker}, Andrew C. and {Coulton}, William and {Coupon}, Jean and {Garmilla}, Jose and {Huang}, Song and {Krughoff}, K. Simon and {Lang}, Dustin and {Leauthaud}, Alexie and {Lim}, Kian-Tat and {Lust}, Nate B. and {MacArthur}, Lauren A. and {Mandelbaum}, Rachel and {Miyatake}, Hironao and {Miyazaki}, Satoshi and {Murata}, Ryoma and {More}, Surhud and {Okura}, Yuki and {Owen}, Russell and {Swinbank}, John D. and {Strauss}, Michael A. and {Yamada}, Yoshihiko and {Yamanoi}, Hitomi},
        title = "{The Hyper Suprime-Cam software pipeline}",
      journal = {\pasj},
         year = 2018,
        month = jan,
       volume = {70},
          eid = {S5},
        pages = {S5},
          doi = {10.1093/pasj/psx080},
archivePrefix = {arXiv},
       eprint = {1705.06766},
 primaryClass = {astro-ph.IM},
       adsurl = {https://ui.adsabs.harvard.edu/abs/2018PASJ...70S...5B}
}

@ARTICLE{Shankman2017,
       author = {{Shankman}, Cory and {Kavelaars}, J.~J. and {Bannister}, Michele T. and {Gladman}, Brett J. and {Lawler}, Samantha M. and {Chen}, Ying-Tung and {Jakubik}, Marian and {Kaib}, Nathan and {Alexandersen}, Mike and {Gwyn}, Stephen D.~J. and {Petit}, Jean-Marc and {Volk}, Kathryn},
        title = "{OSSOS. VI. Striking Biases in the Detection of Large Semimajor Axis Trans-Neptunian Objects}",
      journal = {\aj},
         year = 2017,
        month = aug,
       volume = {154},
       number = {2},
          eid = {50},
        pages = {50},
          doi = {10.3847/1538-3881/aa7aed},
archivePrefix = {arXiv},
       eprint = {1706.05348},
 primaryClass = {astro-ph.EP},
       adsurl = {https://ui.adsabs.harvard.edu/abs/2017AJ....154...50S}
}

@ARTICLE{Petit2017,
       author = {{Petit}, J.-M. and {Kavelaars}, J.~J. and {Gladman}, B.~J. and {Jones}, R.~L. and {Parker}, J. Wm. and {Bieryla}, A. and {Van Laerhoven}, C. and {Pike}, R.~E. and {Nicholson}, P. and {Ashby}, M.~L.~N. and {Lawler}, S.~M.},
        title = "{The Canada-France Ecliptic Plane Survey (CFEPS){\textemdash}High-latitude Component}",
      journal = {\aj},
         year = 2017,
        month = may,
       volume = {153},
       number = {5},
          eid = {236},
        pages = {236},
          doi = {10.3847/1538-3881/aa6aa5},
archivePrefix = {arXiv},
       eprint = {1608.02873},
 primaryClass = {astro-ph.EP},
       adsurl = {https://ui.adsabs.harvard.edu/abs/2017AJ....153..236P}
}

@ARTICLE{Alexandersen2016,
       author = {{Alexandersen}, Mike and {Gladman}, Brett and {Kavelaars}, J.~J. and {Petit}, Jean-Marc and {Gwyn}, Stephen D.~J. and {Shankman}, Cory J. and {Pike}, Rosemary E.},
        title = "{A Carefully Characterized and Tracked Trans-Neptunian Survey: The Size distribution of the Plutinos and the Number of Neptunian Trojans}",
      journal = {\aj},
         year = 2016,
        month = nov,
       volume = {152},
       number = {5},
          eid = {111},
        pages = {111},
          doi = {10.3847/0004-6256/152/5/111},
archivePrefix = {arXiv},
       eprint = {1411.7953},
 primaryClass = {astro-ph.EP},
       adsurl = {https://ui.adsabs.harvard.edu/abs/2016AJ....152..111A}
}

@ARTICLE{Kaib2016,
       author = {{Kaib}, Nathan A. and {Sheppard}, Scott S.},
        title = "{Tracking Neptune{\textquoteright}s Migration History through High-perihelion Resonant Trans-Neptunian Objects}",
      journal = {\aj},
         year = 2016,
        month = nov,
       volume = {152},
       number = {5},
          eid = {133},
        pages = {133},
          doi = {10.3847/0004-6256/152/5/133},
archivePrefix = {arXiv},
       eprint = {1607.01777},
 primaryClass = {astro-ph.EP},
       adsurl = {https://ui.adsabs.harvard.edu/abs/2016AJ....152..133K}
}

@ARTICLE{Bannister2016,
       author = {{Bannister}, Michele T. and {Kavelaars}, J.~J. and {Petit}, Jean-Marc and {Gladman}, Brett J. and {Gwyn}, Stephen D.~J. and {Chen}, Ying-Tung and {Volk}, Kathryn and {Alexandersen}, Mike and {Benecchi}, Susan D. and {Delsanti}, Audrey and {Fraser}, Wesley C. and {Granvik}, Mikael and {Grundy}, Will M. and {Guilbert-Lepoutre}, Aur{\'e}lie and {Hestroffer}, Daniel and {Ip}, Wing-Huen and {Jakubik}, Marian and {Jones}, R. Lynne and {Kaib}, Nathan and {Kavelaars}, Catherine F. and {Lacerda}, Pedro and {Lawler}, Samantha and {Lehner}, Matthew J. and {Lin}, Hsing Wen and {Lister}, Tim and {Lykawka}, Patryk Sofia and {Monty}, Stephanie and {Marsset}, Michael and {Murray-Clay}, Ruth and {Noll}, Keith S. and {Parker}, Alex and {Pike}, Rosemary E. and {Rousselot}, Philippe and {Rusk}, David and {Schwamb}, Megan E. and {Shankman}, Cory and {Sicardy}, Bruno and {Vernazza}, Pierre and {Wang}, Shiang-Yu},
        title = "{The Outer Solar System Origins Survey. I. Design and First-quarter Discoveries}",
      journal = {\aj},
         year = 2016,
        month = sep,
       volume = {152},
       number = {3},
          eid = {70},
        pages = {70},
          doi = {10.3847/0004-6256/152/3/70},
archivePrefix = {arXiv},
       eprint = {1511.02895},
 primaryClass = {astro-ph.EP},
       adsurl = {https://ui.adsabs.harvard.edu/abs/2016AJ....152...70B}
}

@ARTICLE{Volk2016,
       author = {{Volk}, Kathryn and {Murray-Clay}, Ruth and {Gladman}, Brett and {Lawler}, Samantha and {Bannister}, Michele T. and {Kavelaars}, J.~J. and {Petit}, Jean-Marc and {Gwyn}, Stephen and {Alexandersen}, Mike and {Chen}, Ying-Tung and {Lykawka}, Patryk Sofia and {Ip}, Wing and {Lin}, Hsing Wen},
        title = "{OSSOS III{\textemdash}Resonant Trans-Neptunian Populations: Constraints from the first quarter of the Outer Solar System Origins Survey}",
      journal = {\aj},
         year = 2016,
        month = jul,
       volume = {152},
       number = {1},
          eid = {23},
        pages = {23},
          doi = {10.3847/0004-6256/152/1/23},
archivePrefix = {arXiv},
       eprint = {1604.08177},
 primaryClass = {astro-ph.EP},
       adsurl = {https://ui.adsabs.harvard.edu/abs/2016AJ....152...23V}
}

@ARTICLE{Fraser2016,
       author = {{Fraser}, Wesley and {Alexandersen}, Mike and {Schwamb}, Megan E. and {Marsset}, Micha{\"e}l and {Pike}, Rosemary E. and {Kavelaars}, J.~J. and {Bannister}, Michele T. and {Benecchi}, Susan and {Delsanti}, Audrey},
        title = "{TRIPPy: Trailed Image Photometry in Python}",
      journal = {\aj},
         year = 2016,
        month = jun,
       volume = {151},
       number = {6},
          eid = {158},
        pages = {158},
          doi = {10.3847/0004-6256/151/6/158},
archivePrefix = {arXiv},
       eprint = {1604.00031},
 primaryClass = {astro-ph.IM},
       adsurl = {https://ui.adsabs.harvard.edu/abs/2016AJ....151..158F}
}

@INCOLLECTION{Kluyver2016,
       author = {{Kluyver}, Thomas and {Ragan-Kelley}, Benjain and {P{\'e}rez}, Fernando and {Granger}, Brian and {Bussonnier}, Matthias and {Frederic}, Jonathan and {Kelley}, Kyle and {Hamrick}, Jessica and {Grout}, Jason and {Corlay}, Sylvain and {Ivanov}, Paul and {Avila}, Dami{\'a}n and {Abdalla}, Safia and {Willing}, Carol and {Jupyter Development Team}},
        title = "{Jupyter Notebooks{\textemdash}a publishing format for reproducible computational workflows}",
    booktitle = {IOS Press},
         year = 2016,
        pages = {87-90},
          doi = {10.3233/978-1-61499-649-1-87},
       adsurl = {https://ui.adsabs.harvard.edu/abs/2016ppap.book...87K}
}

@software{Droettboom2016,
       author = {{Droettboom}, Michael and {Hunter}, John and {Caswell}, Thomas A and {Firing}, Eric and {Hedegaard Nielsen}, Jens and {Elson}, Phil and {Root}, Benjamin and {Dale}, Darren and {Lee}, Jae-Joon and {Sepp{\"a}nen}, Jouni K. and {McDougall}, Damon and {Straw}, Andrew and {May}, Ryan and {Varoquaux}, Nelle and {Yu}, Tony S and {Ma}, Eric and {Moad}, Charlie and {Silvester}, Steven and {Gohlke}, Christoph and {W{\"u}rtz}, Peter and {Hisch}, Thomas and {Ariza}, Federico and {Cimarron} and {Thomas}, Ian and {Evans}, James and {Ivanov}, Paul and {Whitaker}, Jeff and {Hobson}, Paul and {Mdehoon} and {Giuca}, Matt},
        title = "{matplotlib: matplotlib v1.5.1}",
         year = 2016,
        month = jan,
          eid = {10.5281/zenodo.44579},
          doi = {10.5281/zenodo.44579},
      version = {v1.5.1},
    publisher = {Zenodo},
       adsurl = {https://ui.adsabs.harvard.edu/abs/2016zndo.....44579D}
}

@ARTICLE{Dones2015,
       author = {{Dones}, Luke and {Brasser}, Ramon and {Kaib}, Nathan and {Rickman}, Hans},
        title = "{Origin and Evolution of the Cometary Reservoirs}",
      journal = {\ssr},
         year = 2015,
        month = dec,
       volume = {197},
       number = {1-4},
        pages = {191-269},
          doi = {10.1007/s11214-015-0223-2},
       adsurl = {https://ui.adsabs.harvard.edu/abs/2015SSRv..197..191D}
}

@ARTICLE{Spergel2015,
       author = {{Spergel}, D. and {Gehrels}, N. and {Baltay}, C. and {Bennett}, D. and {Breckinridge}, J. and {Donahue}, M. and {Dressler}, A. and {Gaudi}, B.~S. and {Greene}, T. and {Guyon}, O. and {Hirata}, C. and {Kalirai}, J. and {Kasdin}, N.~J. and {Macintosh}, B. and {Moos}, W. and {Perlmutter}, S. and {Postman}, M. and {Rauscher}, B. and {Rhodes}, J. and {Wang}, Y. and {Weinberg}, D. and {Benford}, D. and {Hudson}, M. and {Jeong}, W.-S. and {Mellier}, Y. and {Traub}, W. and {Yamada}, T. and {Capak}, P. and {Colbert}, J. and {Masters}, D. and {Penny}, M. and {Savransky}, D. and {Stern}, D. and {Zimmerman}, N. and {Barry}, R. and {Bartusek}, L. and {Carpenter}, K. and {Cheng}, E. and {Content}, D. and {Dekens}, F. and {Demers}, R. and {Grady}, K. and {Jackson}, C. and {Kuan}, G. and {Kruk}, J. and {Melton}, M. and {Nemati}, B. and {Parvin}, B. and {Poberezhskiy}, I. and {Peddie}, C. and {Ruffa}, J. and {Wallace}, J.~K. and {Whipple}, A. and {Wollack}, E. and {Zhao}, F.},
        title = "{Wide-Field InfrarRed Survey Telescope-Astrophysics Focused Telescope Assets WFIRST-AFTA 2015 Report}",
      journal = {arXiv e-prints},
         year = 2015,
        month = mar,
          eid = {arXiv:1503.03757},
        pages = {arXiv:1503.03757},
          doi = {10.48550/arXiv.1503.03757},
archivePrefix = {arXiv},
       eprint = {1503.03757},
 primaryClass = {astro-ph.IM},
       adsurl = {https://ui.adsabs.harvard.edu/abs/2015arXiv150303757S}
}

@ARTICLE{Brasil2014,
       author = {{Brasil}, P.~I.~O. and {Gomes}, R.~S. and {Soares}, J.~S.},
        title = "{Dynamical formation of detached trans-Neptunian objects close to the 2:5 and 1:3 mean motion resonances with Neptune}",
      journal = {\aap},
         year = 2014,
        month = apr,
       volume = {564},
          eid = {A44},
        pages = {A44},
          doi = {10.1051/0004-6361/201322041},
archivePrefix = {arXiv},
       eprint = {1405.3249},
 primaryClass = {astro-ph.EP},
       adsurl = {https://ui.adsabs.harvard.edu/abs/2014A&A...564A..44B}
}

@ARTICLE{Lawler2013,
       author = {{Lawler}, S.~M. and {Gladman}, B.},
        title = "{Plutino Detection Biases, Including the Kozai Resonance}",
      journal = {\aj},
         year = 2013,
        month = jul,
       volume = {146},
       number = {1},
          eid = {6},
        pages = {6},
          doi = {10.1088/0004-6256/146/1/6},
archivePrefix = {arXiv},
       eprint = {1305.1662},
 primaryClass = {astro-ph.EP},
       adsurl = {https://ui.adsabs.harvard.edu/abs/2013AJ....146....6L}
}

@ARTICLE{Brasser2013,
       author = {{Brasser}, R. and {Morbidelli}, A.},
        title = "{Oort cloud and Scattered Disc formation during a late dynamical instability in the Solar System}",
      journal = {\icarus},
         year = 2013,
        month = jul,
       volume = {225},
       number = {1},
        pages = {40-49},
          doi = {10.1016/j.icarus.2013.03.012},
archivePrefix = {arXiv},
       eprint = {1303.3098},
 primaryClass = {astro-ph.EP},
       adsurl = {https://ui.adsabs.harvard.edu/abs/2013Icar..225...40B}
}

@ARTICLE{Magnier2013,
       author = {{Magnier}, E.~A. and {Schlafly}, E. and {Finkbeiner}, D. and {Juric}, M. and {Tonry}, J.~L. and {Burgett}, W.~S. and {Chambers}, K.~C. and {Flewelling}, H.~A. and {Kaiser}, N. and {Kudritzki}, R.-P. and {Morgan}, J.~S. and {Price}, P.~A. and {Sweeney}, W.~E. and {Stubbs}, C.~W.},
        title = "{The Pan-STARRS 1 Photometric Reference Ladder, Release 12.01}",
      journal = {\apjs},
         year = 2013,
        month = apr,
       volume = {205},
       number = {2},
          eid = {20},
        pages = {20},
          doi = {10.1088/0067-0049/205/2/20},
archivePrefix = {arXiv},
       eprint = {1303.3634},
 primaryClass = {astro-ph.IM},
       adsurl = {https://ui.adsabs.harvard.edu/abs/2013ApJS..205...20M}
}

@ARTICLE{Schlafly2012,
       author = {{Schlafly}, E.~F. and {Finkbeiner}, D.~P. and {Juri{\'c}}, M. and {Magnier}, E.~A. and {Burgett}, W.~S. and {Chambers}, K.~C. and {Grav}, T. and {Hodapp}, K.~W. and {Kaiser}, N. and {Kudritzki}, R.-P. and {Martin}, N.~F. and {Morgan}, J.~S. and {Price}, P.~A. and {Rix}, H.-W. and {Stubbs}, C.~W. and {Tonry}, J.~L. and {Wainscoat}, R.~J.},
        title = "{Photometric Calibration of the First 1.5 Years of the Pan-STARRS1 Survey}",
      journal = {\apj},
         year = 2012,
        month = sep,
       volume = {756},
       number = {2},
          eid = {158},
        pages = {158},
          doi = {10.1088/0004-637X/756/2/158},
archivePrefix = {arXiv},
       eprint = {1201.2208},
 primaryClass = {astro-ph.IM},
       adsurl = {https://ui.adsabs.harvard.edu/abs/2012ApJ...756..158S}
}

@ARTICLE{Gallardo2012,
       author = {{Gallardo}, Tabar{\'e} and {Hugo}, Gast{\'o}n and {Pais}, Pablo},
        title = "{Survey of Kozai dynamics beyond Neptune}",
      journal = {\icarus},
         year = 2012,
        month = aug,
       volume = {220},
       number = {2},
        pages = {392-403},
          doi = {10.1016/j.icarus.2012.05.025},
archivePrefix = {arXiv},
       eprint = {1205.4935},
 primaryClass = {astro-ph.EP},
       adsurl = {https://ui.adsabs.harvard.edu/abs/2012Icar..220..392G}
}

@ARTICLE{Gwyn2012,
       author = {{Gwyn}, Stephen D.~J. and {Hill}, Norman and {Kavelaars}, J.~J.},
        title = "{SSOS: A Moving-Object Image Search Tool for Asteroid Precovery}",
      journal = {\pasp},
         year = 2012,
        month = jun,
       volume = {124},
       number = {916},
        pages = {579},
          doi = {10.1086/666462},
archivePrefix = {arXiv},
       eprint = {1204.4182},
 primaryClass = {astro-ph.EP},
       adsurl = {https://ui.adsabs.harvard.edu/abs/2012PASP..124..579G}
}

@ARTICLE{Tonry2012,
       author = {{Tonry}, J.~L. and {Stubbs}, C.~W. and {Lykke}, K.~R. and {Doherty}, P. and {Shivvers}, I.~S. and {Burgett}, W.~S. and {Chambers}, K.~C. and {Hodapp}, K.~W. and {Kaiser}, N. and {Kudritzki}, R.-P. and {Magnier}, E.~A. and {Morgan}, J.~S. and {Price}, P.~A. and {Wainscoat}, R.~J.},
        title = "{The Pan-STARRS1 Photometric System}",
      journal = {\apj},
         year = 2012,
        month = may,
       volume = {750},
       number = {2},
          eid = {99},
        pages = {99},
          doi = {10.1088/0004-637X/750/2/99},
archivePrefix = {arXiv},
       eprint = {1203.0297},
 primaryClass = {astro-ph.IM},
       adsurl = {https://ui.adsabs.harvard.edu/abs/2012ApJ...750...99T}
}

@ARTICLE{Rein2012,
       author = {{Rein}, H. and {Liu}, S.-F.},
        title = "{REBOUND: an open-source multi-purpose N-body code for collisional dynamics}",
      journal = {\aap},
         year = 2012,
        month = jan,
       volume = {537},
          eid = {A128},
        pages = {A128},
          doi = {10.1051/0004-6361/201118085},
archivePrefix = {arXiv},
       eprint = {1110.4876},
 primaryClass = {astro-ph.EP},
       adsurl = {https://ui.adsabs.harvard.edu/abs/2012A&A...537A.128R}
}

@ARTICLE{Petit2011,
       author = {{Petit}, J.-M. and {Kavelaars}, J.~J. and {Gladman}, B.~J. and {Jones}, R.~L. and {Parker}, J. Wm. and {Van Laerhoven}, C. and {Nicholson}, P. and {Mars}, G. and {Rousselot}, P. and {Mousis}, O. and {Marsden}, B. and {Bieryla}, A. and {Taylor}, M. and {Ashby}, M.~L.~N. and {Benavidez}, P. and {Campo Bagatin}, A. and {Bernabeu}, G.},
        title = "{The Canada-France Ecliptic Plane Survey{\textemdash}Full Data Release: The Orbital Structure of the Kuiper Belt}",
      journal = {\aj},
         year = 2011,
        month = oct,
       volume = {142},
       number = {4},
          eid = {131},
        pages = {131},
          doi = {10.1088/0004-6256/142/4/131},
archivePrefix = {arXiv},
       eprint = {1108.4836},
 primaryClass = {astro-ph.EP},
       adsurl = {https://ui.adsabs.harvard.edu/abs/2011AJ....142..131P}
}

@ARTICLE{Dressler2011,
       author = {{Dressler}, Alan and {Bigelow}, Bruce and {Hare}, Tyson and {Sutin}, Brian and {Thompson}, Ian and {Burley}, Greg and {Epps}, Harland and {Oemler}, Jr., Augustus and {Bagish}, Alan and {Birk}, Christoph and {Clardy}, Ken and {Gunnels}, Steve and {Kelson}, Daniel and {Shectman}, Stephen and {Osip}, David},
        title = "{IMACS: The Inamori-Magellan Areal Camera and Spectrograph on Magellan-Baade}",
      journal = {\pasp},
         year = 2011,
        month = mar,
       volume = {123},
       number = {901},
        pages = {288},
          doi = {10.1086/658908},
       adsurl = {https://ui.adsabs.harvard.edu/abs/2011PASP..123..288D}
}

@ARTICLE{Sheppard2010,
       author = {{Sheppard}, Scott S. and {Trujillo}, Chadwick A.},
        title = "{The Size Distribution of the Neptune Trojans and the Missing Intermediate-sized Planetesimals}",
      journal = {\apjl},
         year = 2010,
        month = nov,
       volume = {723},
       number = {2},
        pages = {L233-L237},
          doi = {10.1088/2041-8205/723/2/L233},
archivePrefix = {arXiv},
       eprint = {1009.5990},
 primaryClass = {astro-ph.EP},
       adsurl = {https://ui.adsabs.harvard.edu/abs/2010ApJ...723L.233S}
}

@ARTICLE{Lang2010,
       author = {{Lang}, Dustin and {Hogg}, David W. and {Mierle}, Keir and {Blanton}, Michael and {Roweis}, Sam},
        title = "{Astrometry.net: Blind Astrometric Calibration of Arbitrary Astronomical Images}",
      journal = {\aj},
         year = 2010,
        month = may,
       volume = {139},
       number = {5},
        pages = {1782-1800},
          doi = {10.1088/0004-6256/139/5/1782},
archivePrefix = {arXiv},
       eprint = {0910.2233},
 primaryClass = {astro-ph.IM},
       adsurl = {https://ui.adsabs.harvard.edu/abs/2010AJ....139.1782L}
}

@ARTICLE{Snodgrass2010,
       author = {{Snodgrass}, C. and {Carry}, B. and {Dumas}, C. and {Hainaut}, O.},
        title = "{Characterisation of candidate members of (136108) Haumea's family}",
      journal = {\aap},
         year = 2010,
        month = feb,
       volume = {511},
          eid = {A72},
        pages = {A72},
          doi = {10.1051/0004-6361/200913031},
archivePrefix = {arXiv},
       eprint = {0912.3171},
 primaryClass = {astro-ph.EP},
       adsurl = {https://ui.adsabs.harvard.edu/abs/2010A&A...511A..72S}
}

@INPROCEEDINGS{Gaudet2009,
       author = {{Gaudet}, S. and {Dowler}, P. and {Goliath}, S. and {Hill}, N. and {Kavelaars}, J.~J. and {Peddle}, M. and {Pritchet}, C. and {Schade}, D.},
        title = "{The Canadian Advanced Network For Astronomical Research}",
    booktitle = {Astronomical Data Analysis Software and Systems XVIII},
         year = 2009,
       editor = {{Bohlender}, D.~A. and {Durand}, D. and {Dowler}, P.},
       series = {Astronomical Society of the Pacific Conference Series},
       volume = {411},
        month = sep,
        pages = {185},
       adsurl = {https://ui.adsabs.harvard.edu/abs/2009ASPC..411..185G}
}

@ARTICLE{Gladman2009,
       author = {{Gladman}, Brett J. and {Davis}, Donald R. and {Neese}, Carol and {Jedicke}, Robert and {Williams}, Gareth and {Kavelaars}, J.~J. and {Petit}, Jean-Marc and {Scholl}, Hans and {Holman}, Matthew and {Warrington}, Ben and {Esquerdo}, Gil and {Tricarico}, Pasquale},
        title = "{On the asteroid belt's orbital and size distribution}",
      journal = {\icarus},
         year = 2009,
        month = jul,
       volume = {202},
       number = {1},
        pages = {104-118},
          doi = {10.1016/j.icarus.2009.02.012},
       adsurl = {https://ui.adsabs.harvard.edu/abs/2009Icar..202..104G}
}

@ARTICLE{Kaib2008,
       author = {{Kaib}, Nathan A. and {Quinn}, Thomas},
        title = "{The formation of the Oort cloud in open cluster environments}",
      journal = {\icarus},
         year = 2008,
        month = sep,
       volume = {197},
       number = {1},
        pages = {221-238},
          doi = {10.1016/j.icarus.2008.03.020},
archivePrefix = {arXiv},
       eprint = {0707.4515},
 primaryClass = {astro-ph},
       adsurl = {https://ui.adsabs.harvard.edu/abs/2008Icar..197..221K}
}

@ARTICLE{Gwyn2008,
       author = {{Gwyn}, Stephen D.~J.},
        title = "{MegaPipe: The MegaCam Image Stacking Pipeline at the Canadian Astronomical Data Centre}",
      journal = {\pasp},
         year = 2008,
        month = feb,
       volume = {120},
       number = {864},
        pages = {212},
          doi = {10.1086/526794},
archivePrefix = {arXiv},
       eprint = {0710.0370},
 primaryClass = {astro-ph},
       adsurl = {https://ui.adsabs.harvard.edu/abs/2008PASP..120..212G}
}

@INCOLLECTION{Gladman2008,
       author = {{Gladman}, B. and {Marsden}, B.~G. and {Vanlaerhoven}, C.},
        title = "{Nomenclature in the Outer Solar System}",
    booktitle = {The Solar System Beyond Neptune},
         year = 2008,
       editor = {{Barucci}, M.~A. and {Boehnhardt}, H. and {Cruikshank}, D.~P. and {Morbidelli}, A. and {Dotson}, Renee},
        pages = {43-57},
       adsurl = {https://ui.adsabs.harvard.edu/abs/2008ssbn.book...43G}
}

@ARTICLE{Ragozzine2007,
       author = {{Ragozzine}, D. and {Brown}, M.~E.},
        title = "{Candidate Members and Age Estimate of the Family of Kuiper Belt Object 2003 EL61}",
      journal = {\aj},
         year = 2007,
        month = dec,
       volume = {134},
       number = {6},
        pages = {2160-2167},
          doi = {10.1086/522334},
archivePrefix = {arXiv},
       eprint = {0709.0328},
 primaryClass = {astro-ph},
       adsurl = {https://ui.adsabs.harvard.edu/abs/2007AJ....134.2160R}
}

@ARTICLE{Lykawka2007,
       author = {{Lykawka}, Patryk Sofia and {Mukai}, Tadashi},
        title = "{Resonance sticking in the scattered disk}",
      journal = {\icarus},
         year = 2007,
        month = dec,
       volume = {192},
       number = {1},
        pages = {238-247},
          doi = {10.1016/j.icarus.2007.06.007},
archivePrefix = {arXiv},
       eprint = {0707.4301},
 primaryClass = {astro-ph},
       adsurl = {https://ui.adsabs.harvard.edu/abs/2007Icar..192..238L}
}

@ARTICLE{Hunter2007,
       author = {{Hunter}, John D.},
        title = "{Matplotlib: A 2D Graphics Environment}",
      journal = {Computing in Science and Engineering},
         year = 2007,
        month = jan,
       volume = {9},
       number = {3},
        pages = {90-95},
          doi = {10.1109/MCSE.2007.55},
       adsurl = {https://ui.adsabs.harvard.edu/abs/2007CSE.....9...90H}
}

@ARTICLE{Jordi2006,
       author = {{Jordi}, K. and {Grebel}, E.~K. and {Ammon}, K.},
        title = "{Empirical color transformations between SDSS photometry and other photometric systems}",
      journal = {\aap},
         year = 2006,
        month = dec,
       volume = {460},
       number = {1},
        pages = {339-347},
          doi = {10.1051/0004-6361:20066082},
archivePrefix = {arXiv},
       eprint = {astro-ph/0609121},
 primaryClass = {astro-ph},
       adsurl = {https://ui.adsabs.harvard.edu/abs/2006A&A...460..339J}
}

@ARTICLE{Jones2006,
       author = {{Jones}, R.~L. and {Gladman}, B. and {Petit}, J.-M. and {Rousselot}, P. and {Mousis}, O. and {Kavelaars}, J.~J. and {Campo Bagatin}, A. and {Bernabeu}, G. and {Benavidez}, P. and {Parker}, J. Wm. and {Nicholson}, P. and {Holman}, M. and {Grav}, T. and {Doressoundiram}, A. and {Veillet}, C. and {Scholl}, H. and {Mars}, G.},
        title = "{The CFEPS Kuiper Belt Survey: Strategy and presurvey results}",
      journal = {\icarus},
         year = 2006,
        month = dec,
       volume = {185},
       number = {2},
        pages = {508-522},
          doi = {10.1016/j.icarus.2006.07.024},
archivePrefix = {arXiv},
       eprint = {astro-ph/0510826},
 primaryClass = {astro-ph},
       adsurl = {https://ui.adsabs.harvard.edu/abs/2006Icar..185..508J}
}

@ARTICLE{Brasser2006,
       author = {{Brasser}, R. and {Duncan}, M.~J. and {Levison}, H.~F.},
        title = "{Embedded star clusters and the formation of the Oort Cloud}",
      journal = {\icarus},
         year = 2006,
        month = sep,
       volume = {184},
       number = {1},
        pages = {59-82},
          doi = {10.1016/j.icarus.2006.04.010},
       adsurl = {https://ui.adsabs.harvard.edu/abs/2006Icar..184...59B}
}

@ARTICLE{Gallardo2006,
       author = {{Gallardo}, Tabar{\'e}},
        title = "{The occurrence of high-order resonances and Kozai mechanism in the scattered disk}",
      journal = {\icarus},
         year = 2006,
        month = mar,
       volume = {181},
       number = {1},
        pages = {205-217},
          doi = {10.1016/j.icarus.2005.11.011},
       adsurl = {https://ui.adsabs.harvard.edu/abs/2006Icar..181..205G}
}

@ARTICLE{Jester2005,
       author = {{Jester}, Sebastian and {Schneider}, Donald P. and {Richards}, Gordon T. and {Green}, Richard F. and {Schmidt}, Maarten and {Hall}, Patrick B. and {Strauss}, Michael A. and {Vanden Berk}, Daniel E. and {Stoughton}, Chris and {Gunn}, James E. and {Brinkmann}, Jon and {Kent}, Stephen M. and {Smith}, J. Allyn and {Tucker}, Douglas L. and {Yanny}, Brian},
        title = "{The Sloan Digital Sky Survey View of the Palomar-Green Bright Quasar Survey}",
      journal = {\aj},
         year = 2005,
        month = sep,
       volume = {130},
       number = {3},
        pages = {873-895},
          doi = {10.1086/432466},
archivePrefix = {arXiv},
       eprint = {astro-ph/0506022},
 primaryClass = {astro-ph},
       adsurl = {https://ui.adsabs.harvard.edu/abs/2005AJ....130..873J}
}

@ARTICLE{Bilir2005,
       author = {{Bilir}, S. and {Karaali}, S. and {Tun{\c{c}}el}, S.},
        title = "{Absolute magnitudes for late-type dwarf stars for Sloan photometry}",
      journal = {Astronomische Nachrichten},
         year = 2005,
        month = jun,
       volume = {326},
       number = {5},
        pages = {321-331},
          doi = {10.1002/asna.200510358},
archivePrefix = {arXiv},
       eprint = {astro-ph/0503415},
 primaryClass = {astro-ph},
       adsurl = {https://ui.adsabs.harvard.edu/abs/2005AN....326..321B}
}

@ARTICLE{Tsiganis2005,
       author = {{Tsiganis}, K. and {Gomes}, R. and {Morbidelli}, A. and {Levison}, H.~F.},
        title = "{Origin of the orbital architecture of the giant planets of the Solar System}",
      journal = {\nat},
         year = 2005,
        month = may,
       volume = {435},
       number = {7041},
        pages = {459-461},
          doi = {10.1038/nature03539},
       adsurl = {https://ui.adsabs.harvard.edu/abs/2005Natur.435..459T}
}

@ARTICLE{Hook2004,
       author = {{Hook}, I.~M. and {J{\o}rgensen}, Inger and {Allington-Smith}, J.~R. and {Davies}, R.~L. and {Metcalfe}, N. and {Murowinski}, R.~G. and {Crampton}, D.},
        title = "{The Gemini-North Multi-Object Spectrograph: Performance in Imaging, Long-Slit, and Multi-Object Spectroscopic Modes}",
      journal = {\pasp},
         year = 2004,
        month = may,
       volume = {116},
       number = {819},
        pages = {425-440},
          doi = {10.1086/383624},
       adsurl = {https://ui.adsabs.harvard.edu/abs/2004PASP..116..425H}
}

@ARTICLE{Petit2004,
       author = {{Petit}, J.-M. and {Holman}, M. and {Scholl}, H. and {Kavelaars}, J. and {Gladman}, B.},
        title = "{A highly automated moving object detection package}",
      journal = {\mnras},
         year = 2004,
        month = jan,
       volume = {347},
       number = {2},
        pages = {471-480},
          doi = {10.1111/j.1365-2966.2004.07217.x},
       adsurl = {https://ui.adsabs.harvard.edu/abs/2004MNRAS.347..471P}
}

@INPROCEEDINGS{Boulade2003,
       author = {{Boulade}, Olivier and {Charlot}, Xavier and {Abbon}, P. and {Aune}, Stephan and {Borgeaud}, Pierre and {Carton}, Pierre-Henri and {Carty}, M. and {Da Costa}, J. and {Deschamps}, H. and {Desforge}, D. and {Eppell{\'e}}, Dominique and {Gallais}, Pascal and {Gosset}, L. and {Granelli}, Remy and {Gros}, Michel and {de Kat}, Jean and {Loiseau}, Denis and {Ritou}, J.-. and {Rouss{\'e}}, Jean Y. and {Starzynski}, Pierre and {Vignal}, Nicolas and {Vigroux}, Laurent G.},
        title = "{MegaCam: the new Canada-France-Hawaii Telescope wide-field imaging camera}",
    booktitle = {Instrument Design and Performance for Optical/Infrared Ground-based Telescopes},
         year = 2003,
       editor = {{Iye}, Masanori and {Moorwood}, Alan F.~M.},
       series = {Society of Photo-Optical Instrumentation Engineers (SPIE) Conference Series},
       volume = {4841},
        month = mar,
        pages = {72-81},
          doi = {10.1117/12.459890},
       adsurl = {https://ui.adsabs.harvard.edu/abs/2003SPIE.4841...72B}
}

@ARTICLE{Gomes2003,
       author = {{Gomes}, Rodney S.},
        title = "{The origin of the Kuiper Belt high-inclination population}",
      journal = {\icarus},
         year = 2003,
        month = feb,
       volume = {161},
       number = {2},
        pages = {404-418},
          doi = {10.1016/S0019-1035(02)00056-8},
       adsurl = {https://ui.adsabs.harvard.edu/abs/2003Icar..161..404G}
}

@ARTICLE{Boulade1998,
       author = {{Boulade}, O.},
        title = "{Wide field imaging at CFHT: the MEGACAM project}",
      journal = {Experimental Astronomy},
         year = 1998,
        month = jan,
       volume = {8},
       number = {1},
        pages = {25-29},
          doi = {10.1023/A:1008036104161},
       adsurl = {https://ui.adsabs.harvard.edu/abs/1998ExA.....8...25B}
}

@INCOLLECTION{Malhotra1997,
       author = {{Malhotra}, R. and {Williams}, J.~G.},
        title = "{Pluto's Heliocentric Orbit}",
    booktitle = {Pluto and Charon},
         year = 1997,
       editor = {{Stern}, S. Alan and {Tholen}, David J.},
        pages = {127},
       adsurl = {https://ui.adsabs.harvard.edu/abs/1997plch.book..127M}
}

@ARTICLE{Bertin1996,
       author = {{Bertin}, E. and {Arnouts}, S.},
        title = "{SExtractor: Software for source extraction.}",
      journal = {\aaps},
         year = 1996,
        month = jun,
       volume = {117},
        pages = {393-404},
          doi = {10.1051/aas:1996164},
       adsurl = {https://ui.adsabs.harvard.edu/abs/1996A&AS..117..393B}
}

@ARTICLE{Malhotra1993,
       author = {{Malhotra}, Renu},
        title = "{The origin of Pluto's peculiar orbit}",
      journal = {\nat},
         year = 1993,
        month = oct,
       volume = {365},
       number = {6449},
        pages = {819-821},
          doi = {10.1038/365819a0},
       adsurl = {https://ui.adsabs.harvard.edu/abs/1993Natur.365..819M}
}

@ARTICLE{Kozai1985,
       author = {{Kozai}, Y.},
        title = "{Secular Perturbations of Resonant Asteroids}",
      journal = {Celestial Mechanics},
         year = 1985,
        month = may,
       volume = {36},
       number = {1},
        pages = {47-69},
          doi = {10.1007/BF01241042},
       adsurl = {https://ui.adsabs.harvard.edu/abs/1985CeMec..36...47K}
}

@ARTICLE{Lidov1962,
       author = {{Lidov}, M.~L.},
        title = "{The evolution of orbits of artificial satellites of planets under the action of gravitational perturbations of external bodies}",
      journal = {\planss},
         year = 1962,
        month = oct,
       volume = {9},
       number = {10},
        pages = {719-759},
          doi = {10.1016/0032-0633(62)90129-0},
       adsurl = {https://ui.adsabs.harvard.edu/abs/1962P&SS....9..719L}
}

@ARTICLE{Kozai1962,
       author = {{Kozai}, Yoshihide},
        title = "{Secular Perturbations of Asteroids with High Inclination and Eccentricity.}",
      journal = {\aj},
         year = 1962,
        month = jun,
       volume = {67},
        pages = {579},
          doi = {10.1086/108876},
       adsurl = {https://ui.adsabs.harvard.edu/abs/1962AJ.....67R.579K}
}

@ARTICLE{Bernardinelli2025,
       author = {{Bernardinelli}, Pedro H. and {Bernstein}, Gary M. and {Abbott}, T.~M.~C. and {Aguena}, M. and {Allam}, S.~S. and {Brooks}, D. and {Carnero Rosell}, A. and {Carretero}, J. and {da Costa}, L.~N. and {Pereira}, M.~E.~S. and {Davis}, T.~M. and {De Vicente}, J. and {Desai}, S. and {Diehl}, H.~T. and {Doel}, P. and {Everett}, S. and {Flaugher}, B. and {Frieman}, J. and {Garc{\'\i}a-Bellido}, J. and {Gaztanaga}, E. and {Gruendl}, R.~A. and {Gutierrez}, G. and {Herner}, K. and {Hinton}, S.~R. and {Hollowood}, D.~L. and {Honscheid}, K. and {James}, D.~J. and {Kuehn}, K. and {Lahav}, O. and {Lee}, S. and {Marshall}, J.~L. and {Mena-Fern{\'a}ndez}, J. and {Miquel}, R. and {Myles}, J. and {Plazas Malag{\'o}n}, A.~A. and {Samuroff}, S. and {Sanchez}, E. and {Santiago}, B. and {Sevilla-Noarbe}, I. and {Smith}, M. and {Suchyta}, E. and {Tarle}, G. and {Tucker}, D.~L. and {Vikram}, V. and {Walker}, A.~R. and {Weaverdyck}, N. and {DES Collaboration}},
        title = "{Photometry of Outer Solar System Objects from the Dark Energy Survey. II. A Joint Analysis of Trans-Neptunian Absolute Magnitudes, Colors, Light Curves and Dynamics}",
      journal = {\aj},
         year = 2025,
        month = jun,
       volume = {169},
       number = {6},
          eid = {305},
        pages = {305},
          doi = {10.3847/1538-3881/adc459},
archivePrefix = {arXiv},
       eprint = {2501.01551},
 primaryClass = {astro-ph.EP},
       adsurl = {https://ui.adsabs.harvard.edu/abs/2025AJ....169..305B}
}

@ARTICLE{Spencer2026,
       author = {{Spencer}, Dallin and {Volk}, Kat and {Ragozzine}, Darin and {Malhotra}, Renu},
        title = "{Small Body Dynamics with SBDynT: Proper Elements and Chaos Analysis}",
      journal = {AJ in press},
         year = 2026,
        month = mar,
          eid = {arXiv:2603.27099},
        pages = {arXiv:2603.27099},
archivePrefix = {arXiv},
       eprint = {2603.27099},
 primaryClass = {astro-ph.EP},
       adsurl = {https://ui.adsabs.harvard.edu/abs/2026arXiv260327099S}
}

@ARTICLE{DraperInPrep,
       author = {{Draper}, J. and {Lawler, S.} and {MoreLiDOPeople, A.}},
        title = "{Mixed Mode}",
      journal = {In Prep},
         year = 2026
}

@ARTICLE{Pike2017-DetailsResonantStructureNiceModelKuiperBelt-Predictions,
       author = {{Pike}, R.~E. and {Lawler}, S.~M.},
        title = "{Details of Resonant Structures within a Nice Model Kuiper Belt: Predictions for High-perihelion TNO Detections}",
      journal = {\aj},
         year = 2017,
        month = oct,
       volume = {154},
       number = {4},
          eid = {171},
        pages = {171},
          doi = {10.3847/1538-3881/aa8b65},
archivePrefix = {arXiv},
       eprint = {1709.03699},
 primaryClass = {astro-ph.EP},
       adsurl = {https://ui.adsabs.harvard.edu/abs/2017AJ....154..171P}
}

@INPROCEEDINGS{Volk:2025dda,
       author = {{Volk}, Kathryn and {Malhotra}, Renu and {Spencer}, Dallin and {Ragozzine}, Darin},
        title = "{Detailed Dynamical Classification of TNOs with Machine Learning}",
    booktitle = {AAS/Division of Dynamical Astronomy Meeting},
         year = 2025,
       series = {AAS/Division of Dynamical Astronomy Meeting},
       volume = {56},
        month = jun,
          eid = {402.03},
        pages = {402.03},
       adsurl = {https://ui.adsabs.harvard.edu/abs/2025DDA....5640203V}
}

@ARTICLE{Nesvorny2016a,
       author = {{Nesvorn{\'y}}, David and {Vokrouhlick{\'y}}, David},
        title = "{Neptune's Orbital Migration Was Grainy, Not Smooth}",
      journal = {\apj},
         year = 2016,
        month = jul,
       volume = {825},
       number = {2},
          eid = {94},
        pages = {94},
          doi = {10.3847/0004-637X/825/2/94},
archivePrefix = {arXiv},
       eprint = {1602.06988},
 primaryClass = {astro-ph.EP},
       adsurl = {https://ui.adsabs.harvard.edu/abs/2016ApJ...825...94N}
}

@ARTICLE{Nesvorny2016b,
       author = {{Nesvorn{\'y}}, David and {Vokrouhlick{\'y}}, David and {Roig}, Fernando},
        title = "{The Orbital Distribution of Trans-Neptunian Objects Beyond 50 au}",
      journal = {\apjl},
         year = 2016,
        month = aug,
       volume = {827},
       number = {2},
          eid = {L35},
        pages = {L35},
          doi = {10.3847/2041-8205/827/2/L35},
archivePrefix = {arXiv},
       eprint = {1607.08279},
 primaryClass = {astro-ph.EP},
       adsurl = {https://ui.adsabs.harvard.edu/abs/2016ApJ...827L..35N}
}

@ARTICLE{Dziadura_in_prep,
       author = {{Dziadura}, Karolina and {Spoto}, Federica},
         year = {in prep}
}

@article{ad1954,
  author = {Anderson, T. W. and Darling, D. A.},
  copyright = {Copyright � 1954 American Statistical Association},
  description = {two papers about anderson-darling test},
  issn = {01621459},
  journal = {Journal of the American Statistical Association},
  jstor_articletype = {primary_article},
  jstor_formatteddate = {Dec., 1954},
  number = 268,
  pages = {765--769},
  publisher = {American Statistical Association},
  title = {A Test of Goodness of Fit},
  url = {http://www.jstor.org/stable/2281537},
  volume = 49,
  year = 1954
}

@software{Volk2026Zenodo,
  author       = {Volk, Kathryn and
                  Spencer, Dallin},
  title        = {small-body-dynamics/SBDynT: Small Body Dynamics
                   Tool (SBDynT) v1.0.0
                  },
  month        = mar,
  year         = 2026,
  publisher    = {Zenodo},
  version      = {v1.0.0},
  doi          = {10.5281/zenodo.19138321},
  url          = {https://doi.org/10.5281/zenodo.19138321},
}

\end{document}